\documentclass{aastex631}

\usepackage{soul}
\usepackage{amsmath}
\graphicspath{{./}{figures/}}

\received{\today}
\revised{NOT YET}
\accepted{NOT YET}

\submitjournal{ApJ}

\shorttitle{MHD waves in sunspots with changing shape}
\shortauthors{Albidah et al.}

\graphicspath{{./}{figures/}}
\usepackage{graphicx}

\begin{document}

\title{The temporal and spatial evolution of MHD wave modes in sunspots }



\correspondingauthor{Abdulrahman Albidah }
\email{a.albedah@mu.edu.sa}

\author[0000-0001-7314-1347]{A. B. Albidah}

\affiliation{Department of Mathematics, College Of Science, Majmaah University, P.O Box 66, Majmaah 11952, Saudi Arabia}

\author[0000-0002-0893-7346]{V. Fedun}
\affiliation{Plasma Dynamics Group, Department of Automatic Control and Systems Engineering, The University of Sheffield, Mappin Street, Sheffield, S1 3JD, UK}

\author[0000-0003-2220-5042]{A. A. Aldhafeeri}
\affiliation{Mathematics and Statistic Department, Faculty of Science, King Faisal University, Al-Hassa, P.O. Box 400, Hofuf 31982, Saudi Arabia}

\author[0000-0002-3066-7653]{I. Ballai}
\affiliation{Plasma Dynamics Group, School of Mathematics and Statistics, The University of Sheffield, Hicks Building, Hounsfield Road, Sheffield, S3 7RH, UK}

\author[0000-0002-9155-8039]{D. B. Jess}
\affiliation{Astrophysics Research Centre, School of Mathematics and Physics, Queen’s University Belfast, Belfast, BT7 1NN, UK}
\affiliation{Department of Physics and Astronomy, California State University Northridge, Northridge, CA 91330, USA}

\author[0000-0001-8161-4677]{W. Brevis}
\affiliation{School of Engineering, Pontificia Universidad Cat\'{o}lica de Chile, Chile}

\author[0000-0001-7577-0913]{J. Higham}
\affiliation{School of Environmental Sciences, Department of Geography and Planning, University of Liverpool, Roxby Building, Liverpool, L69 7ZT, UK}

\author[0000-0002-5365-7546]{M. Stangalini}
\affiliation{ASI, Italian Space Agency, Via del Politecnico snc,
00133 Rome, Italy}

\author[0000-0001-5414-0197]{S. S. A. Silva}
\affiliation{Plasma Dynamics Group, Department of Automatic Control and Systems Engineering, The University of Sheffield, Mappin Street, Sheffield, S1 3JD, UK}

\author[0000-0002-9901-8723]{C. D. MacBride}
\affiliation{Astrophysics Research Centre, School of Mathematics and Physics, Queen’s University Belfast, Belfast, BT7 1NN, UK}

\author[0000-0002-9546-2368]{G. Verth}
\affiliation{Plasma Dynamics Group, School of Mathematics and Statistics, The University of Sheffield, Hicks Building, Hounsfield Road, Sheffield, S3 7RH, UK}

\begin{abstract}
Through their lifetime  sunspots undergo a change in their area and shape and, as they decay, they fragment into smaller structures. Here, for the first time we analyze the spatial structure of magnetohydrodynamic (MHD) slow body and fast surface modes in observed umbrae as their cross-sectional shape changes. 
The Proper Orthogonal Decomposition (POD) and Dynamic Mode Decomposition (DMD) techniques were used to analyze 3 and 6 hours SDO/HMI time series of Doppler velocities at the photospheric level of  approximately circular and elliptically shaped sunspots. Each time series were divided equally into time intervals, to evidence the change of the sunspots' shape. To identify physical wave modes, the POD/DMD modes were cross-correlated with a slow body mode model using the exact shape of the umbra, whereas the shape obtained by applying a threshold level of the mean intensity for every time interval. Our results show that the spatial structure of MHD modes are affected, even by apparently small changes of the umbral shape, especially in the case of the higher-order modes. For the datasets used in our study, the optimal time intervals to consider the influence of the change in the shape on the observed MHD modes is 37 - 60 minutes. The choice of these intervals is crucial to properly quantify the energy contribution of each wave mode to the power spectrum.   
\end{abstract}


\section{Introduction} \label{sec:intro}

One of the most rapidly evolving fields in solar physics is the study of MHD waves and oscillations in the solar atmosphere. Waves observed with high accuracy in various wavelengths makes it possible to diagnose plasma parameters, e.g. density, temperature, chemical composition, heating/cooling functions etc. and analyze the magnetic field structure and magnitude. Sunspots support a large variety of MHD waves propagating along and across the magnetic field, making them an ideal location for studying MHD waves given their stability, a relatively simple magnetic configuration that can be observed, and long lifetime \citep[to name but a few]{2015SSRv..190..103J, khomenko2015,stangalini2021novel,stangalini2022large, albidah2020RS,albidahApJ}.

The high-resolution observations of sunspots and the dynamics associated with these magnetic features also allow investigation the effect of particular shapes of sunspots' cross-sections on the nature and morphology of waves. Theoretical models provide a relatively accurate behavior of MHD waves in the magnetic waveguides with simple geometry, i.e. approximately circular (for conciseness, from this point on we will refer to that sunspot as 'circular'), and to a lesser extent, approximately elliptical ('elliptical') cross-sections, and observations seem to recover rather well these properties. However, so far very few studies were dedicated to the theoretical study of MHD modes in waveguides with arbitrary shapes. The identification and analysis of MHD modes in the observations are challenging problem, as well. In fact, current studies assume a stationary magnetic waveguide for the duration of observation. While this assumption is valid for short-lived or transient modes, for waves with characteristic lifetime comparable to the rate of change in the shape and dynamical/thermodynamical state of the waveguide, these changes might become important.  
Recently \cite{albidahApJ} have found that higher-order MHD modes and higher harmonics in the sunspots' umbral regions are strongly affected by the shape of the waveguide. Using the Proper Orthogonal Decomposition   \citep[POD,][]{pearson1901liii} and the Dynamic Mode Decomposition \citep[DMD,][]{schmid2010dynamic} analysis of average umbral shapes, the study by \citet{albidahApJ} was one of the first investigations to identify the presence of higher order modes in sunspots.


The POD and DMD techniques have been widely applied to investigate periodic changes in fluids \citep[i.e.][]{murray2007application,rowley2009spectral,bagheri2013koopman, jovanovic2014sparsity,berry2017application}.
POD assumes the orthogonality in space and offers a clear ranking criteria based on the contribution of modes to the total variance of the signal. In contrast, DMD assumes the orthogonality in time, that is different modes cannot have identical frequencies. 
POD and DMD can be used together to reveal temporally and spatially orthogonal features in solar observations by following the approach developed by \cite{higham2018implications}. When applied to investigate MHD waves,  POD and DMD disentangle the signals and separate the oscillating pattern resulting from wave propagation. 
Previous investigation employing the combined POD/DMD techniques \citep[][]{albidah2020RS,albidahApJ} on solar physics data assumed that the umbral boundary is not changing, which is an idealistic approach. In reality, extended high resolution observations show that the boundary between a sunspot umbra and its penumbra is not stationary, instead its shape can change in time. 

In this work, we study the effect of the change in the shape of the umbral boundary on the nature and morphology of waves identified by applying POD and DMD techniques on Doppler velocity data sets of two distinct sunspots. The identified modes are compared with the model of the irregular shape that correspond to the realistic shape of the sunspot, where the model has been described in the work of \citep{albidahApJ,stangalini2022large}. 
\section{Observations}

\begin{figure}
	\plotone{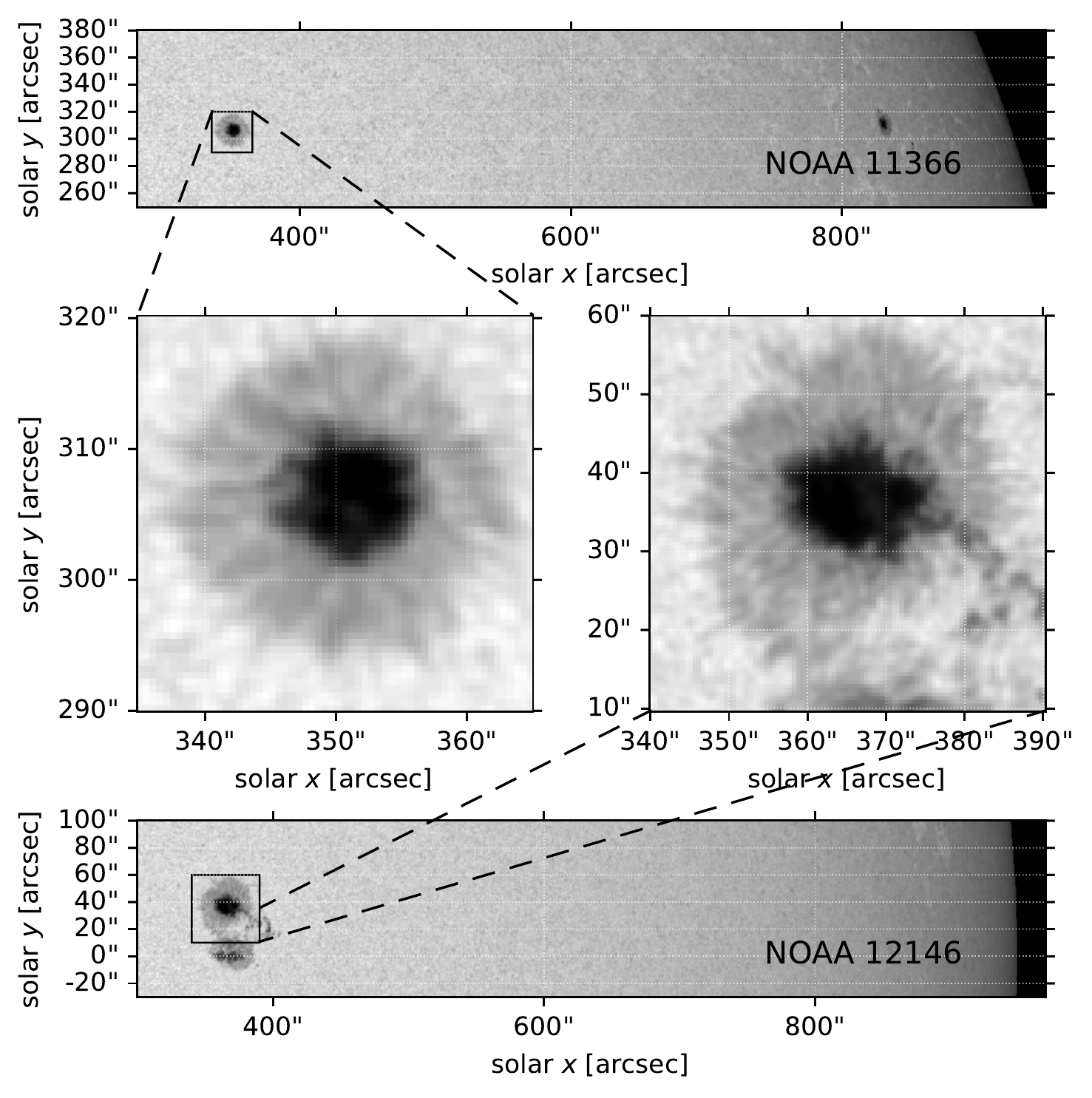}
	\caption{Continuum intensity images from SDO/HMI observations of two different sunspots. The sunspot location on the solar disk during the period of observation is identified with a black rectangle in the top and bottom panels for active region NOAA~11366 and NOAA~12146, respectively. Magnified views of the NOAA~11366 circular cross-section sunspot and the NOAA~12146 elliptical cross-section sunspot are shown in the middle left and right panels, respectively.
\label{fig:continuum}}
\end{figure}

The data used in this study, continuum intensity and Doppler velocity, were obtained from observations by the Helioseismic and Magnetic Imager \citep[HMI;][]{Schou2011} onboard the Solar Dynamics Observatory \citep[SDO;][]{Pesnell2011}. 

The two active regions contain a sunspot with an approximately circular and elliptical cross-sectional shape, respectively.
Data for NOAA~11366 were acquired from 15:00--18:00~UT on 2011 December 10, while data for NOAA~12146 were acquired from 10:00--16:00~UT on 2014 August 24.
The cadence of the measurements was 45~s, which provided 241 images ($\sim$ 3 hours) of NOAA~11366 and 481 images ($\sim$ 6 hours) of NOAA~12146 for both the continuum intensity and the Doppler velocity.
The spatial sampling of all the data products was \(0{\,}.{\!\!}{''}504\) per pixel, which is approximately 356 km on the surface of the Sun.

To account for the rotation of the Sun relative to SDO, the observations of the two active regions were reprojected to the reference frame of an Earth-based observer at 15:00~UT and 10:00~UT respectively, on the same date each active region was observed.
The reprojection was carried out using Version 3.0.0 \citep{sunpy_community2021} of the SunPy open source software package \citep{sunpy_community2020}.
The pixel values of each image are mapped to the new projection and interpolated using a nearest neighbor algorithm.
The motion of the center of the Sun is ignored such that coordinates are always relative to the center of the Sun. Figure~\ref{fig:continuum} shows the sunspots used in this study and identifies their position on the solar disk as would be seen by an Earth-based observer.

\begin{figure}
\centering
\begin{tabular}{ccccc}

\includegraphics[width=50mm]{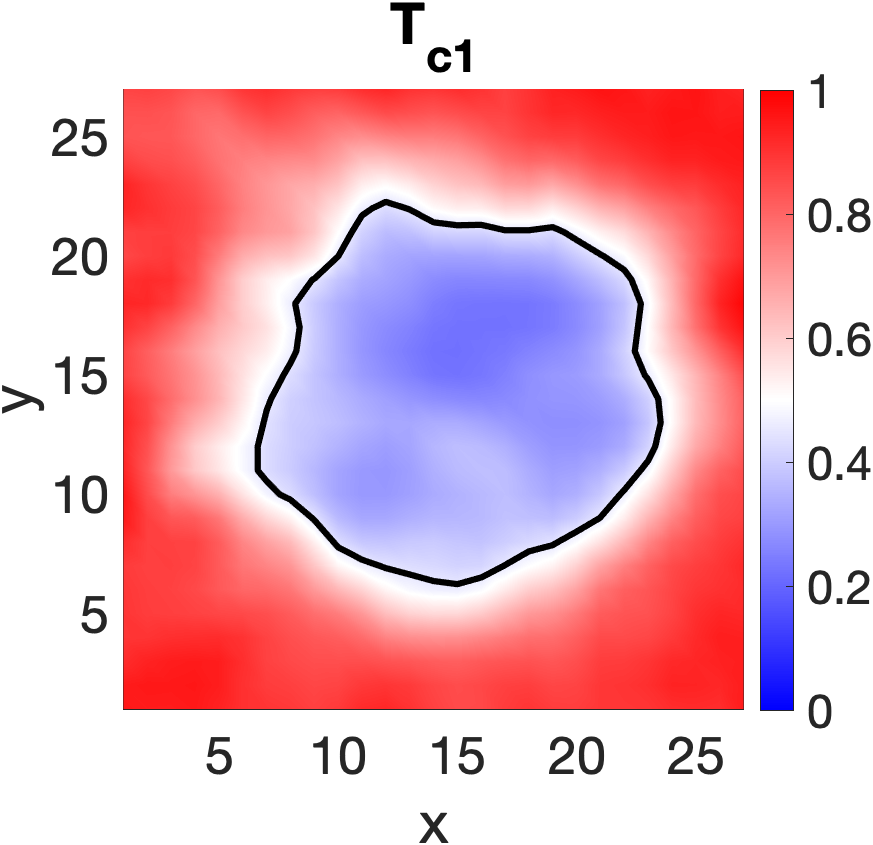}&
\hspace{0em} 

\includegraphics[width=50mm]{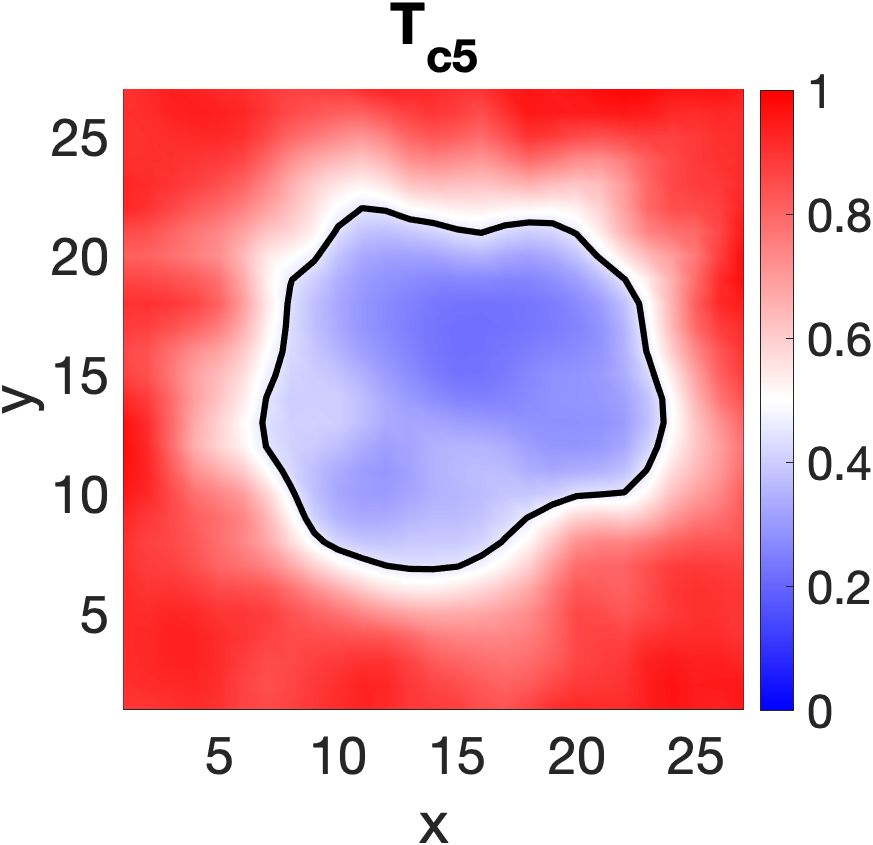}&
\hspace{0em} 

\includegraphics[width=50mm]{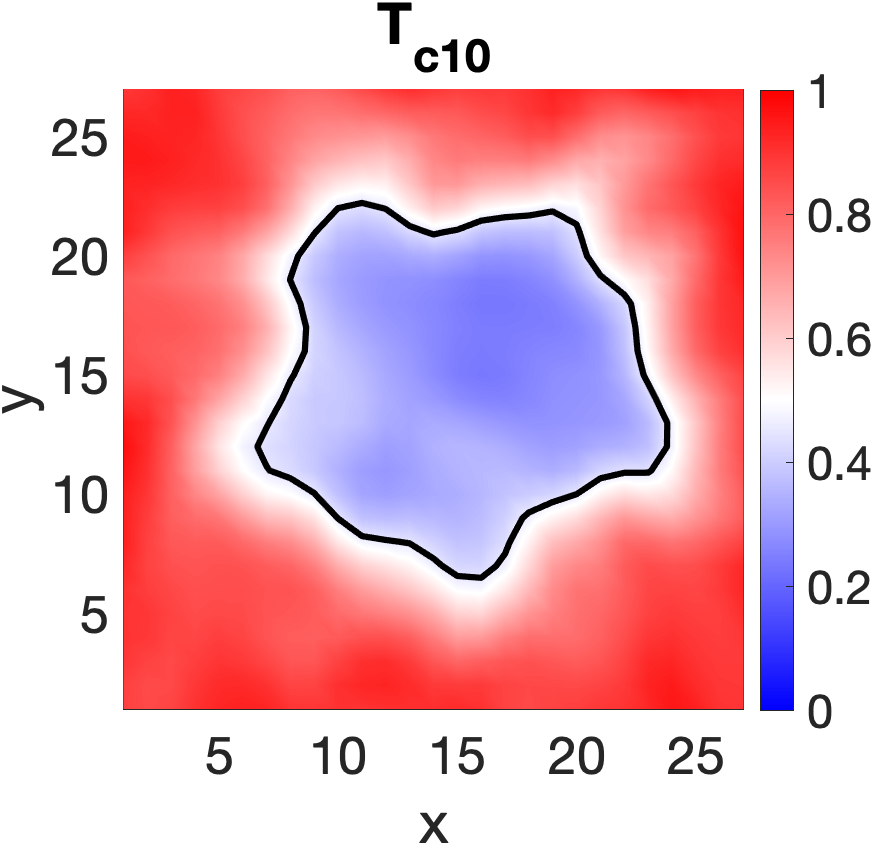}\\

\includegraphics[width=50mm]{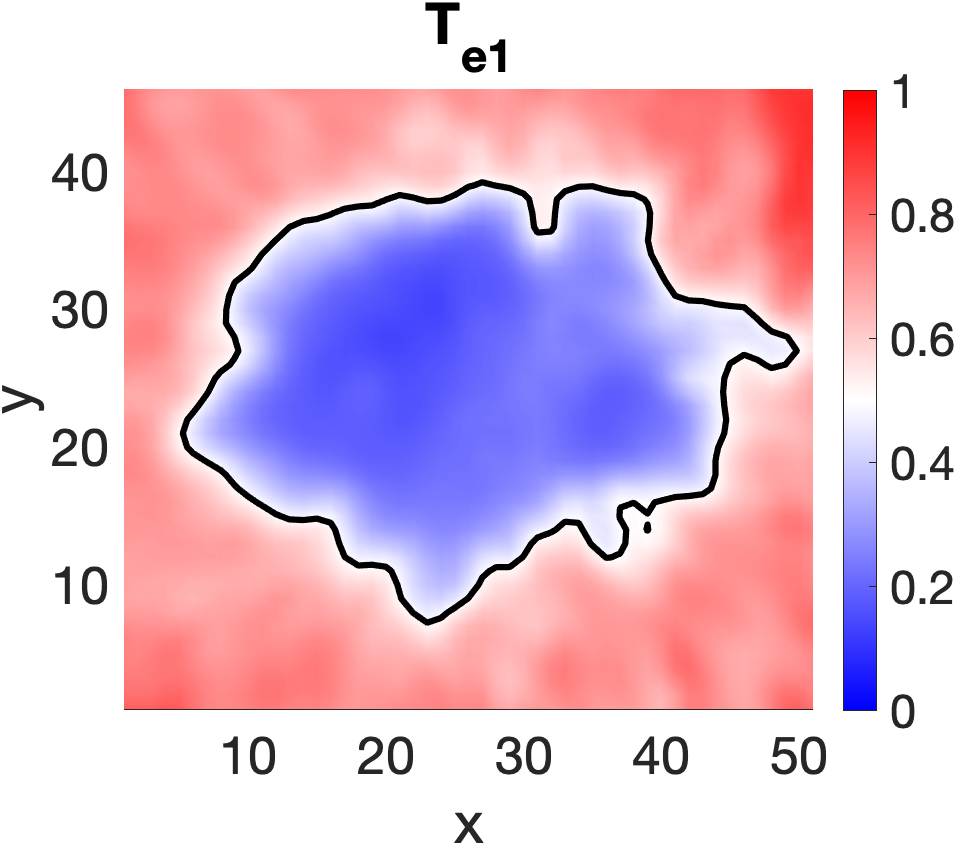}&
\hspace{0em} 
\includegraphics[width=50mm]{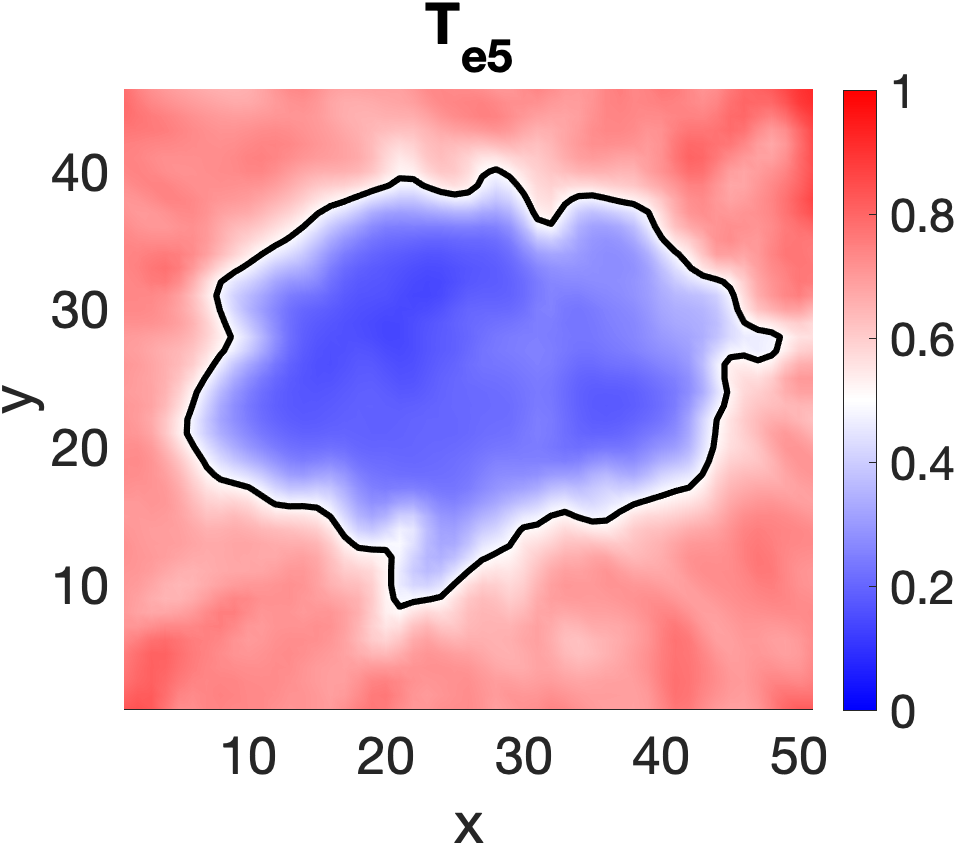}&
\hspace{0em} 
\includegraphics[width=50mm]{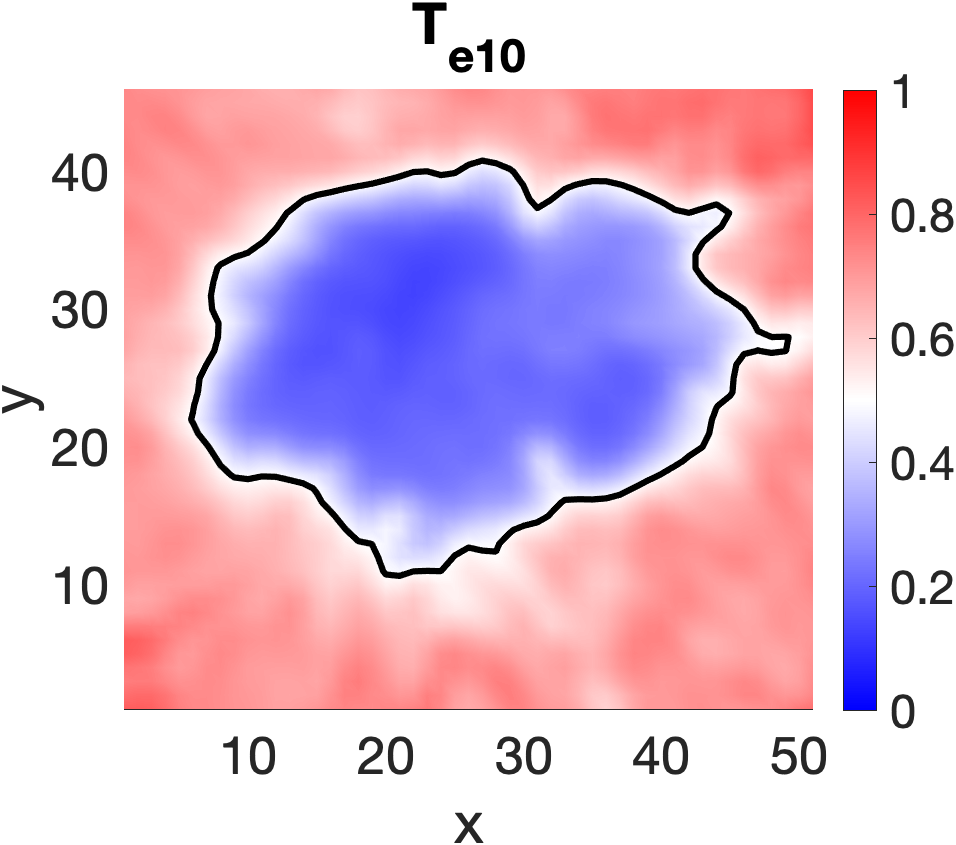}\\
\hspace{0em}

\end{tabular}
\caption{The mean of the continuum intensity time series for three different time intervals for each sunspot are shown for approximately 'circular' sunspot, NOAA 11366, T$_{c1}$, T$_{c5}$ and T$_{c10}$ (upper panels) and the 'elliptical' sunspot, NOAA 12146, T$_{e1}$, T$_{e5}$ and T$_{e10}$ (lower panels). The colour bar displays the magnitude of the mean time series for each time interval. The umbra/penumbra boundary was obtained by taking threshold levels for both sunspots and indicated in black. The spatial scale is given in pixels (one pixel is equal to $0.504^{\prime\prime}$ which is approximately 356 km on the surface of the Sun).} \label{fig: mean}
\end{figure}

\section{Analysis and MHD wave modes identification}

The selected sunspots present changes in their umbral cross-sectional shape over the time interval used for our analysis: 3 and 6 hours for the circular and elliptical sunspots, respectively. In order to facilitate the temporal analysis of the modification in the shape of cross section, we divide the time series data of both sunspots into 10 time intervals (T$_{ci}$ and T$_{ei}$). Here, indexes $c$ and $e$ are used to distinguish between the time intervals of the circular and elliptical sunspots, and $i=1 \dots 10$ which correspond to 10 separate time intervals. In the case of the circular sunspot, every time interval contains 50 images ($\sim$ 37.5 minutes) and it is overlapped with the previous time interval by 20 images ($\sim$ 15 minutes). Thus, the time interval T$_{ci+1}$ is at 20 images ahead of the time interval T$_{ci}$. For the elliptical sunspot, every time interval contains 80 images ($\sim$ 60 minutes) and they are overlapped with the previous time interval by 40 images ($\sim$ 30 minutes), i.e. time interval T$_{ei+1}$ is at 40 images ahead of the time interval T$_{ei}$. For both sunspots, the width of the time intervals were chosen to cover, at least, 5 times the typical period of oscillations, which varies with height, from 5 to 3 minutes, from the photosphere to the chromosphere respectively \citep[see][]{nagashima2007observations,stangalini2011mhd,jess2012propagating,khomenko2015,stangalini2021novel}. 

The boundaries between the umbra and penumbra were constructed by computing the average intensity for every time interval and then applying intensity threshold levels. 
Accordingly, in the case of the circular sunspot we set this intensity level at 0.45, while in the case of the elliptical sunspot we adopted the level at 0.5. Thereby, for each time interval the umbra boundary will have a different shape, as illustrated in Figure \ref{fig: mean} for three different time intervals. 

To obtain the theoretical umbral slow body wave modes we use the observed irregular cross-sectional shape, as was done previously by \citet{stangalini2022large} and \citet{albidahApJ}. This allows us to accurately correlate the modes in observational data with their theoretical counterparts. The governing Helmholtz type equation for the vertical velocity perturbation solved by us \citep[see Equation 14 in][]{albidahApJ} does not assume any long or short wavelength limits and is therefore valid for arbitrary wavelengths. As was done previously by \citet{stangalini2022large} and \citet{albidahApJ} we fix the velocity perturbation to be zero at the boundary which was shown by \cite{aldhafeeri2022comparison} to be a valid approximation for slow body modes in photospheric magnetic waveguides due to the very close proximity of the boundary and outermost nodes. More importantly, for interpreting slow modes inside the umbra of sunspots, it was also shown that this assumption has a negligible effect on the actual spatial structure of the modes \citep[see Figures 2 and 3 in][]{aldhafeeri2022comparison}. This approximation has also been validated by observational data, which show that the Doppler velocity perturbations at the umbra/penumbra boundary for slow body modes are indeed very small relative to the perturbations inside the umbra \citep[see e.g., Figures 1c and 1d in][which show only small differences in the observation and model at the umbra/penumbra boundary]{stangalini2022large}.

The theoretical slow body modes for every time interval are shown in Appendix \ref{sec_app1}, Figures \ref{fig:model_circule} and \ref{fig:model_elliptc}. The modifications in the spatial structure of modes confirm that the higher-order modes are more sensitive to the irregularities in the cross-sectional shape shape of umbrae, i.e. the morphology of higher-order modes is changing as the shape of the waveguide is changing, even when these changes are small.

Next, we applied the combined POD and DMD techniques for every time interval of the HMI Doppler velocity data sets for both sunspots. 
The spatial structures of the first 10 POD Doppler velocity modes are shown in the Appendix \ref{sec_app1}) for each time interval. The analysis was also performed on the intensity and magnetic field data sets, however apart from the first POD modes, the higher order modes could not be distinguished from noise. It has previously been shown that the power of intensity oscillations inside umbral regions is heavily suppressed at the photospheric level compared with the lower chromosphere \citep[see e.g.][]{nagashima2007observations}. Here we found a similar result for the HMI intensity sunspot umbrae data, i.e. the noise level was sufficiently high to prevent extracting physical perturbations. In the HMI magnetic field data series for the sunspots analyzed in this paper the fluctuations only have a maximum amplitude of about 20 G (similar to \cite{rubio2000oscillations}), but the noise level of HMI magnetograms for a 45~s time cadence is approximately 10-15 G. Therefore, it is not surprising that physical POD modes could not be identified as was done for the Doppler velocity data where the signal to noise ratio was much better. We found that the maximum amplitudes for Doppler velocity were between 150 - 200 m s$^{-1}$ compared with the noise level of 13 m s$^{-1}$ for HMI Doppler velocity data with a 45~s time cadence.

To quantify the correlation between the POD modes obtained from every time interval of the HMI Doppler velocity data-sets and the modes predicted by theoretical models, we perform a cross-correlation analysis and compute the integral of the correlations, defined as the summation of the pixels in the correlation matrix as shown in the Appendix \ref{sec_app2}.
This step is taken as guideline to observe and avoid missing the higher correlations since we have too many modes from the POD analysis and different models for each time interval. However, these higher correlations need to be checked and validated in order to be considered. Therefore, the power spectrum of the time coefficient that corresponds to the selected POD modes are calculated to obtain the dominant frequencies. This is followed by obtaining the spatial structure of the DMD modes that correspond to the dominant frequencies.

To analyze the possible MHD modes, we selected the observed modes which presented a good agreement with the theoretical modes, i.e.  having highest correlation. All the possible observed modes displaying a good correlation with the theoretical models are presented in the Appendix in Tables \eqref{Table_C}, for the circular sunspot, and \eqref{Table_E}, for the elliptical sunspot. It is remarkable that the observed MHD modes in the elliptical sunspot have a higher-order number of POD, i.e. have a lower contribution to the total variance along with the time intervals. In contrast, the modes observed in the circular sunspot were observed with a lower-order POD as the first and the second POD which means that the observed modes have the highest contribution to the total variance of the signal.

\begin{figure}
\centering
\begin{tabular}{ccccc}

\includegraphics[width=30mm]{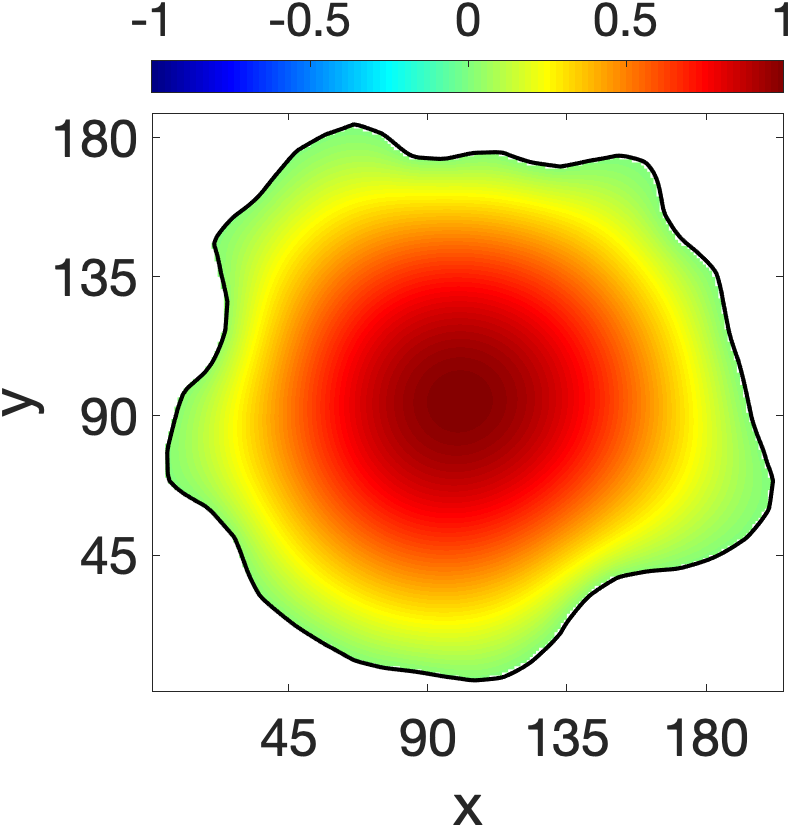}&
\hspace{-.8em}
\includegraphics[width=30mm]{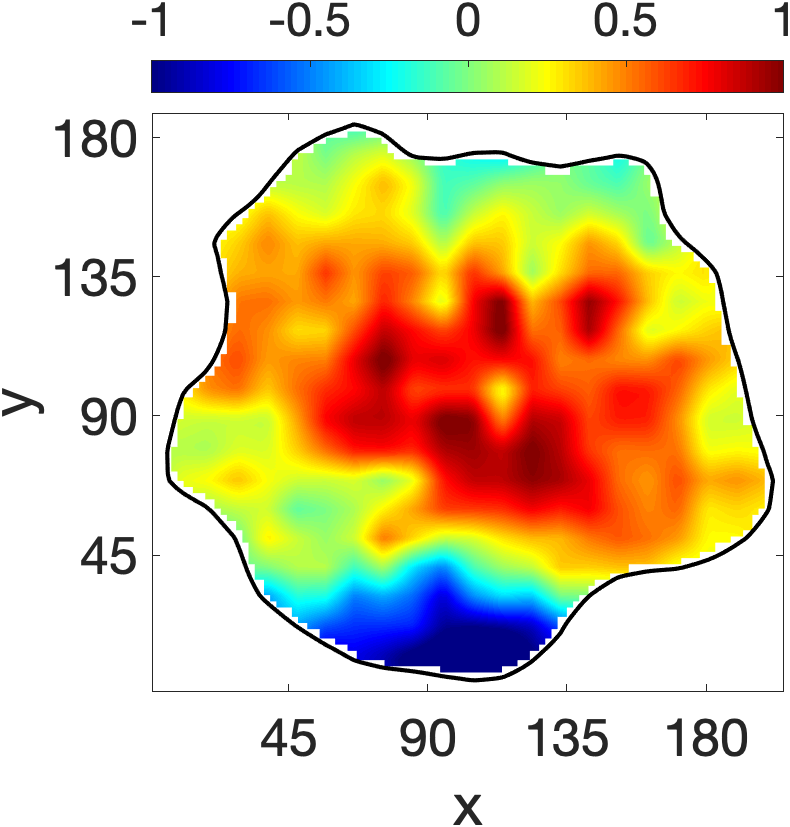}&
\hspace{-.8em}
\includegraphics[width=30mm]{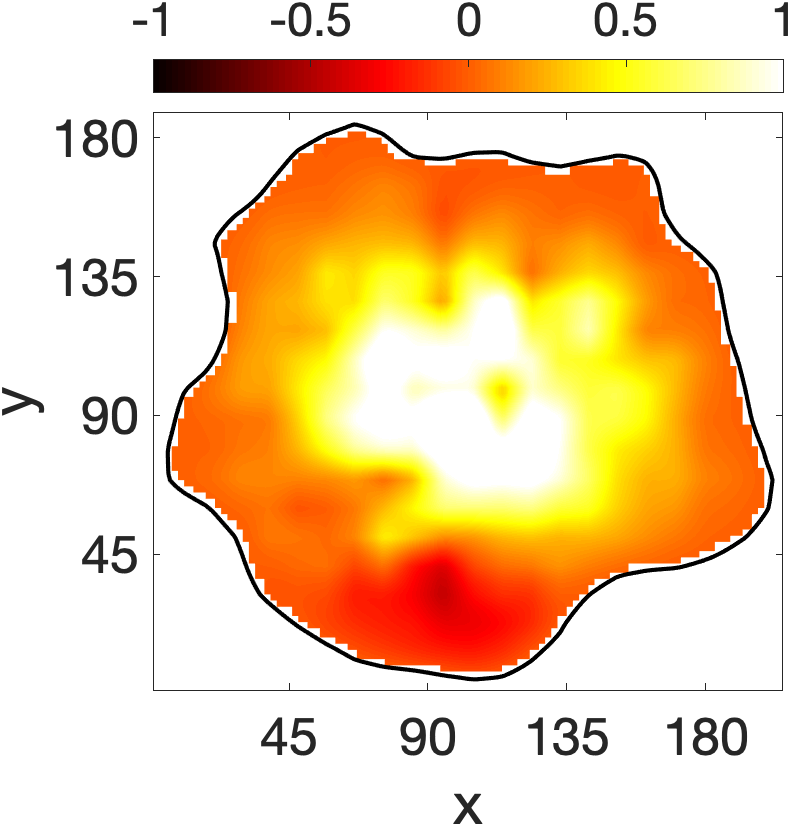}&
\hspace{-.8em}
\includegraphics[width=30mm]{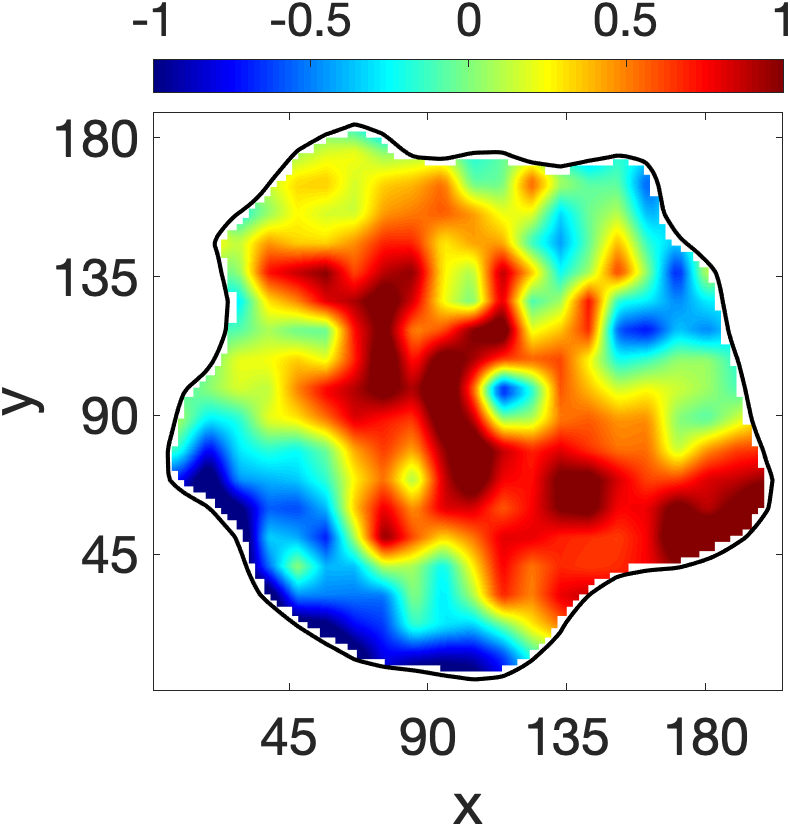}&
\hspace{-.8em}
\includegraphics[width=30mm]{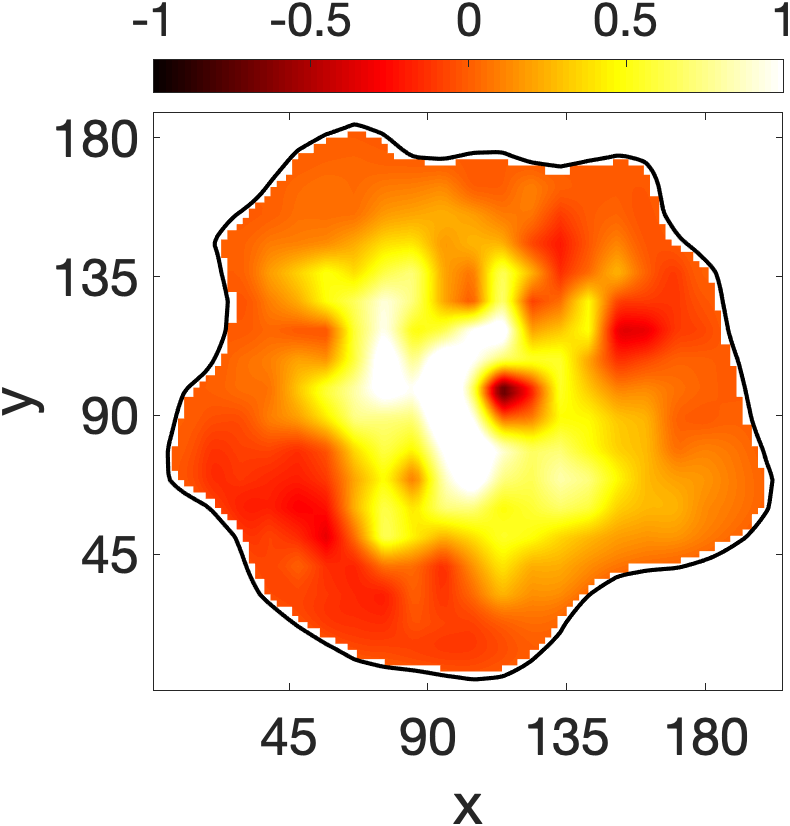}\\

\includegraphics[width=30mm]{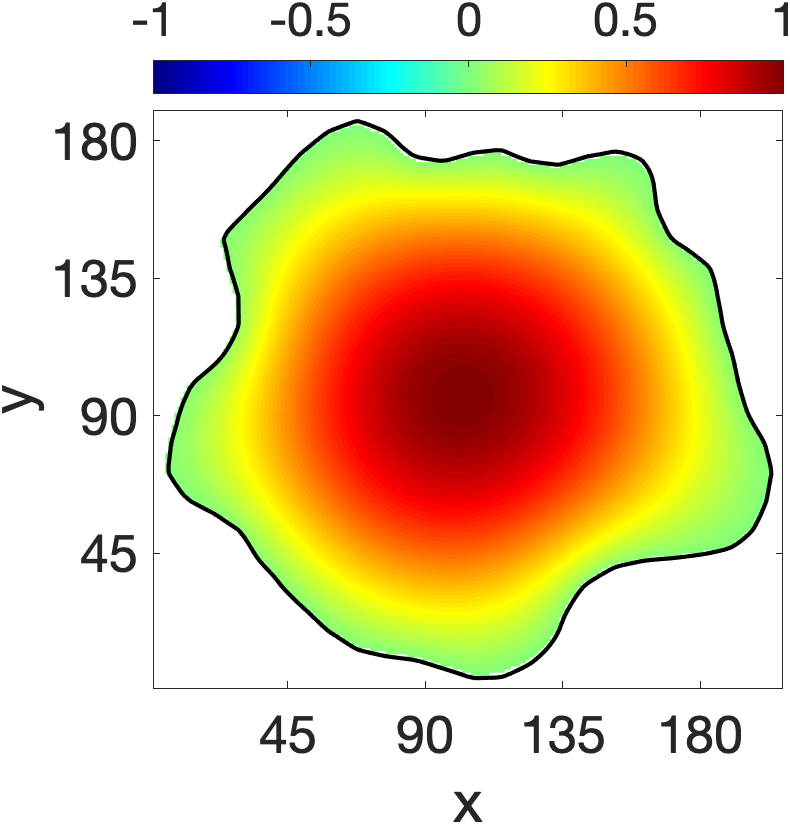}&
\hspace{-.8em}
\includegraphics[width=30mm]{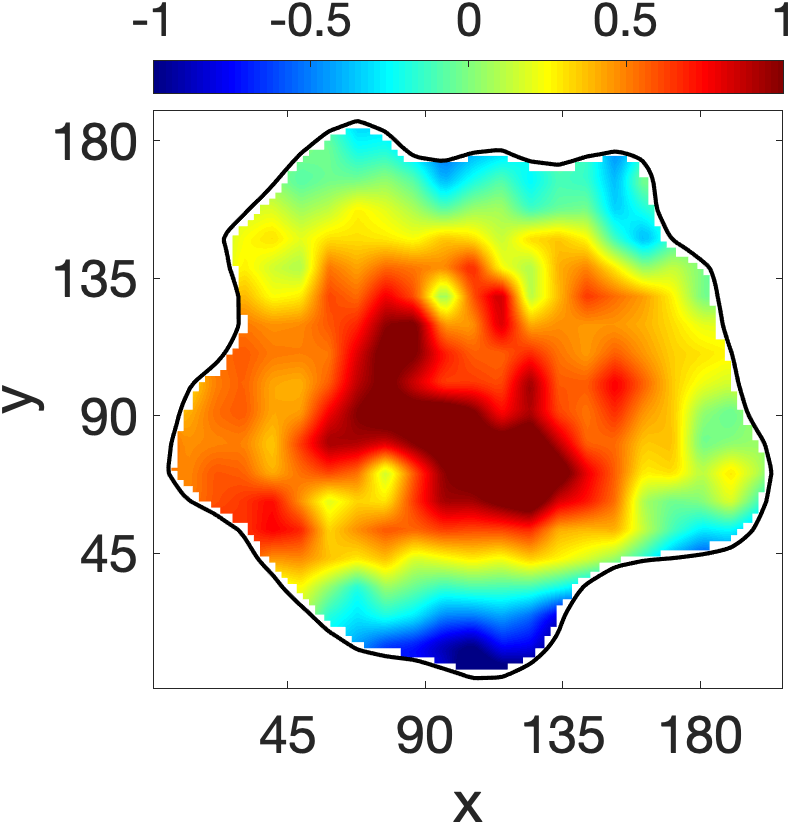}&
\hspace{-.8em}
\includegraphics[width=30mm]{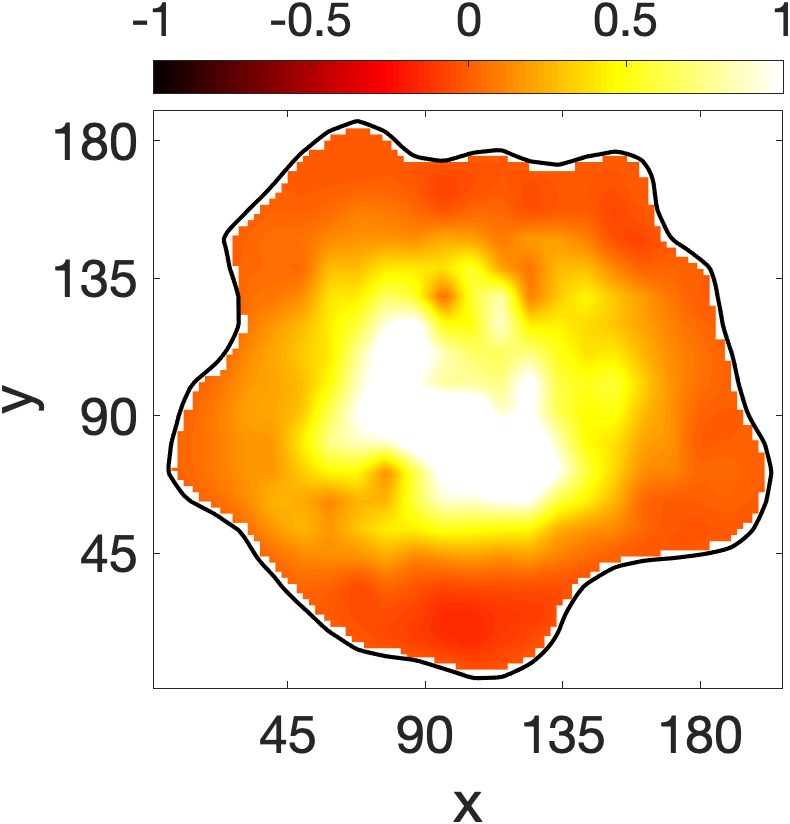}&
\hspace{-.8em}
\includegraphics[width=30mm]{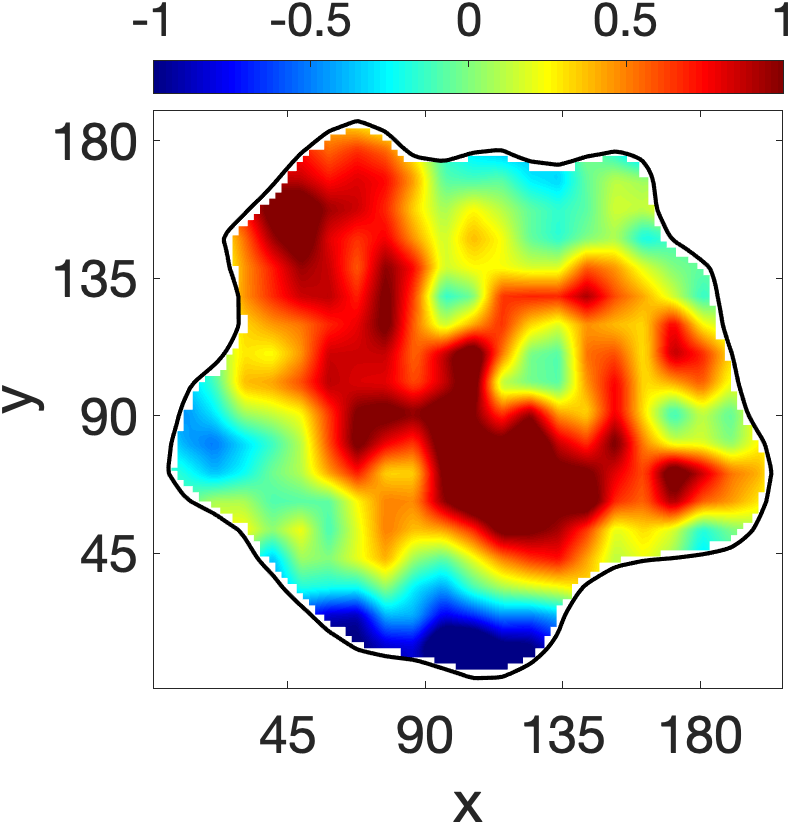}&
\hspace{-.8em}
\includegraphics[width=30mm]{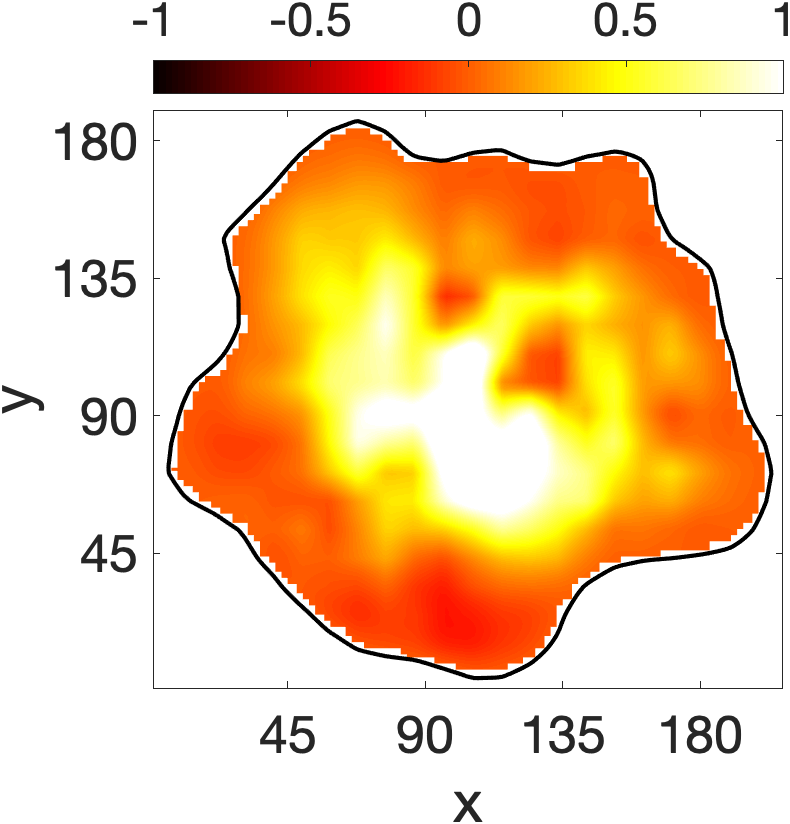}\\

\end{tabular}

\caption{The results of theoretical modeling and the the results of the POD and DMD analysis obtained for the sunspot (NOAA 11366) are presented for the time intervals T$_{c6}$ (top panel) and T$_{c7}$ (bottom panel). The first column displays the spatial structure of the theoretically modelled fundamental slow body sausage mode for the same shape as sunspot (NOAA 11366) umbra shape at the particular time interval. The second column shows the spatial structure of the 1$^{st}$ (top and bottom) POD modes, where the dominant frequency of their time coefficient is in the range between 3.1 to 4.8 mHz. The cross-correlation between the model and the determined POD mode is shown in the third column. The fourth column displays the spatial structure of the DMD modes corresponding to 3 mHz (top) and 4.3 mHz (bottom). Finally, the last column contains the cross-correlation between the model (first column) and the DMD mode (fourth column). The solid black line shows the umbra/penumbra boundary. The same configuration was used for Figures \ref{fig:TEC kink1} and \ref{fig:TEC n=2}. The colourbars in the first column display the density perturbation obtained from the theoretical model, while the second and fourth columns display the magnitude of the Doppler velocity after being analyzed by POD and DMD, respectively. The colourbars in the third and the fifth columns denote the correlation/anti-correlation.} \label{fig:TEC susa}. 
\end{figure}

In the case of the circular sunspot (NOAA 11366), the first MHD wave mode that was identified is the fundamental slow body sausage mode, and it appears as the first POD mode in the time intervals from T$_{c5}$ up to T$_{c7}$. This mode has a lower contribution in T$_{c3}$ and T$_{c4}$, where it appears as the 4$^{th}$ POD mode. In Figure \ref{fig:TEC susa} we show the fundamental slow body sausage mode in two time intervals (T$_{c6}$ and T$_{c7}$, corresponding to the two rows) and the spatial structure of the DMD mode that corresponds to the dominant frequency of the time coefficient of the POD modes, which are in the range between 3.1 to 4.8 mHz. These DMD modes correspond to 3 mHz, in T$_{c6}$, and 4.3 mHz, in T$_{c7}$, respectively. The determined POD and DMD modes are compared with the theoretical model by applying the cross-correlation analysis and they show a good agreement.

\begin{figure}[!t]
\centering
\begin{tabular}{ccccc}

\includegraphics[width=30mm]{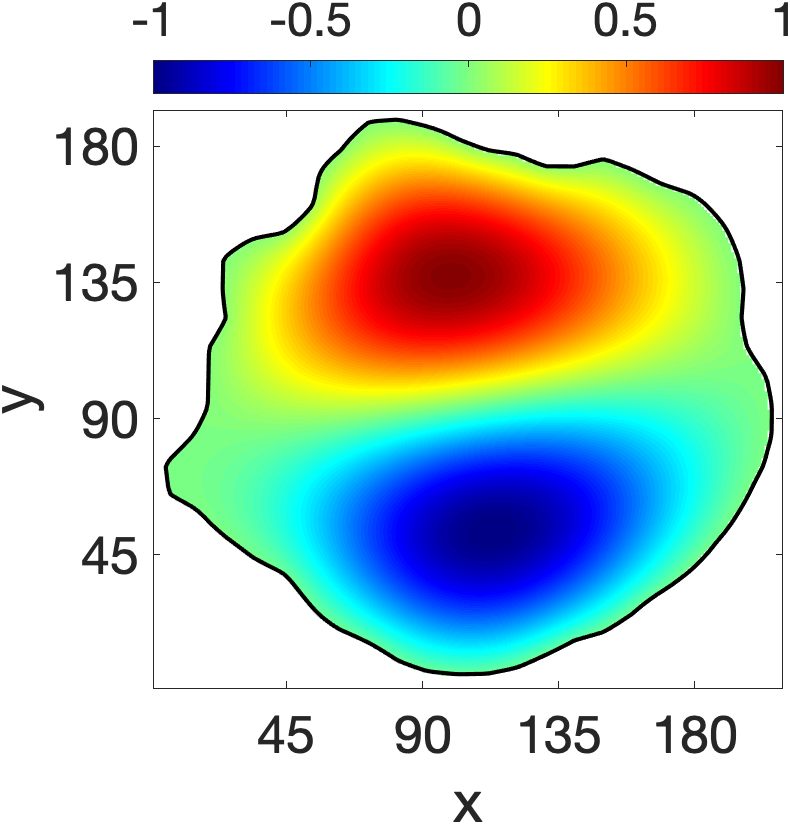}&
\hspace{-.8em}
\includegraphics[width=30mm]{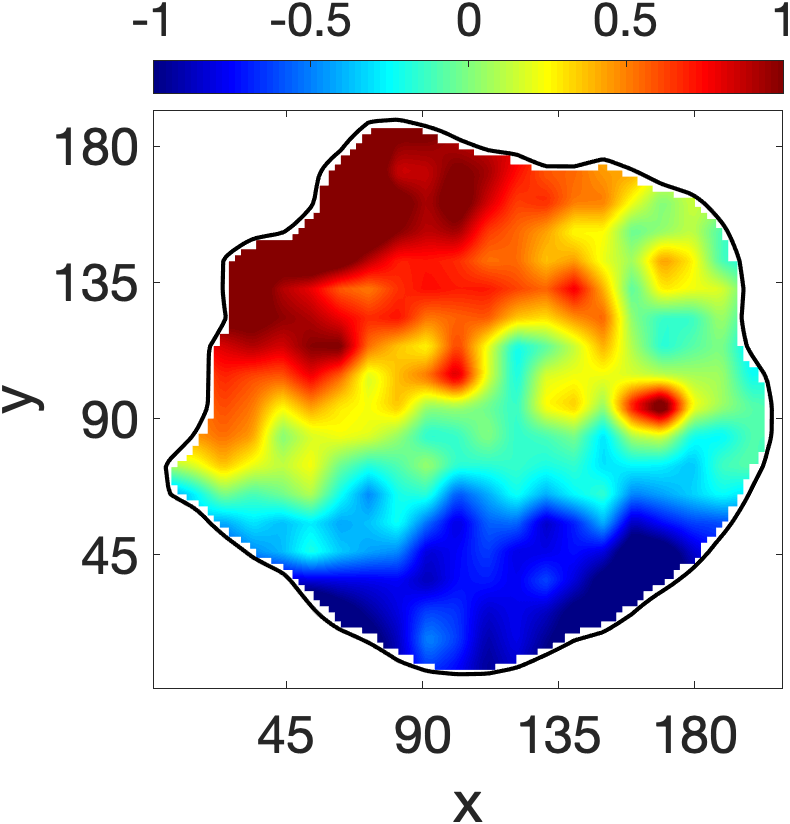}&
\hspace{-.8em}
\includegraphics[width=30mm]{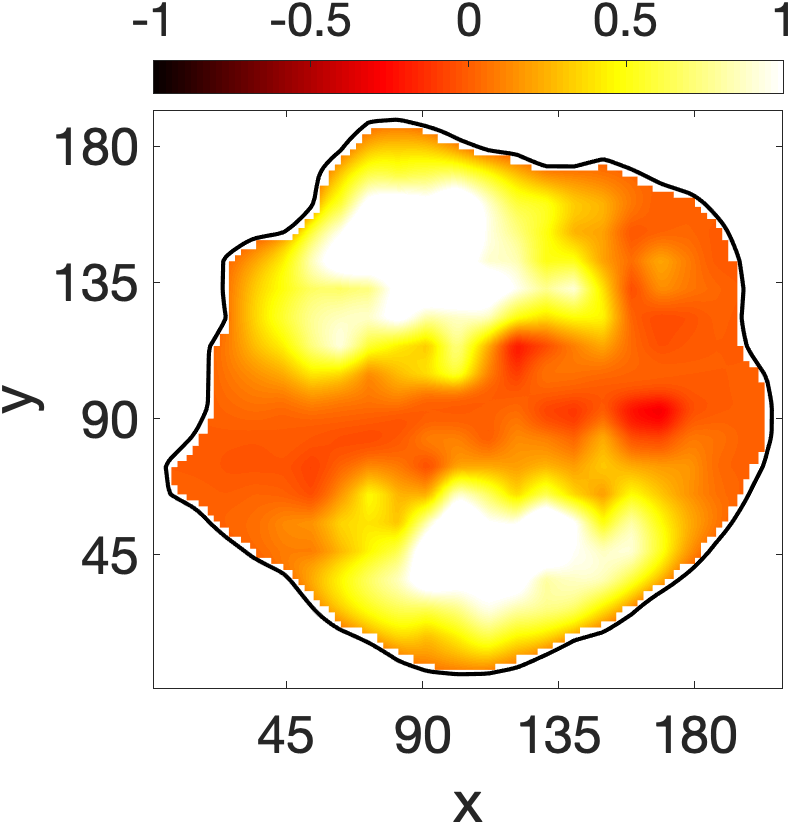}&
\hspace{-.8em}
\includegraphics[width=30mm]{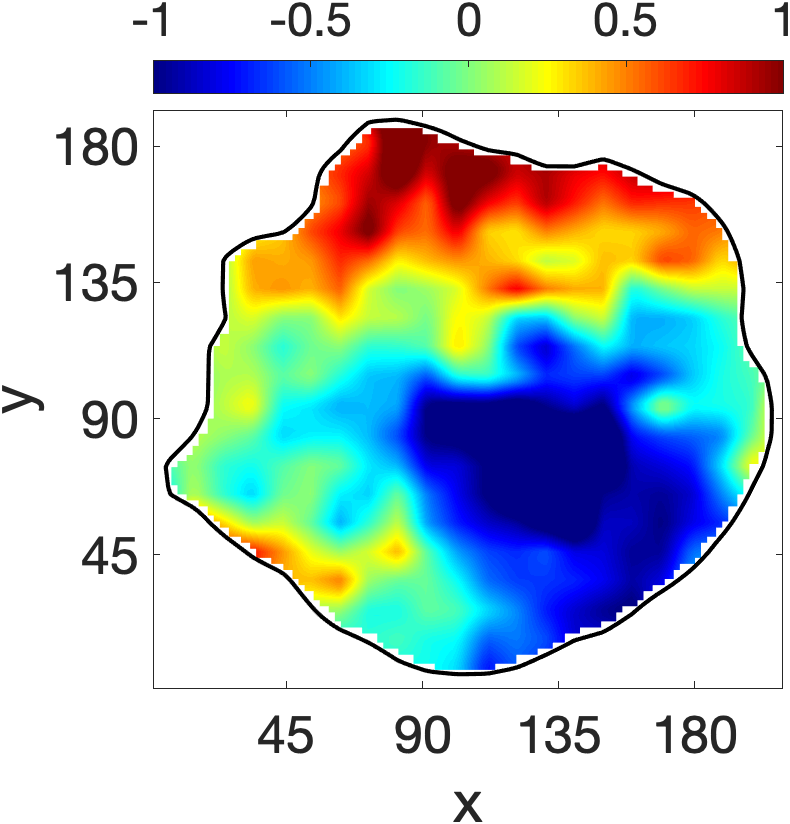}&
\hspace{-.8em}
\includegraphics[width=30mm]{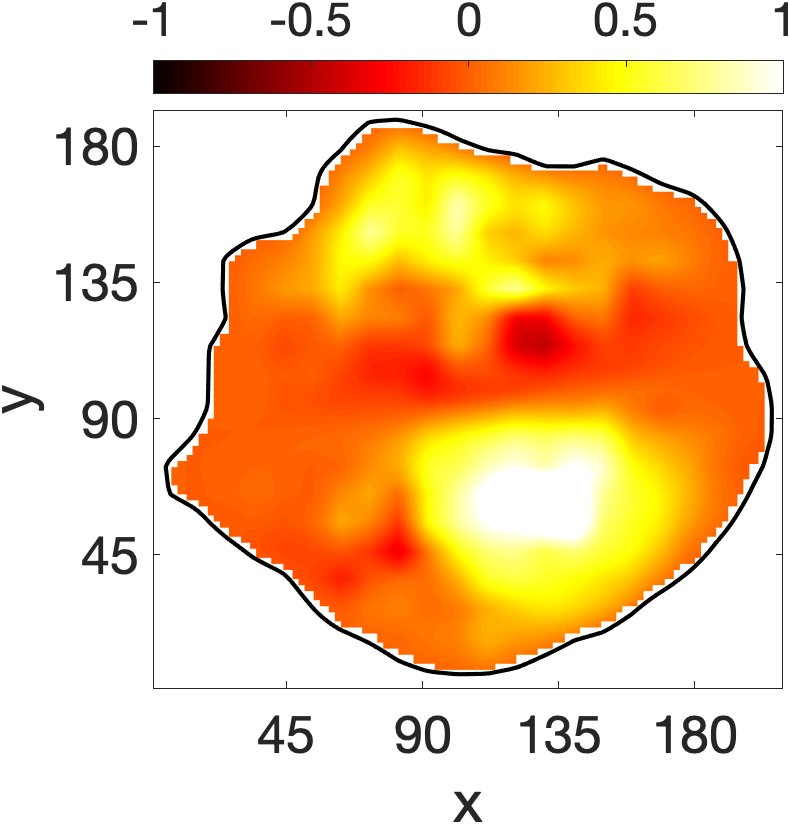}\\

\includegraphics[width=30mm]{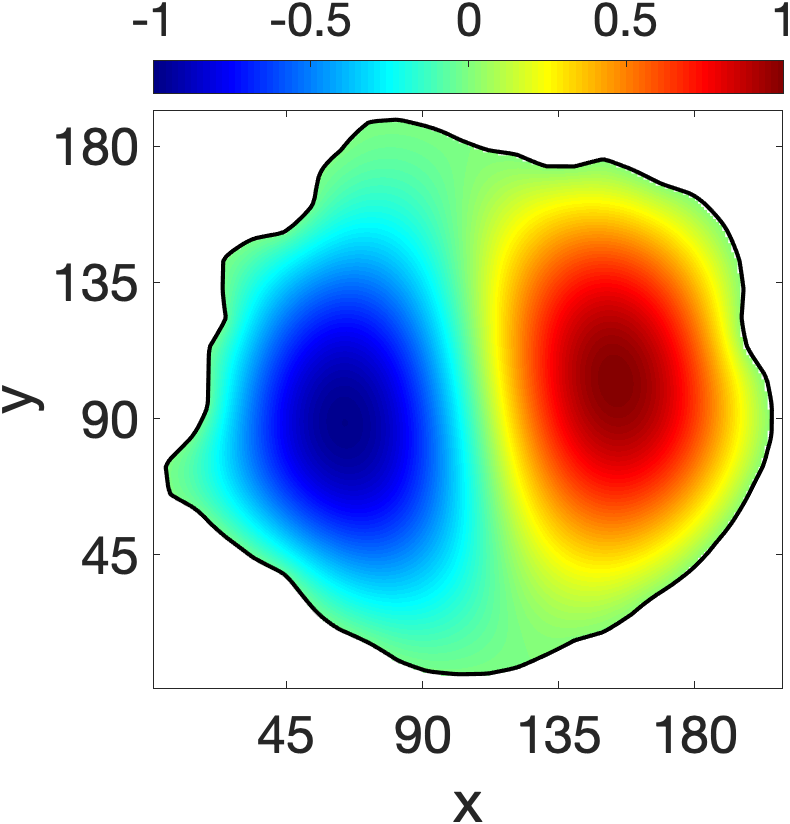}&
\hspace{-.8em}
\includegraphics[width=30mm]{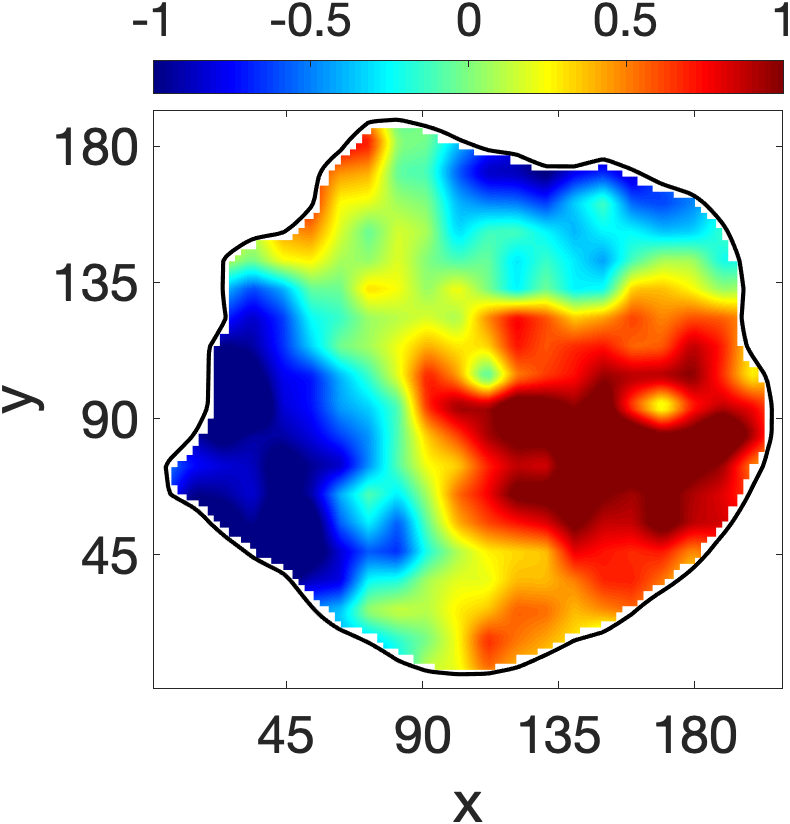}&
\hspace{-.8em}
\includegraphics[width=30mm]{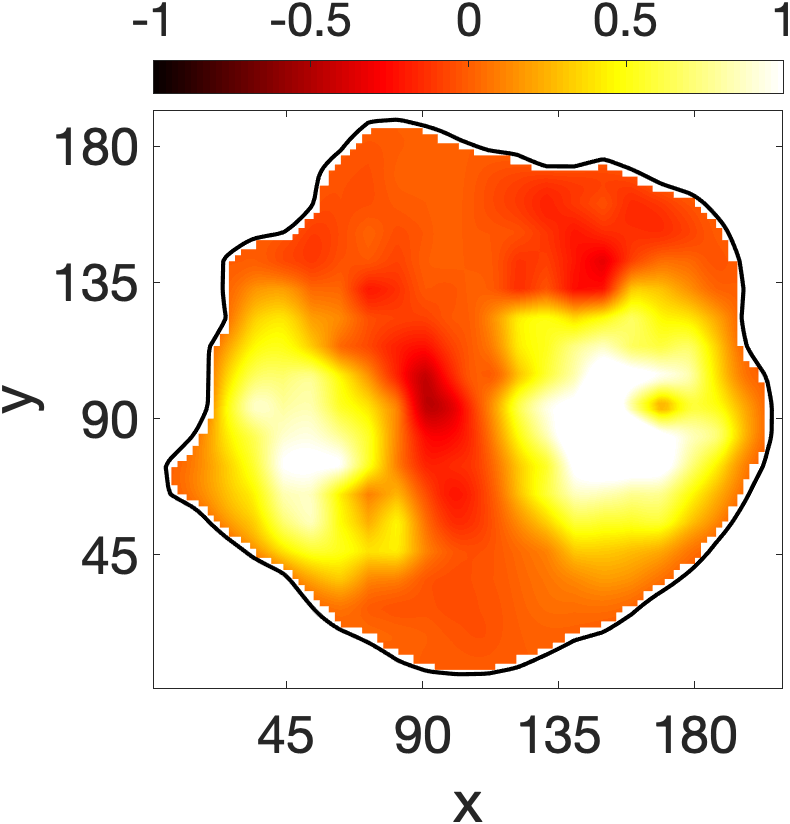}&
\hspace{-.8em}
\includegraphics[width=30mm]{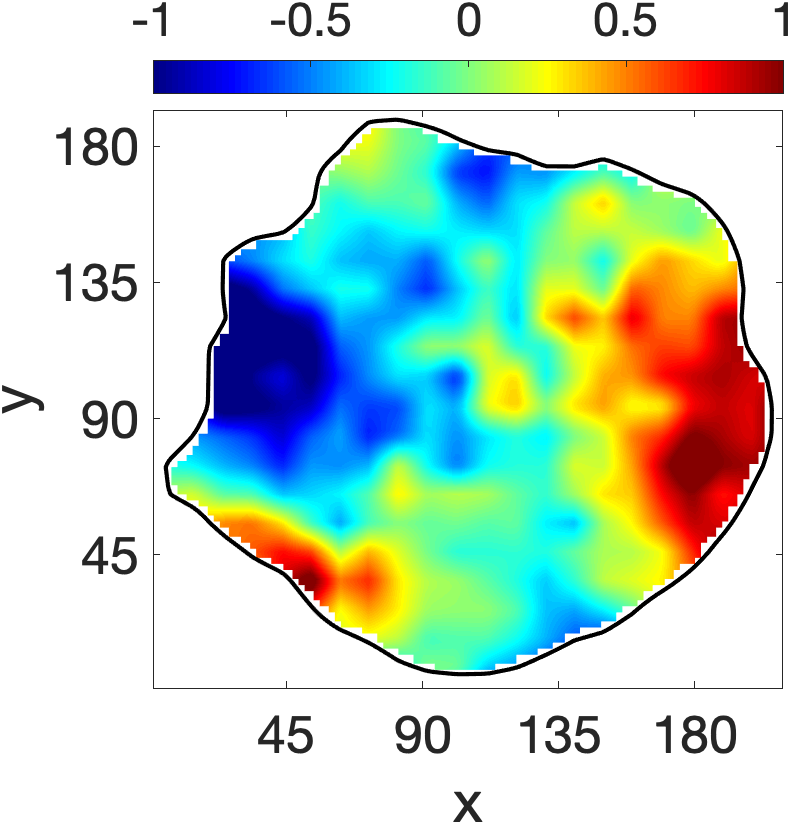}&
\hspace{-.8em}
\includegraphics[width=30mm]{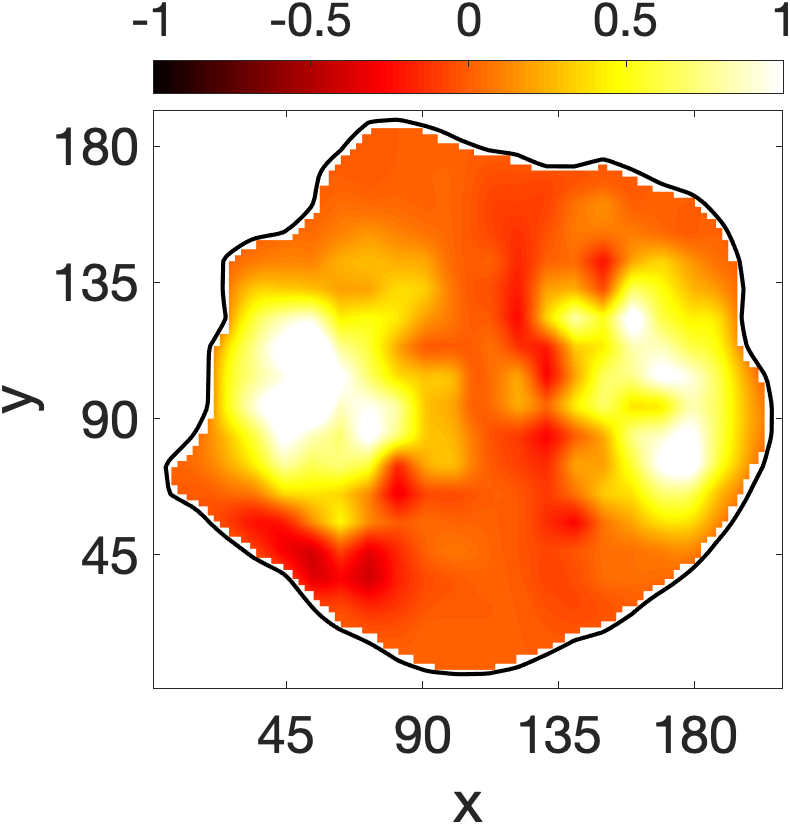}\\

\includegraphics[width=30mm]{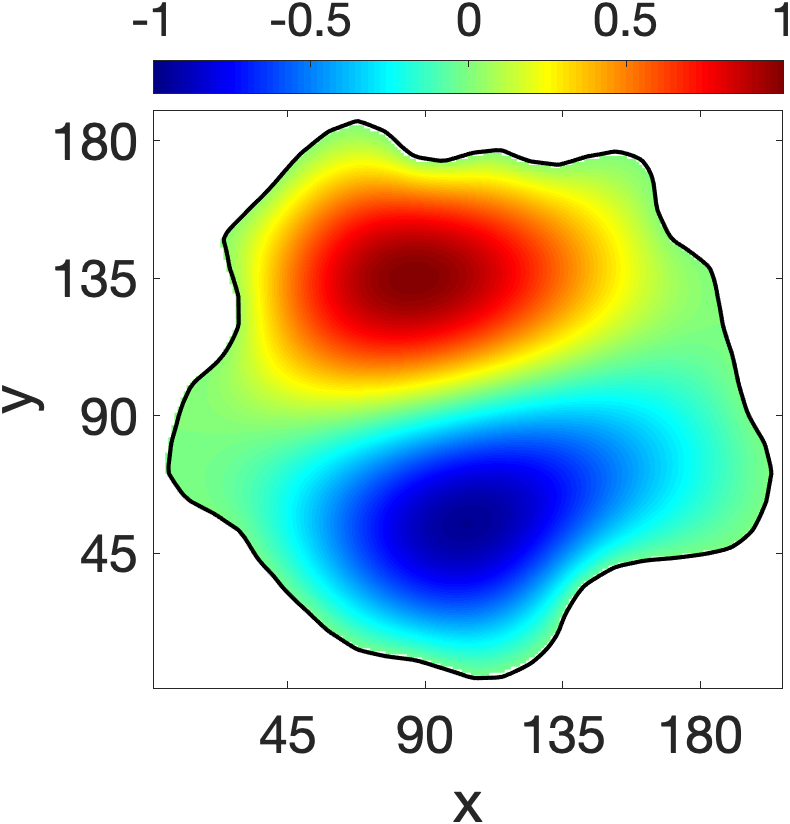}&
\hspace{-.8em}
\includegraphics[width=30mm]{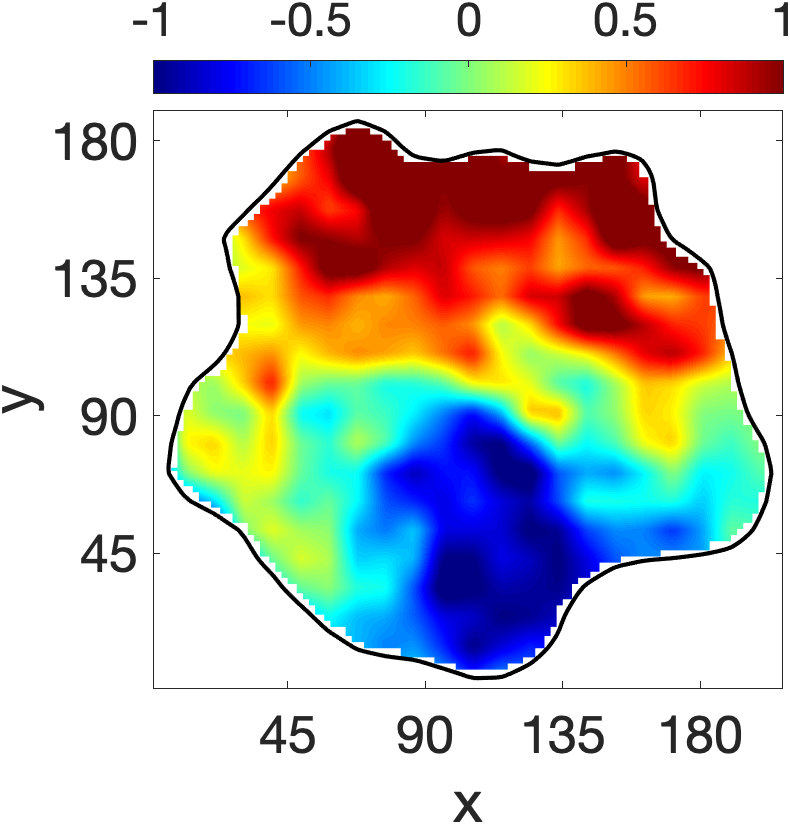}&
\hspace{-.8em}
\includegraphics[width=30mm]{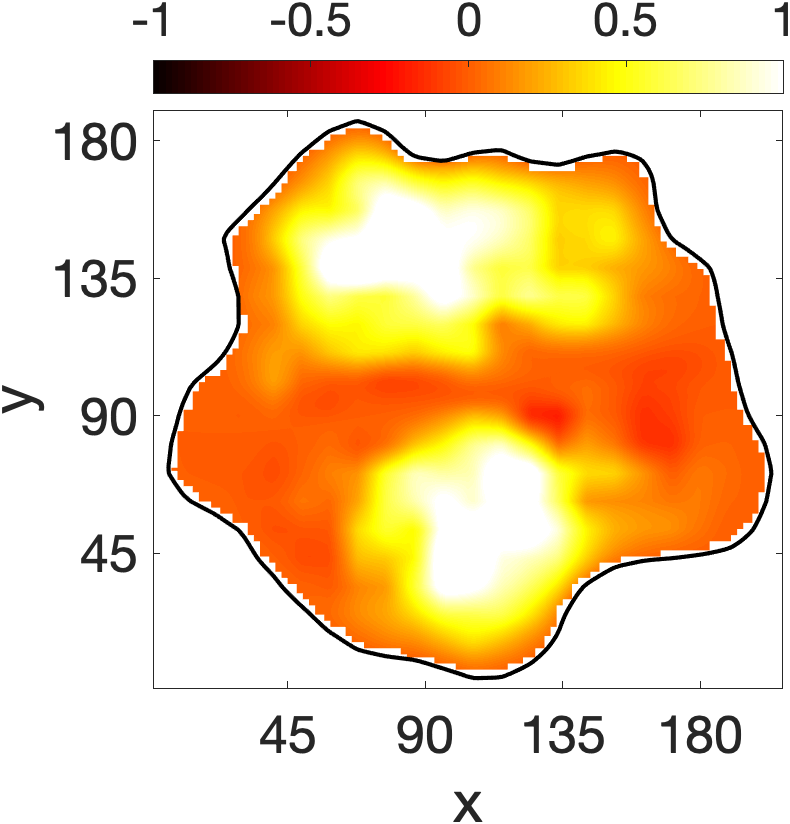}&
\hspace{-.8em}
\includegraphics[width=30mm]{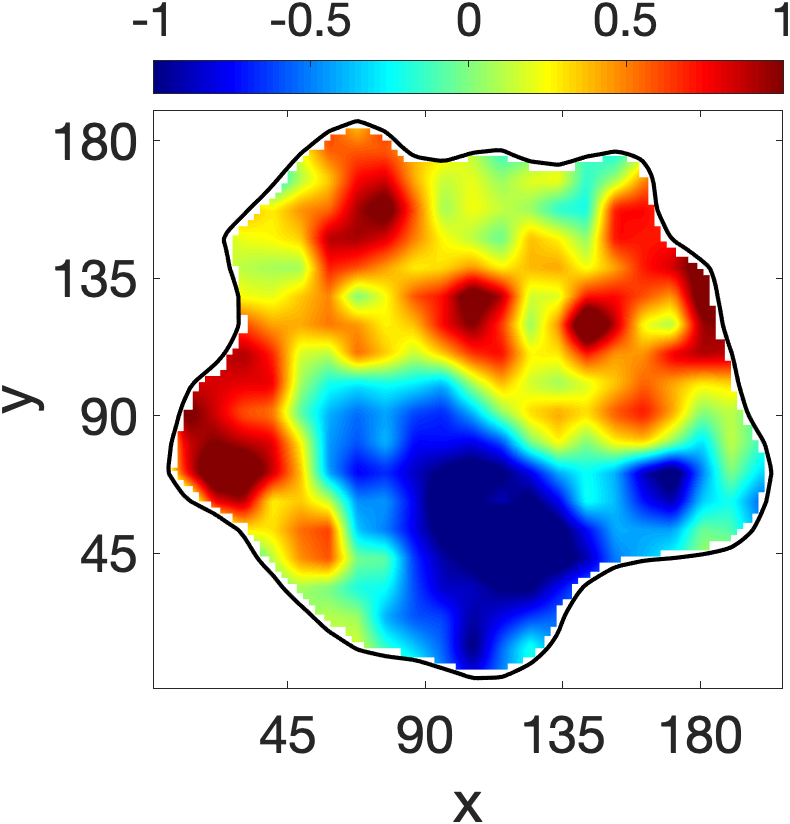}&
\hspace{-.8em}
\includegraphics[width=30mm]{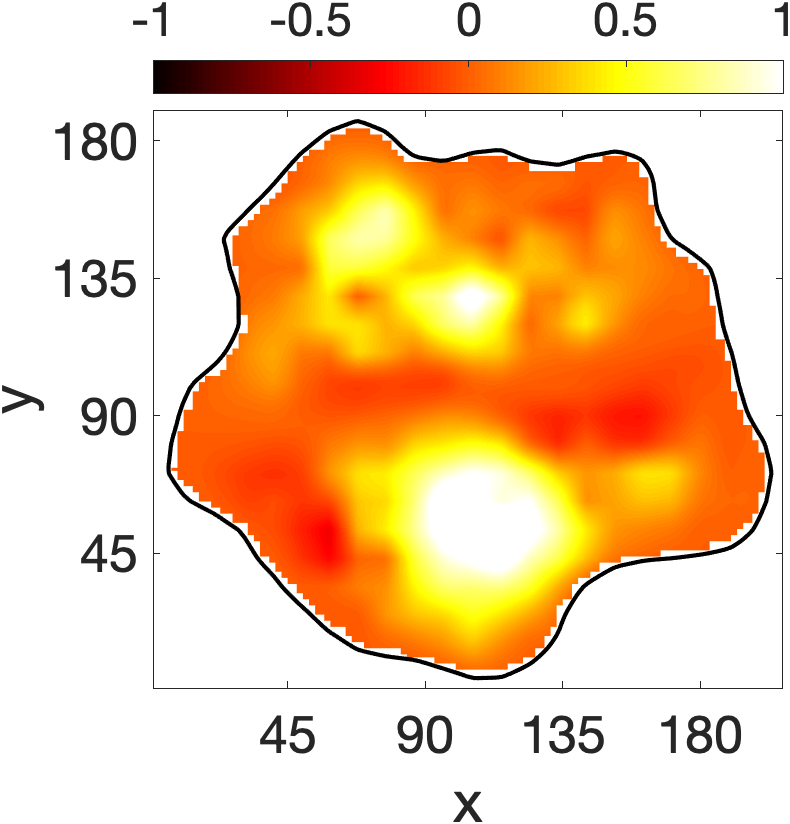}\\

\end{tabular}
\caption{The same as Figure \ref{fig:TEC susa}, but here the first row shows the 1$^{st}$ POD mode and the spatial structure of the DMD mode corresponding to 3.4 mHz at T$_{c1}$. The second row shows the 2$^{nd}$ POD mode and the spatial structure of the DMD mode corresponding to 3.1 mHz at T$_{c1}$. The third row shows the 2$^{nd}$ POD mode and the spatial structure of the DMD mode corresponding to 3.9 mHz at T$_{c7}$. The first column shows the morphology of the theoretical fundamental slow body kink mode corresponding to T$_{c1}$ (M$_3$, top panel and M$_2$ middle panel) and T$_{c7}$ (M$_3$, bottom panel).}

\label{fig:TEC kink1}
\end{figure}

The second MHD mode that we have observed has the azimuthal asymmetry corresponding to a fast surface kink mode and it is shown in Figure \ref{fig:TEC kink1}. These modes have the pattern corresponding to surface waves, as the amplitude increases along the radial direction and attain their maximum at the boundary. Due to the limitations of the theoretical model that is used to describe waves in a waveguide, the only modes that can be determined are the slow body modes. Hence, the  cross-correlation with possible surface modes detected by means of POD/DMD and their direct theoretical counterparts cannot be performed. However, as it was shown in the study by \cite{albidahApJ}, the cross-correlation between the slow body mode and the fast surface mode provides a spatial structure with a pattern closed to the slow body mode. In contrast, the cross-correlation between the slow body mode and the slow surface mode produces a spatial structure closed to a ring for the sausage mode and a broken ring for the kink mode. 

Therefore, the correlation in Figure \ref{fig:TEC kink1} between the slow body modes and the observed modes shows a pattern closed to slow body modes, hence we can identify this mode as being a fast surface kink. The modes that are presented in the first and second row of the second and fourth columns of Figure \ref{fig:TEC kink1} are the 1$^{st}$ and the 2$^{nd}$ POD modes and the DMD modes corresponding to 3.4 and 3.1 mHz, respectively, at T$_{c1}$. The superposition of these two approximately perpendicular kink modes with close  frequencies can provide an apparent rotational motion. The mode that is presented in the third row of Figure \ref{fig:TEC kink1} is the 2$^{nd}$ POD and DMD mode corresponding to 3.9 mHz at T$_{c7}$. The dominant frequencies of the POD modes that have an azimuthal symmetry to the kink mode are in the range between 3 to 4 mHz.

\begin{figure}[!t]
\centering
\begin{tabular}{ccccc}

\includegraphics[width=30mm]{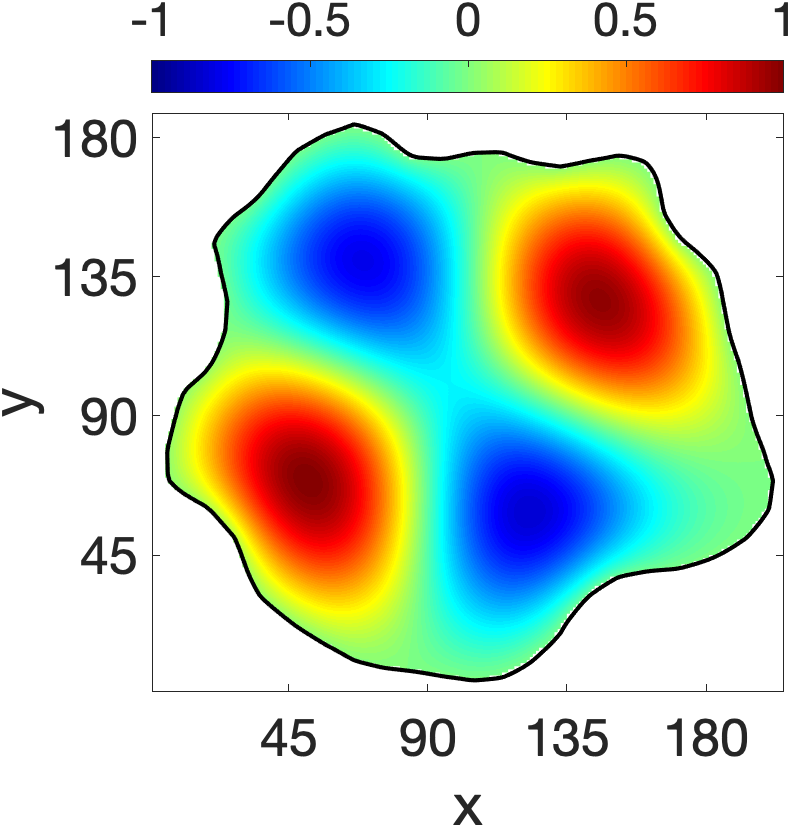}&
\hspace{-.8em}
\includegraphics[width=30mm]{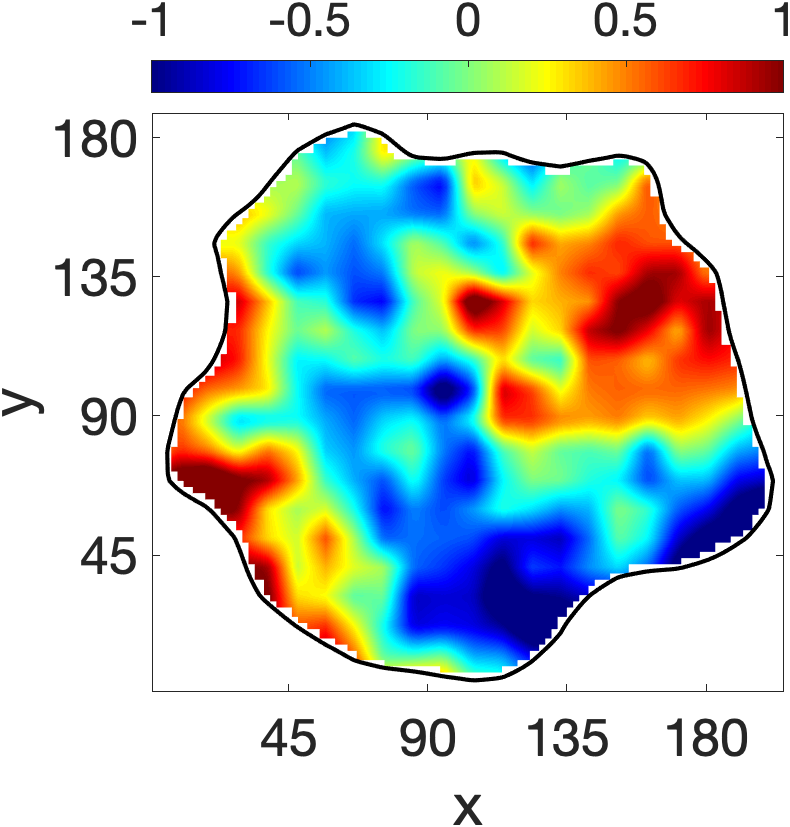}&
\hspace{-.8em}
\includegraphics[width=30mm]{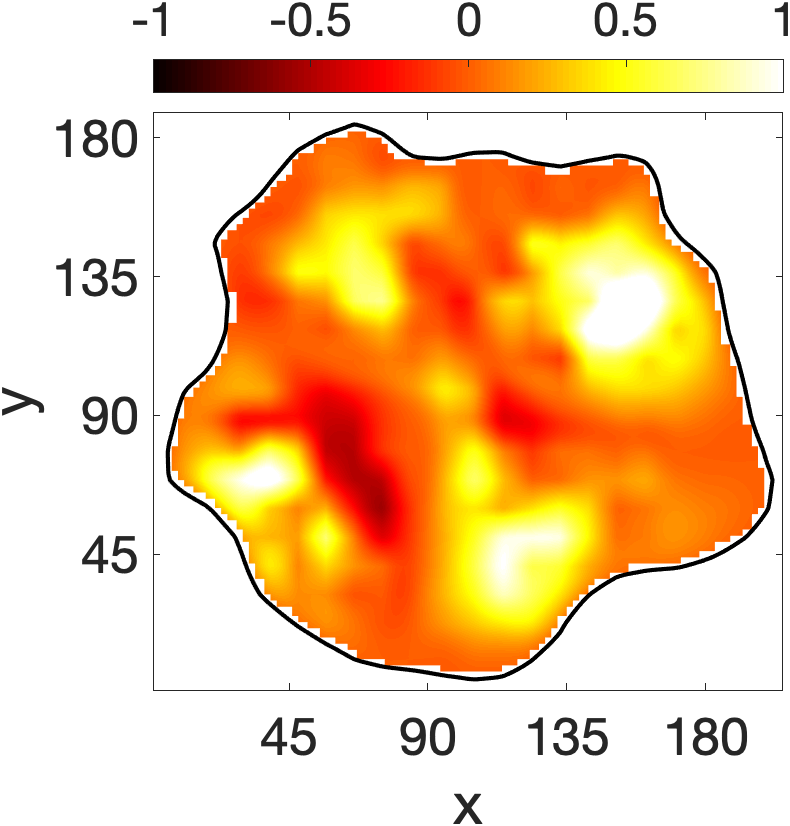}&
\hspace{-.8em}
\includegraphics[width=30mm]{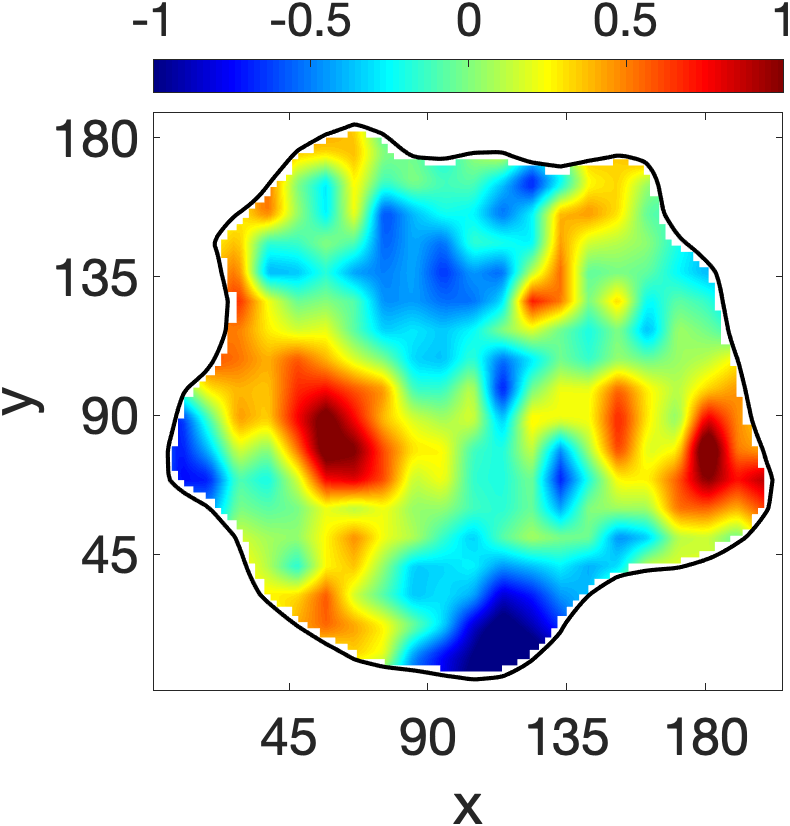}&
\hspace{-.8em}
\includegraphics[width=30mm]{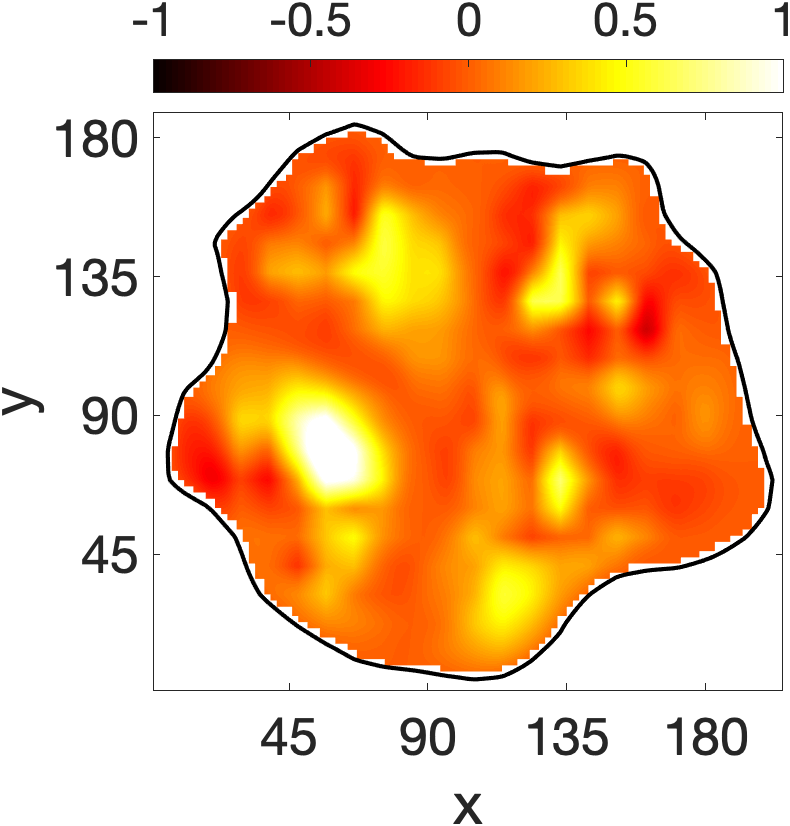}\\

\includegraphics[width=30mm]{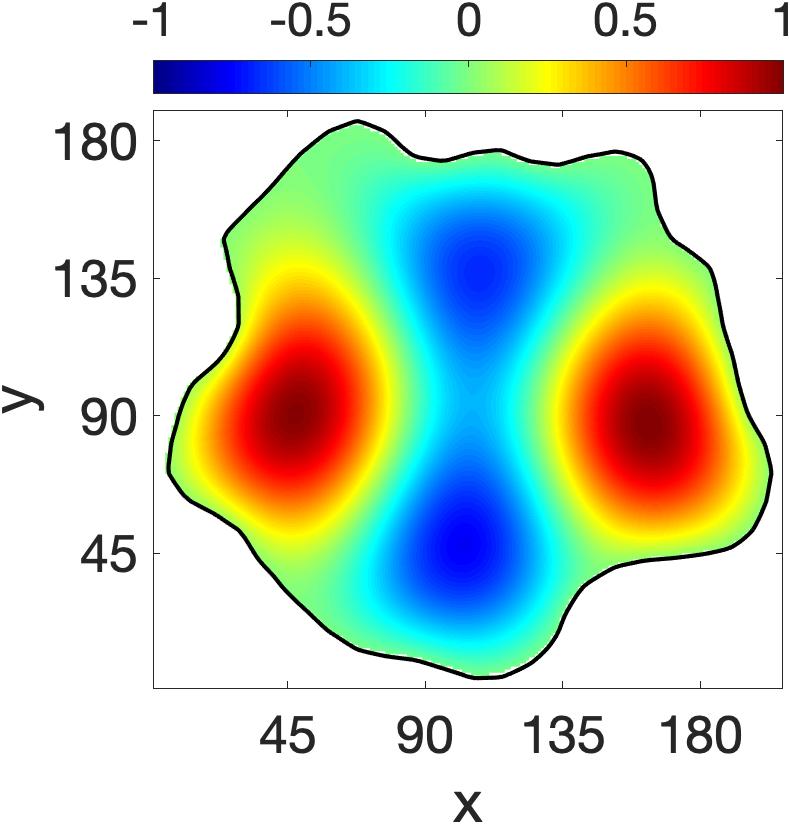}&
\hspace{-.8em}
\includegraphics[width=30mm]{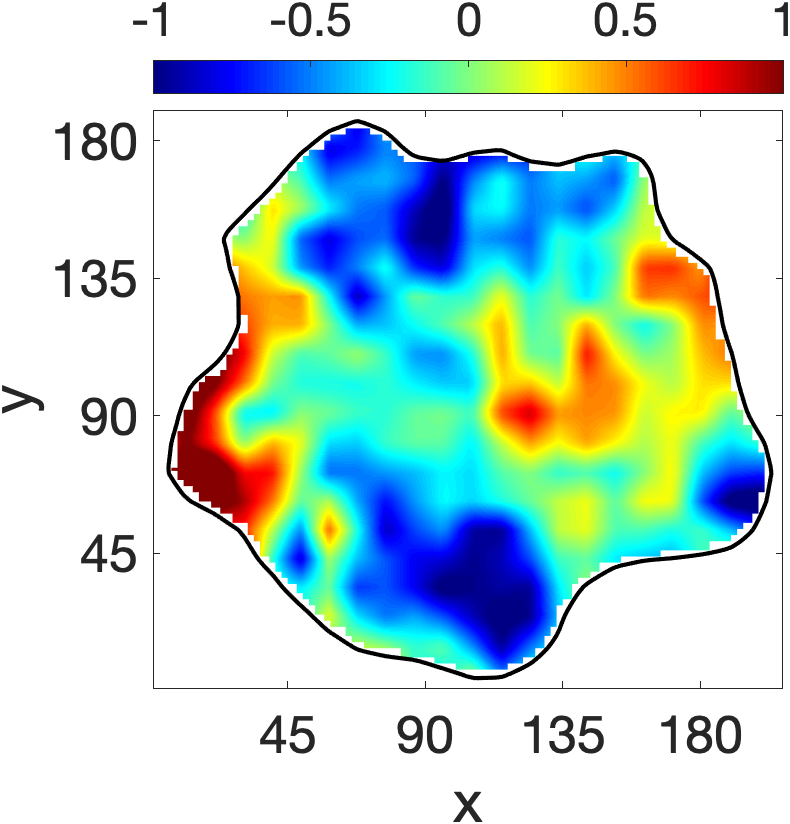}&
\hspace{-.8em}
\includegraphics[width=30mm]{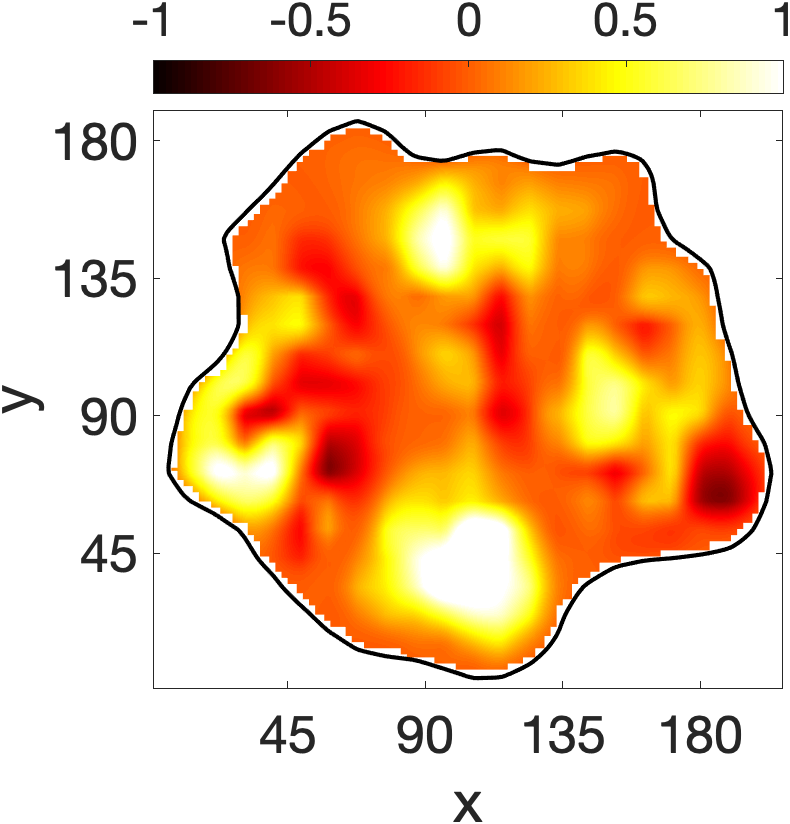}&
\hspace{-.8em}
\includegraphics[width=30mm]{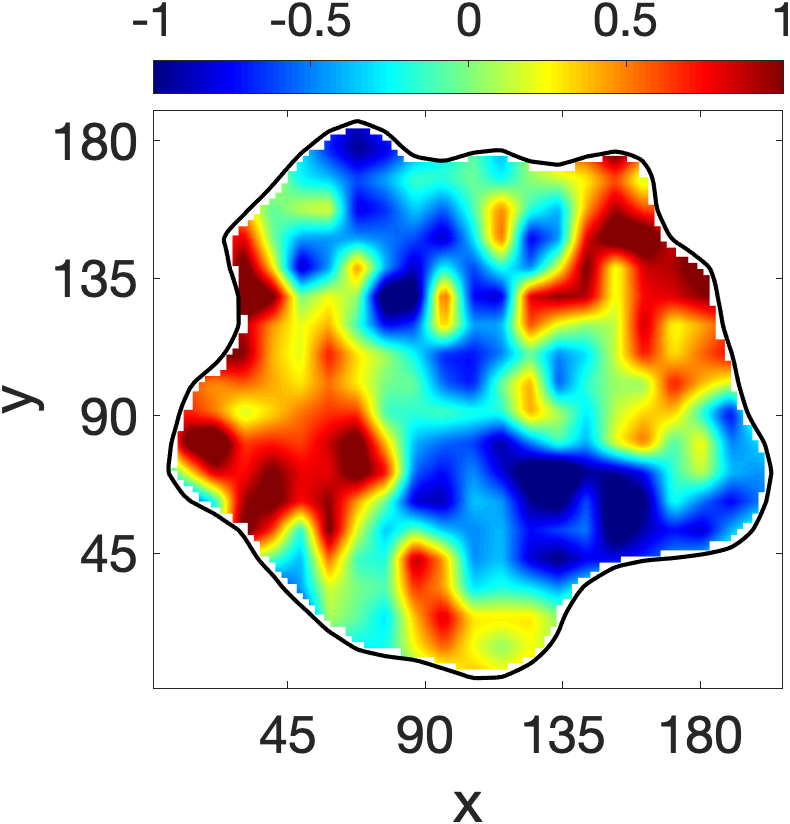}&
\hspace{-.8em}
\includegraphics[width=30mm]{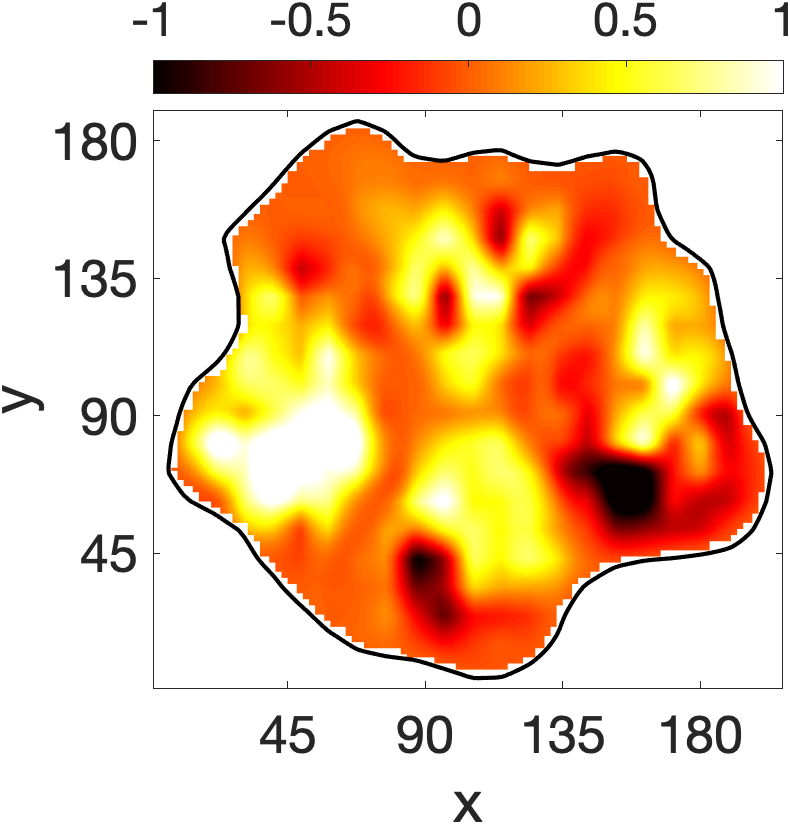}\\

\end{tabular}
\caption{The same as Figure \ref{fig:TEC susa}, but here the first row shows the 8$^{th}$ POD mode and the spatial structure of DMD mode corresponding to 4 mHz at T$_6$. The second row shows the 7$^{th}$ POD and DMD mode corresponding to 3 mHz at T$_7$. The first column shows the slow body fluting ($n=2$) corresponding to T$_6$ (M$_5$, top panel) and T$_7$ (M$_4$, bottom panel).}
\label{fig:TEC n=2}
\end{figure}


The fast surface kink mode is still present in the remaining time intervals,
but with a lower contribution as this mode becomes the 3$^{rd}$ POD mode at T$_{c5}$ and T$_{c6}$ and the 2$^{nd}$ POD at T$_{c7}$. The last MHD wave mode that we have observed in the circular sunspot is the slow body fluting mode ($n=2$), which appears at T$_{c6}$ and T$_{c7}$ (see Figure \ref{fig:TEC n=2}). The first row of Figure \ref{fig:TEC n=2} shows the 8$^{th}$ POD and DMD modes corresponding to 4 mHz at T$_{c6}$. The second row of this figure shows the 7$^{th}$ POD and DMD modes corresponding to a frequency of 3 mHz at T$_{c7}$. 

\begin{figure}
\centering
\begin{tabular}{ccccc}

\includegraphics[width=30mm]{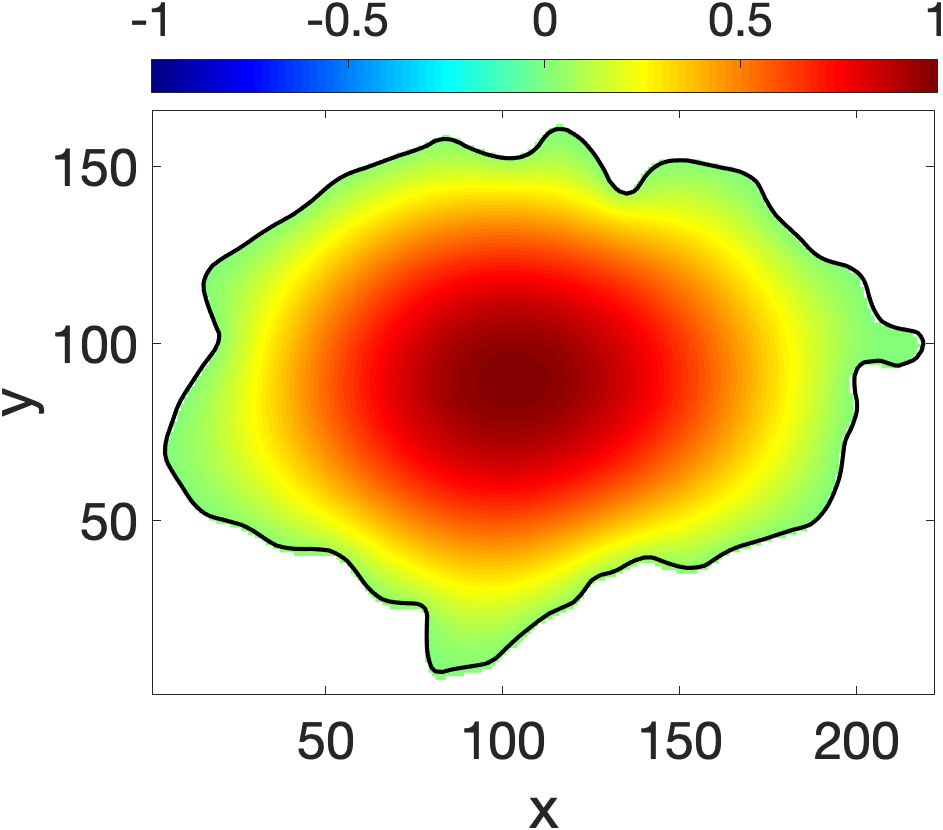}&
\hspace{-.8em}
\includegraphics[width=30mm]{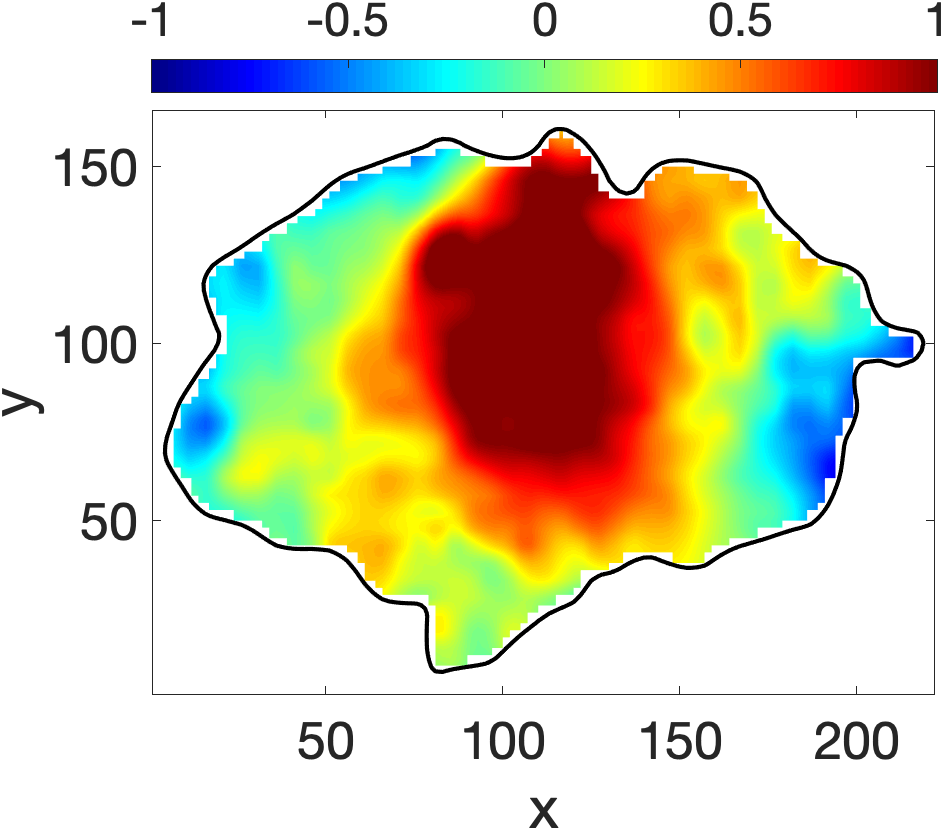}&
\hspace{-.8em}
\includegraphics[width=30mm]{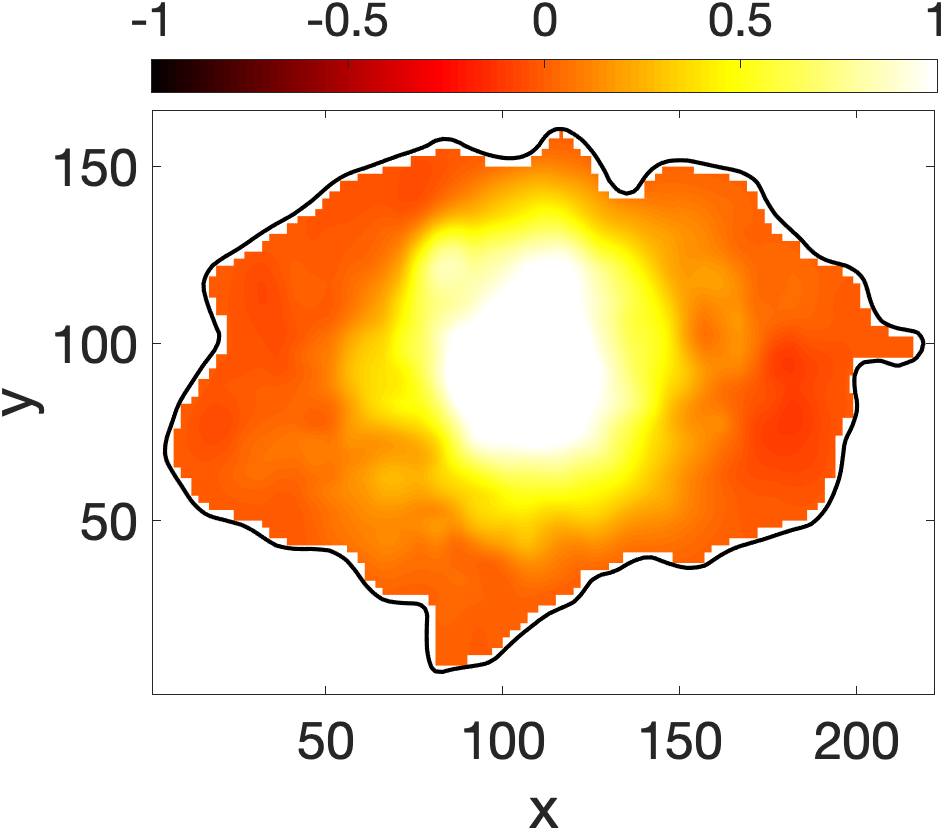}&
\hspace{-.8em}
\includegraphics[width=30mm]{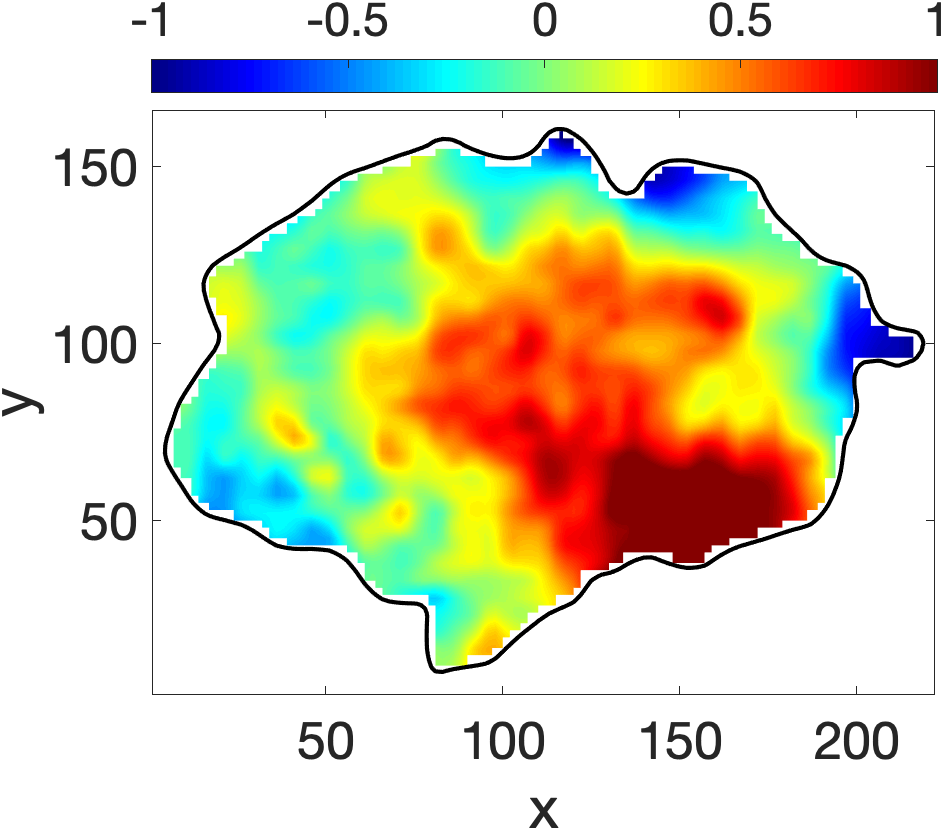}&
\hspace{-.8em}
\includegraphics[width=30mm]{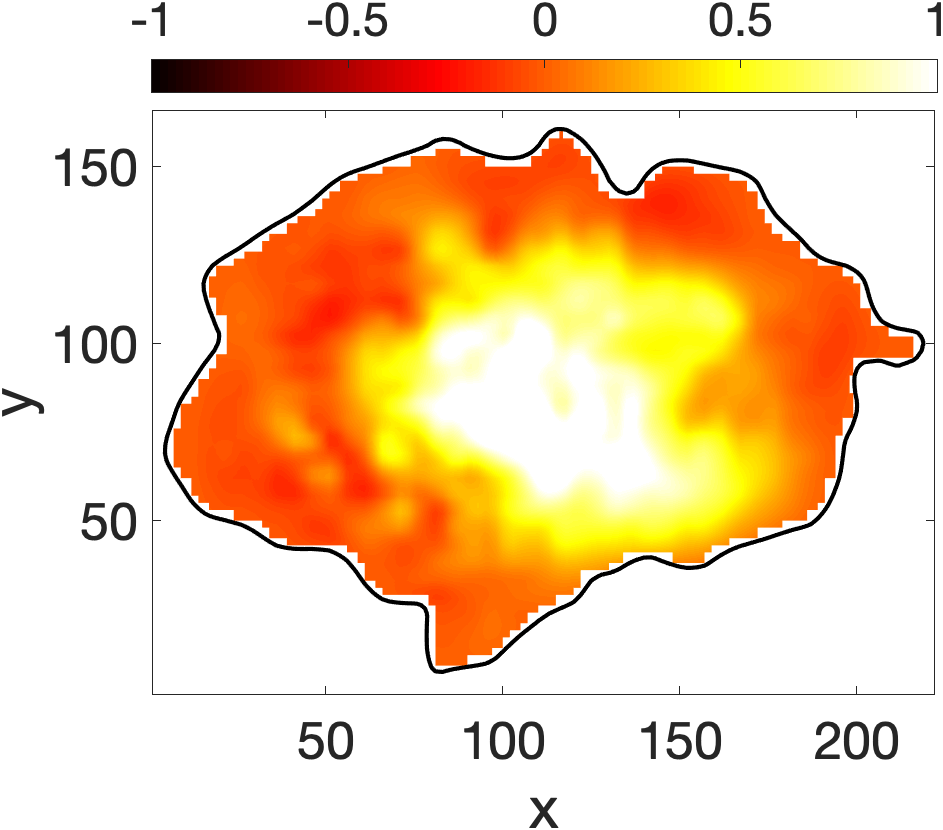}\\

\includegraphics[width=30mm]{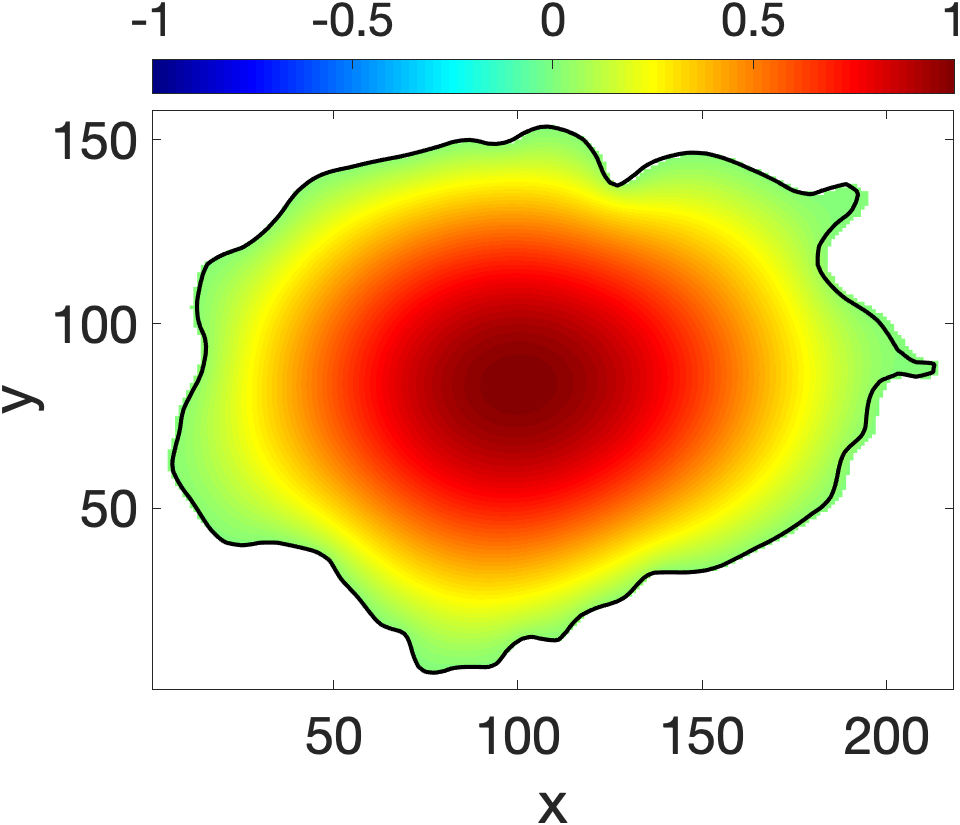}&
\hspace{-.8em}
\includegraphics[width=30mm]{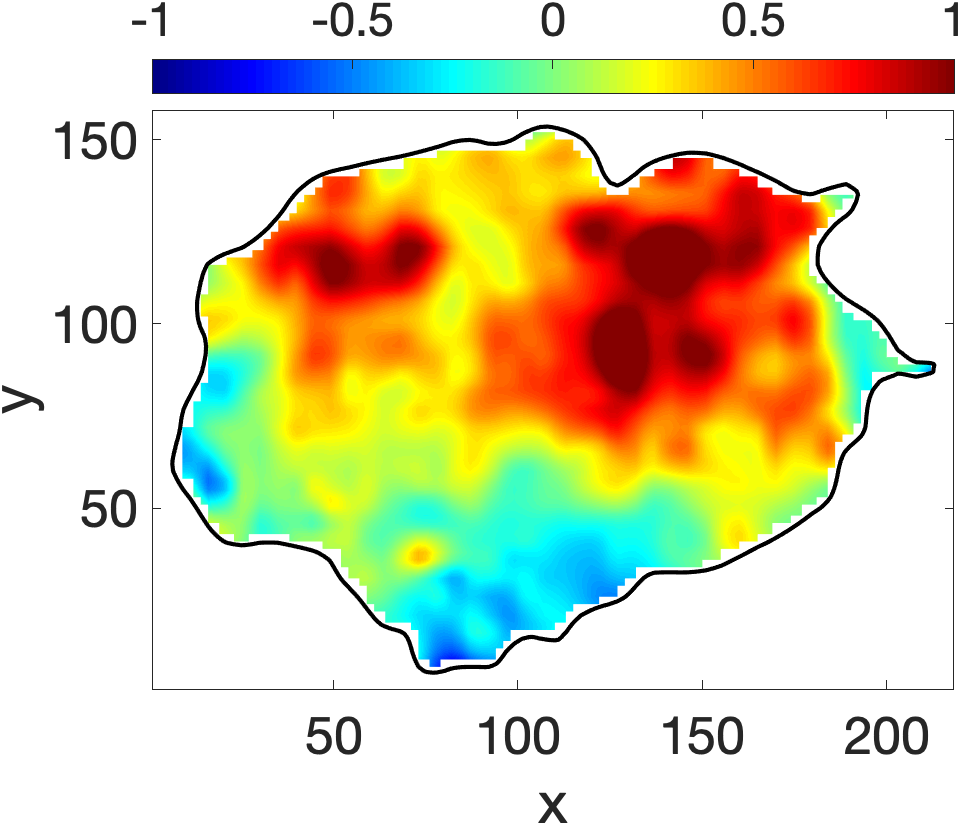}&
\hspace{-.8em}
\includegraphics[width=30mm]{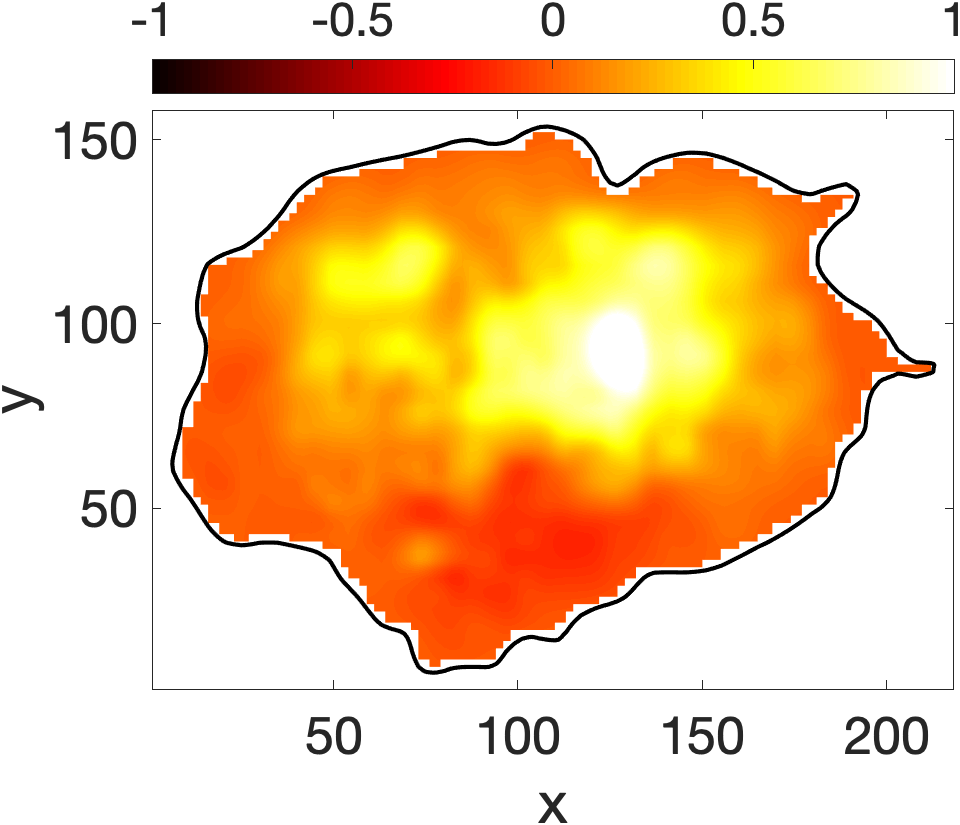}&
\hspace{-.8em}
\includegraphics[width=30mm]{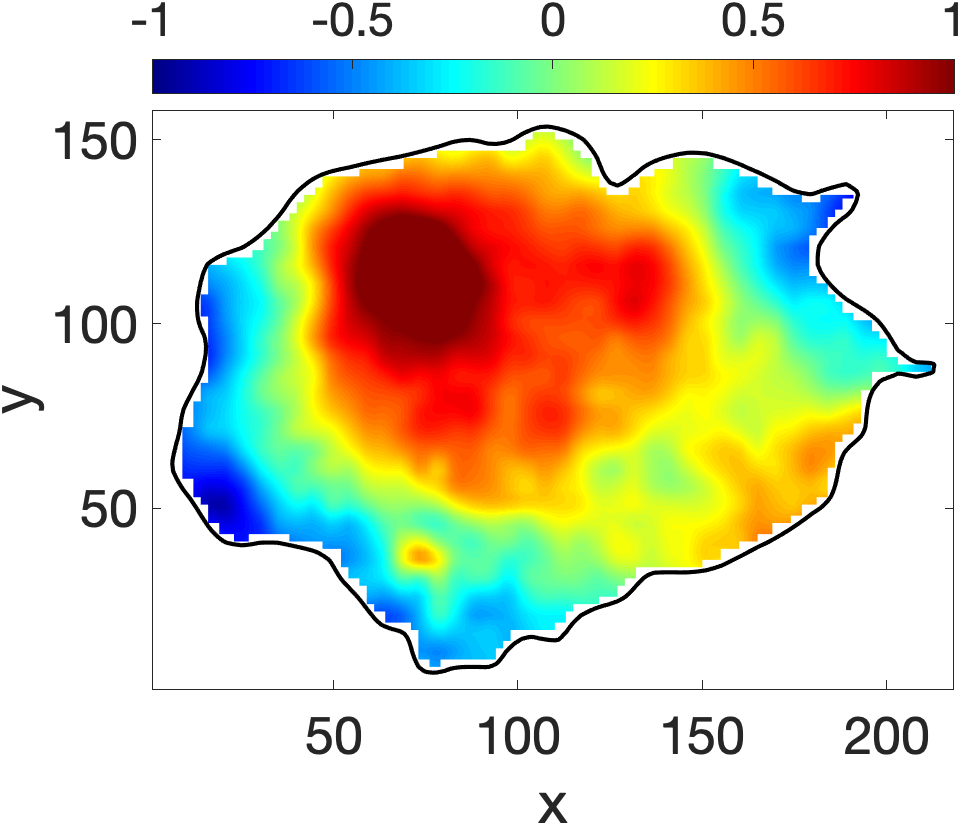}&
\hspace{-.8em}
\includegraphics[width=30mm]{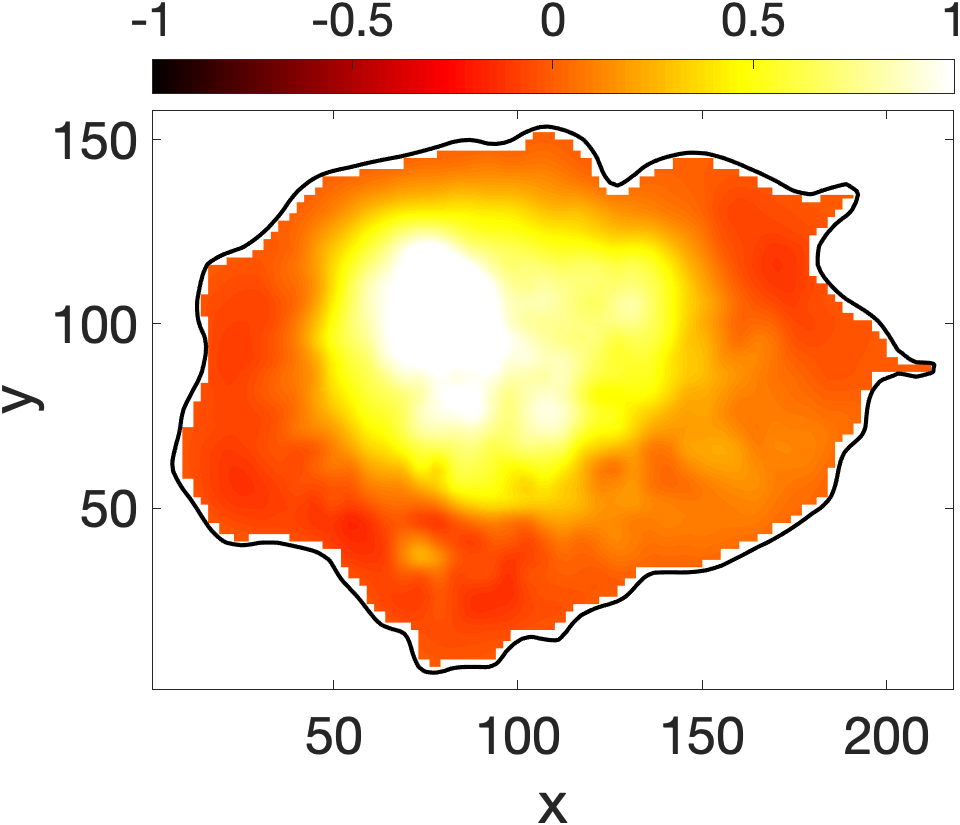}\\

\end{tabular}
\caption{The results of theoretical modeling and the POD and DMD analysis obtained for the sunspot NOAA 12146 are presented for the time intervals T$_{e5}$ (top panel) and T$_{e10}$ (bottom panel). The first column displays the spatial structure of the theoretically modelled fundamental slow body sausage mode for the same shape as the shape of sunspot umbra (NOAA 12146) at two time intervals. The second column shows the spatial structure of the 4$^{th}$ POD mode (top) and the 5$^{th}$ POD mode (bottom), where the dominant frequency of their time coefficient is in the range between 3.3 to 4 mHz. The cross-correlation between the model and the determined POD mode is shown in the third column. The fourth column displays the spatial structure of the DMD modes corresponding to 3.8 mHz (top) and 3.4 mHz (bottom). Finally, the last column contains the cross-correlation between the model (first column) and the DMD mode (fourth column). The solid black line shows the umbra/penumbra boundary. The same configuration was used for Figures \ref{fig:TEE kink1} and \ref{fig:TEE n=2 1}.}
\label{fig:TEE saus}
\end{figure}

\begin{figure}[h]
\centering
\begin{tabular}{ccccc}

\includegraphics[width=30mm]{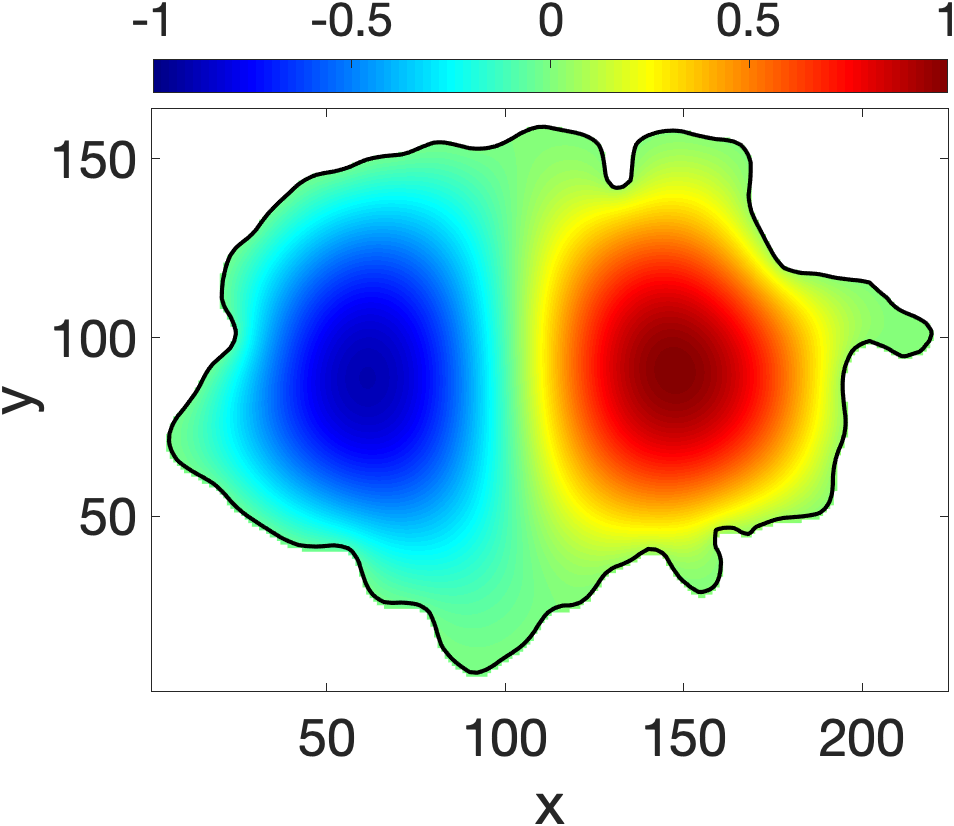}&
\hspace{-.8em}
\includegraphics[width=30mm]{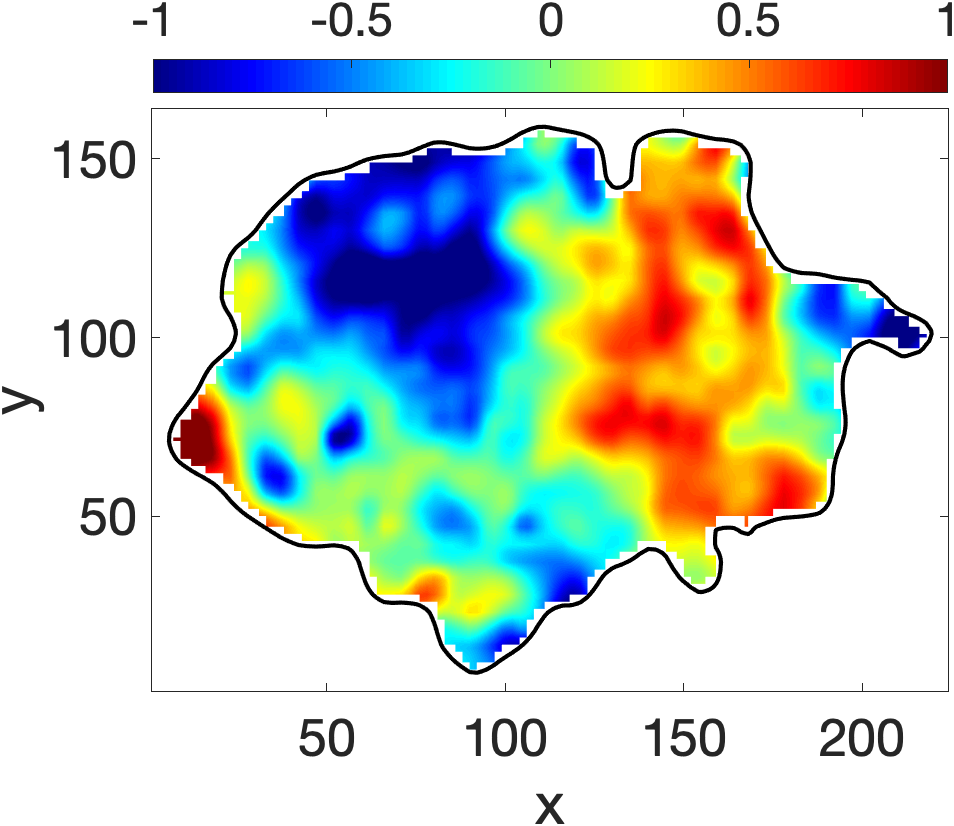}&
\hspace{-.8em}
\includegraphics[width=30mm]{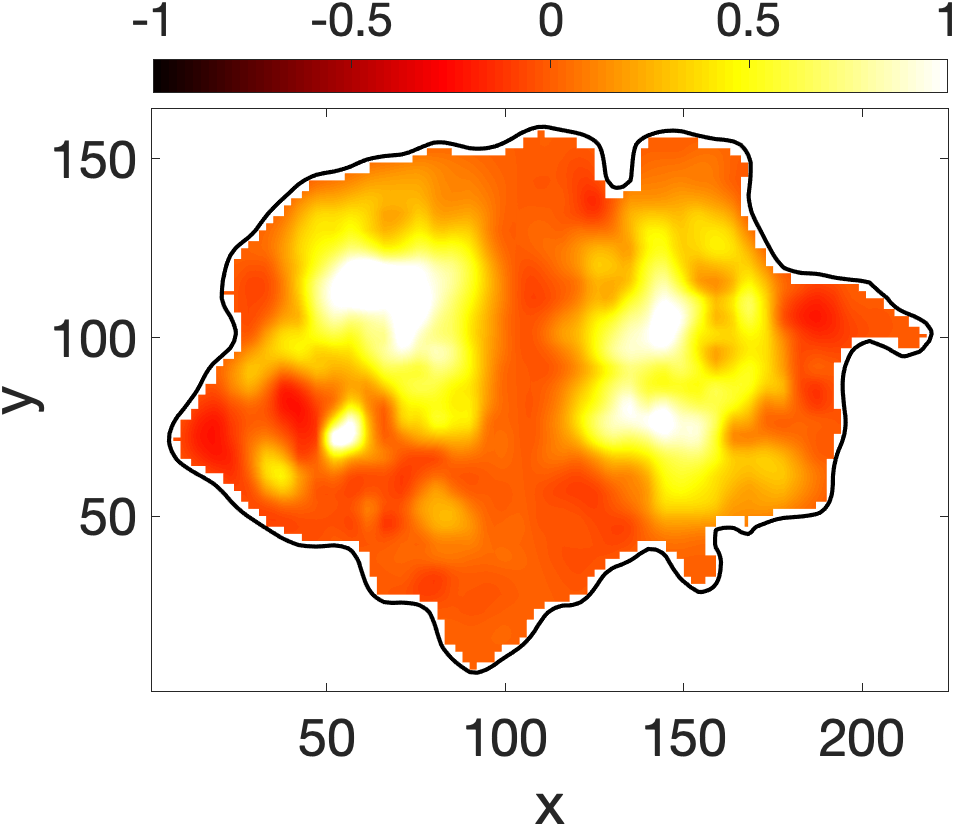}&
\hspace{-.8em}
\includegraphics[width=30mm]{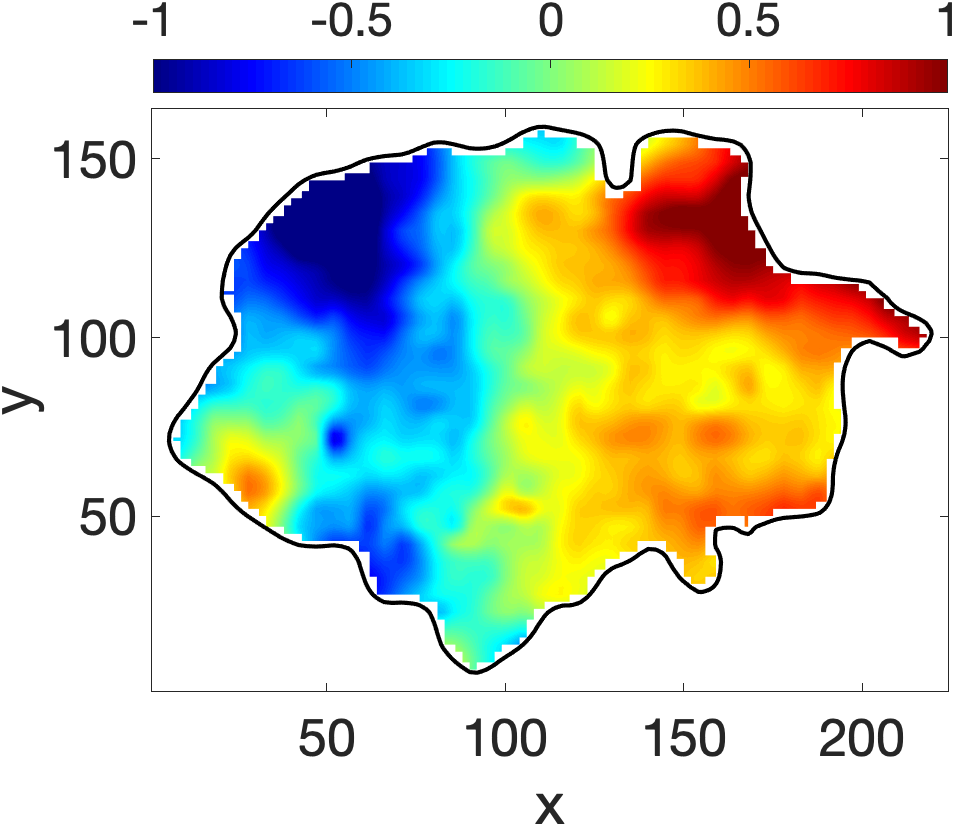}&
\hspace{-.8em}
\includegraphics[width=30mm]{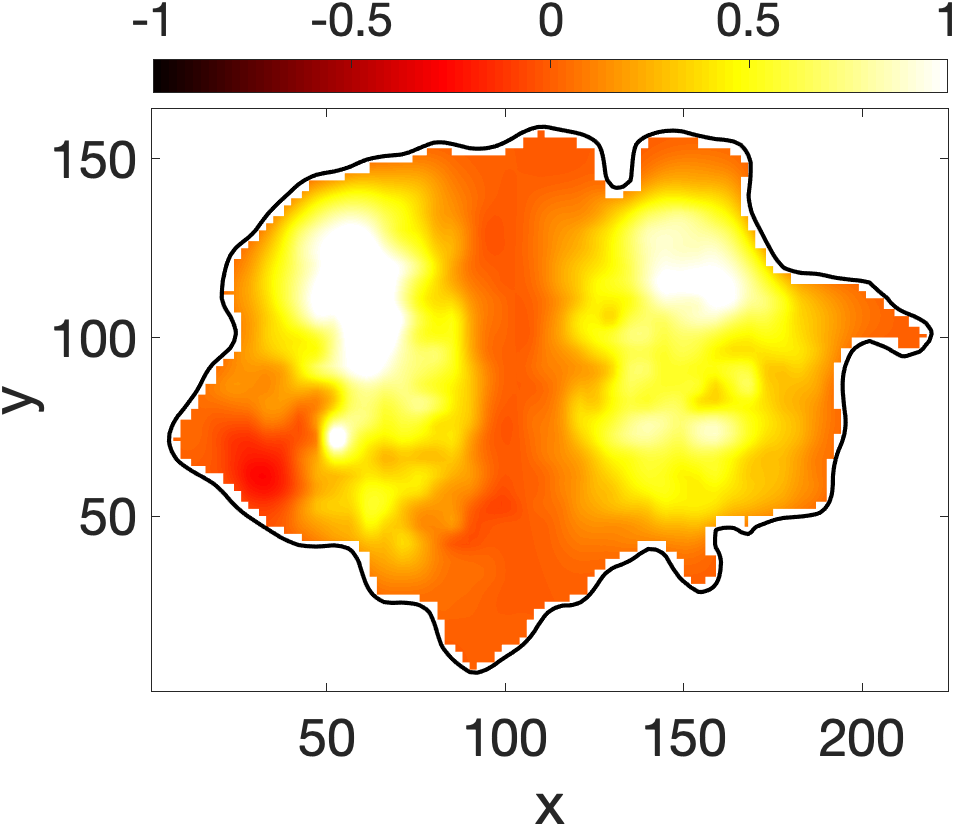}\\

\includegraphics[width=30mm]{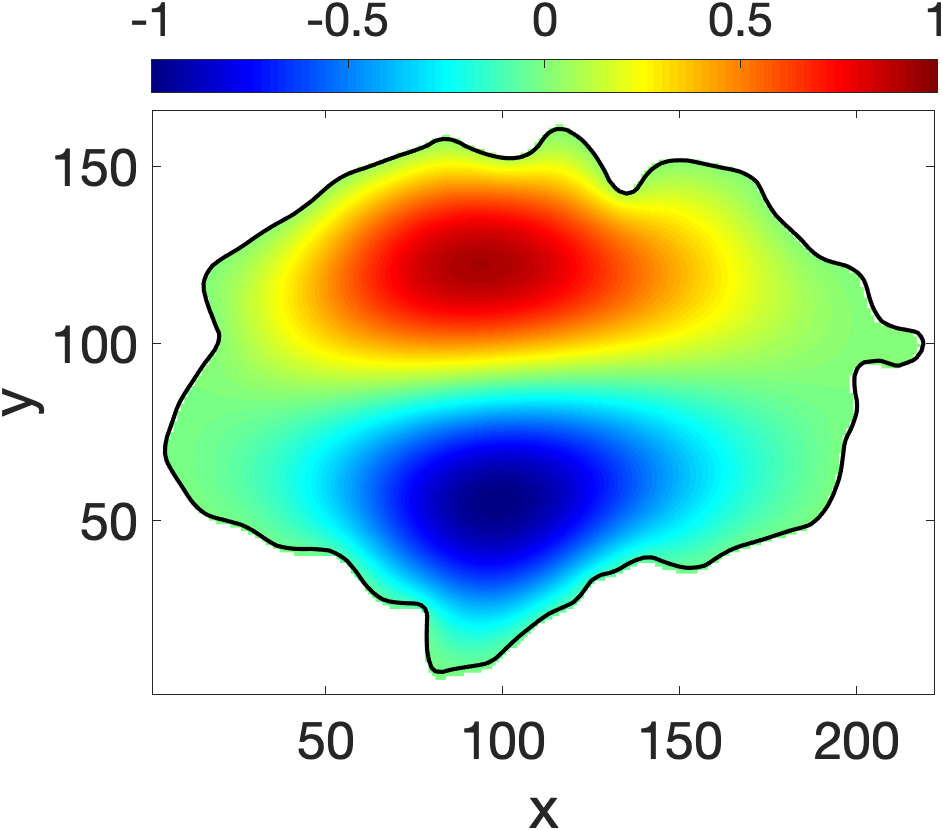}&
\hspace{-.8em}
\includegraphics[width=30mm]{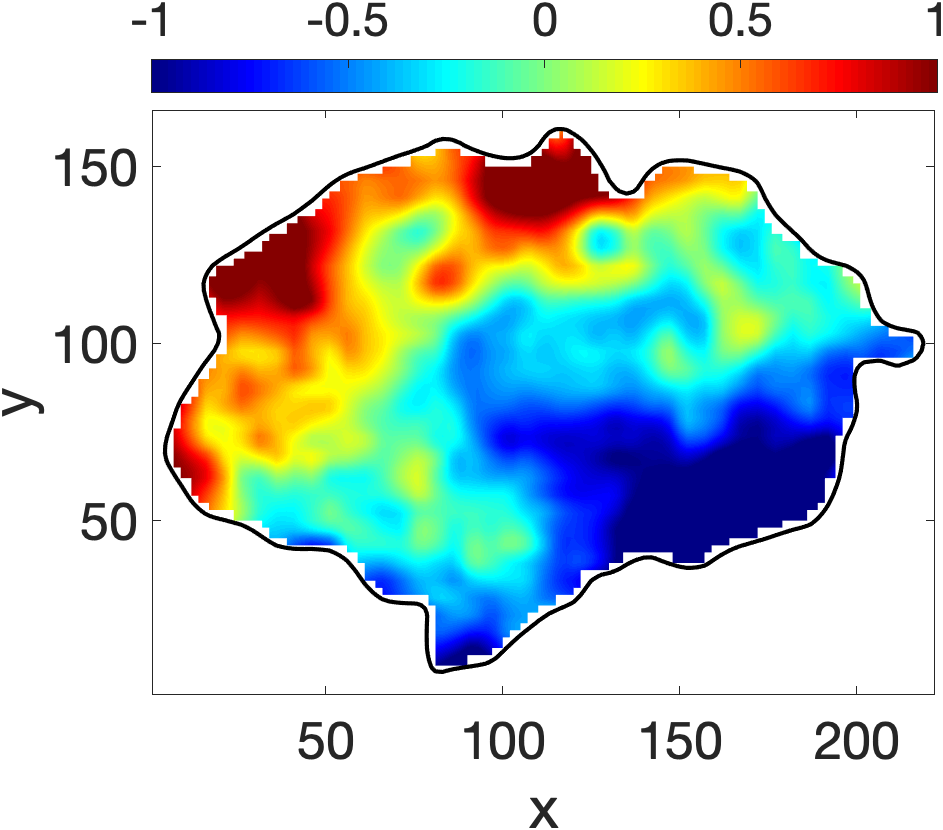}&
\hspace{-.8em}
\includegraphics[width=30mm]{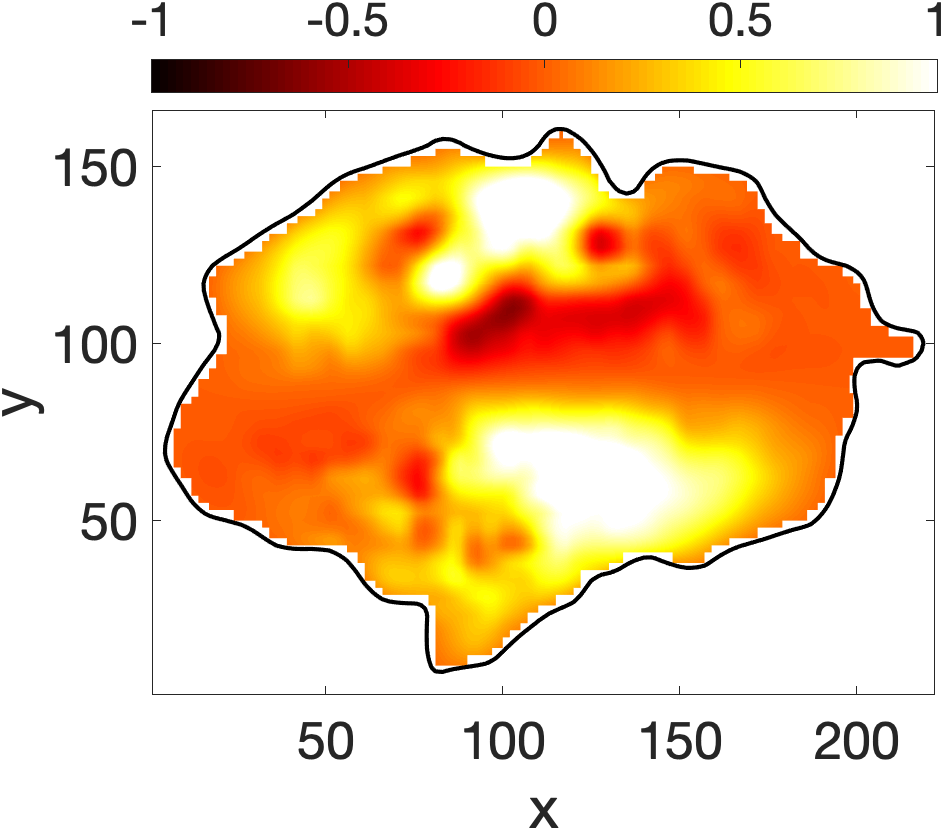}&
\hspace{-.8em}
\includegraphics[width=30mm]{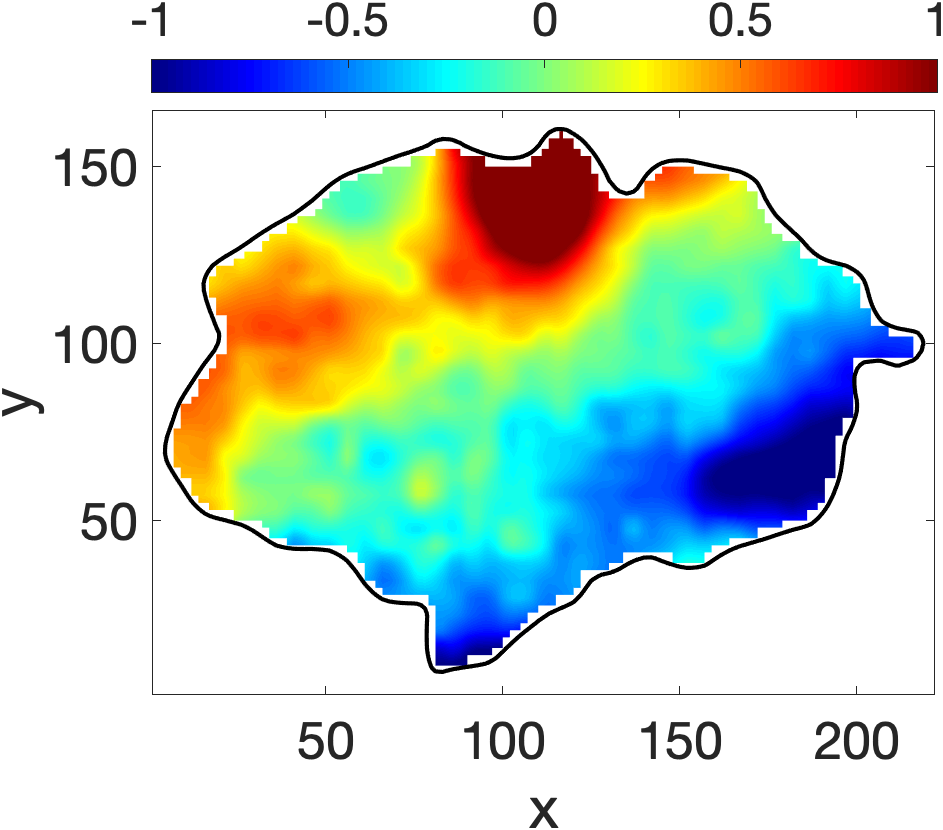}&
\hspace{-.8em}
\includegraphics[width=30mm]{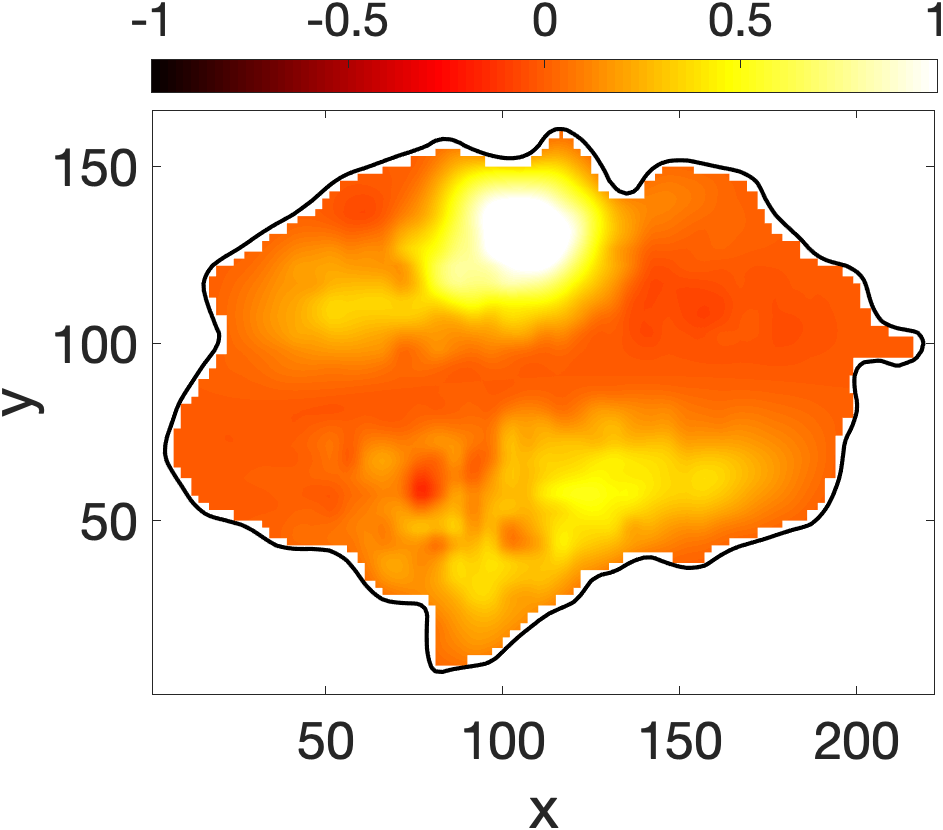}\\

\end{tabular}
  \caption{The same as Figure \ref{fig:TEE saus}, but here the first row shows the 6$^{th}$ POD and the spatial structure of DMD mode corresponding to 3.3 mHz at T$_{e1}$. The second row shows  the 6$^{th}$ POD and the spatial structure of DMD mode corresponding to 3.5 mHz at T$_{e5}$. The first column shows the fundamental slow body kink mode corresponding to T$_{e1}$ (M$_2$, top panel) and T$_{e5}$ (M$_3$, bottom panel).}
\label{fig:TEE kink1}
\end{figure}

\begin{figure}[h]
\centering
\begin{tabular}{ccccc}

\includegraphics[width=30mm]{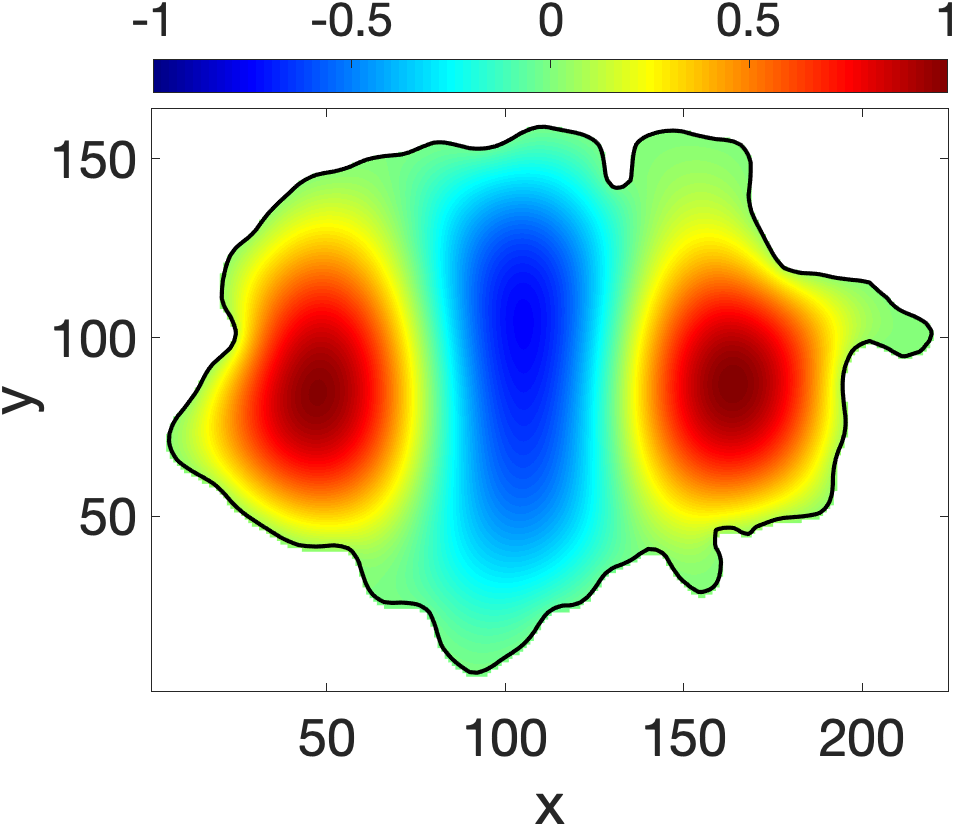}&
\hspace{-.8em}
\includegraphics[width=30mm]{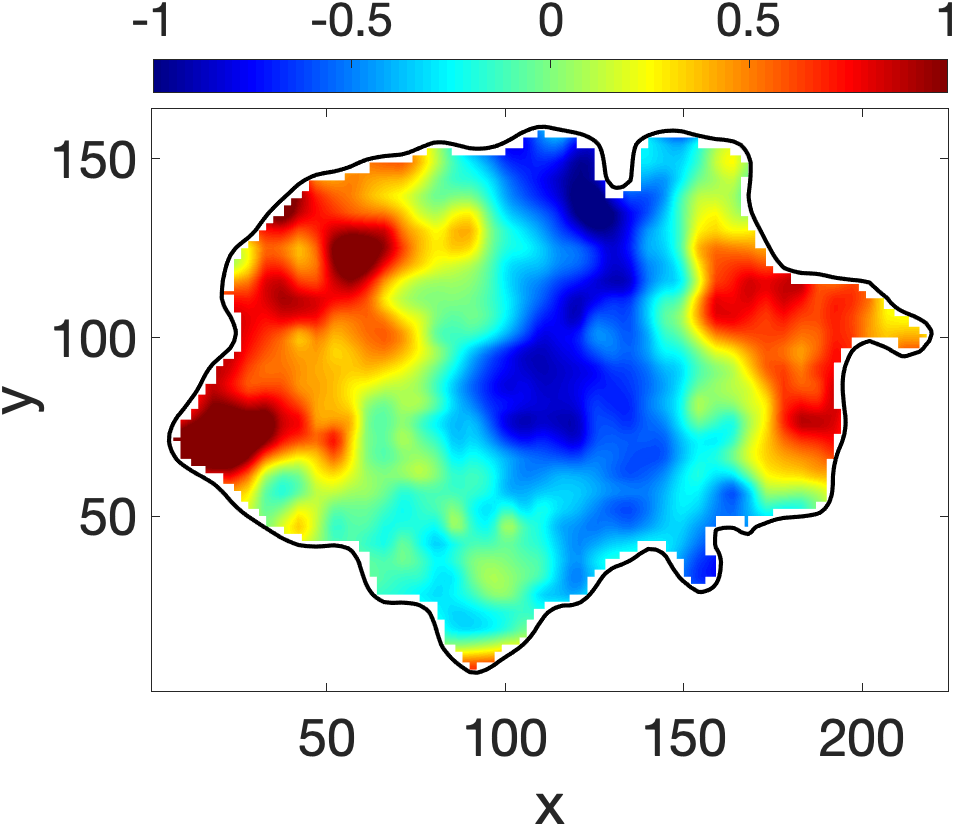}&
\hspace{-.8em}
\includegraphics[width=30mm]{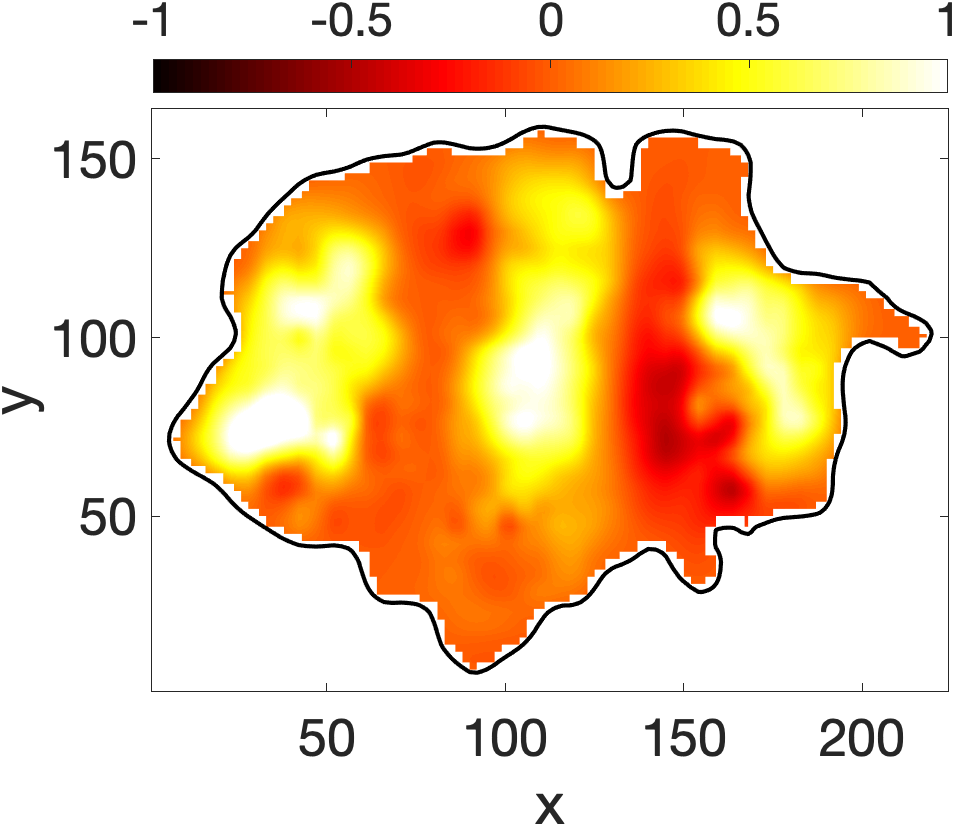}&
\hspace{-.8em}
\includegraphics[width=30mm]{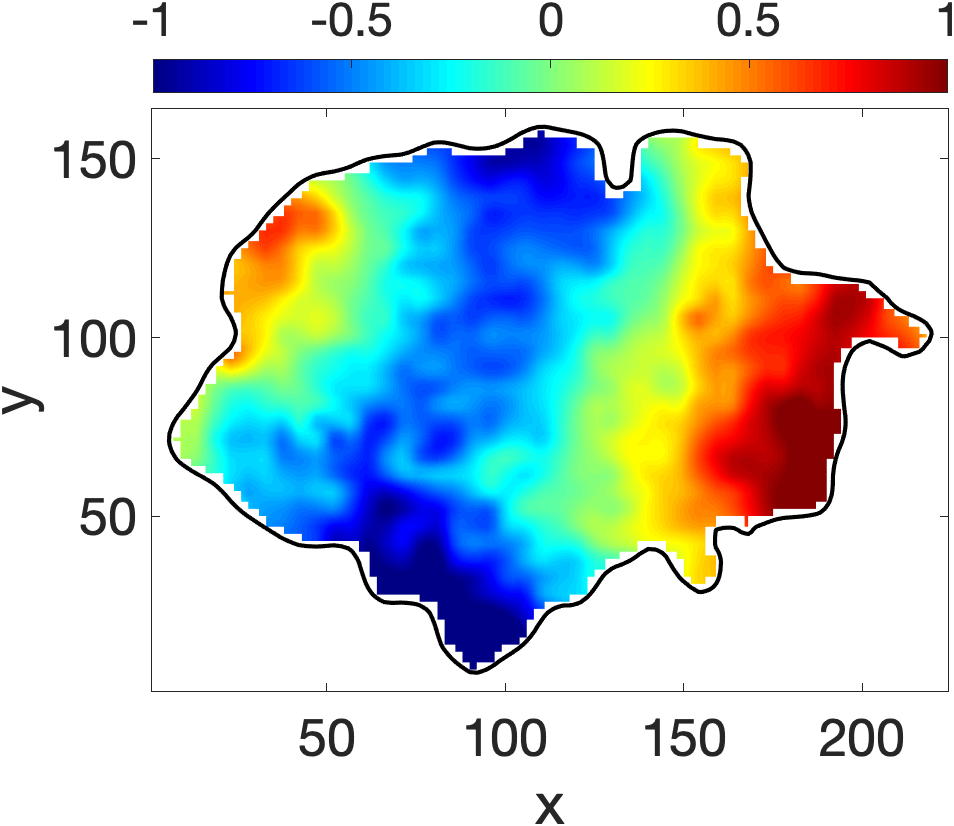}&
\hspace{-.8em}
\includegraphics[width=30mm]{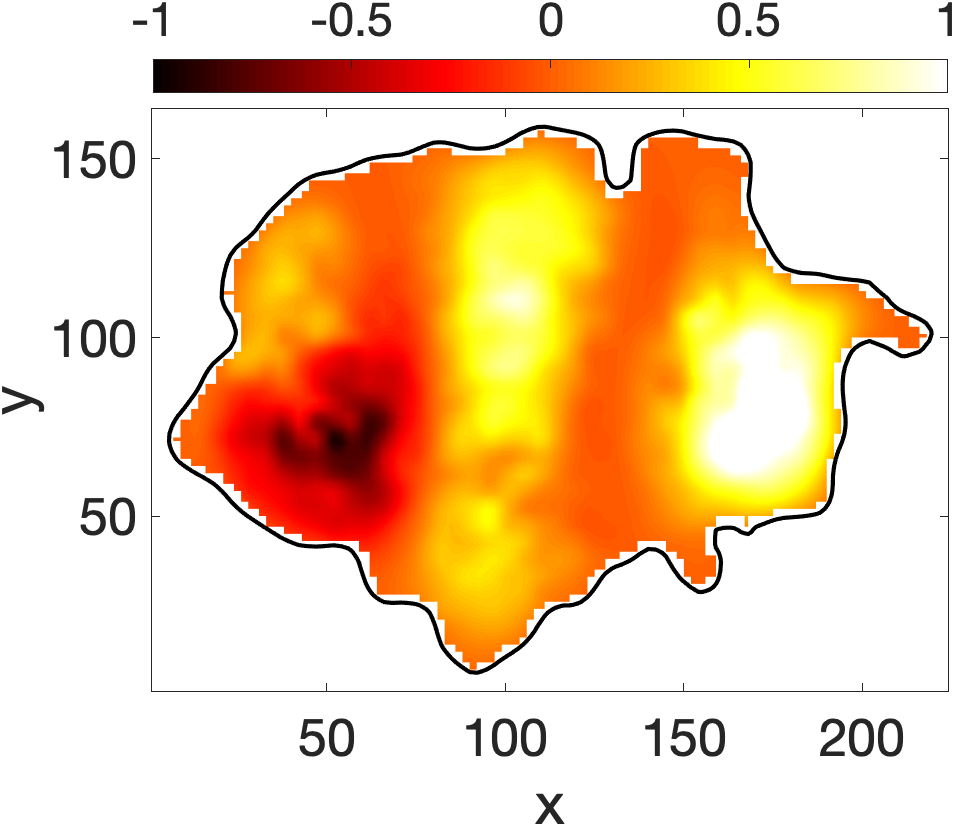}\\

\includegraphics[width=30mm]{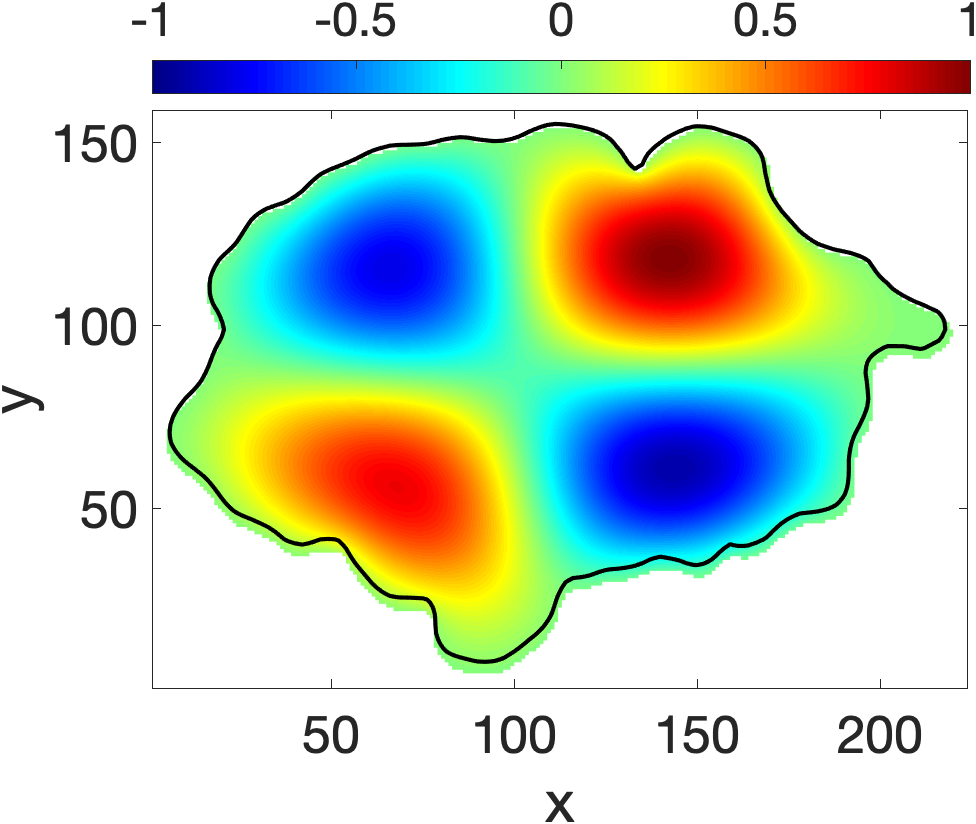}&
\hspace{-.8em}
\includegraphics[width=30mm]{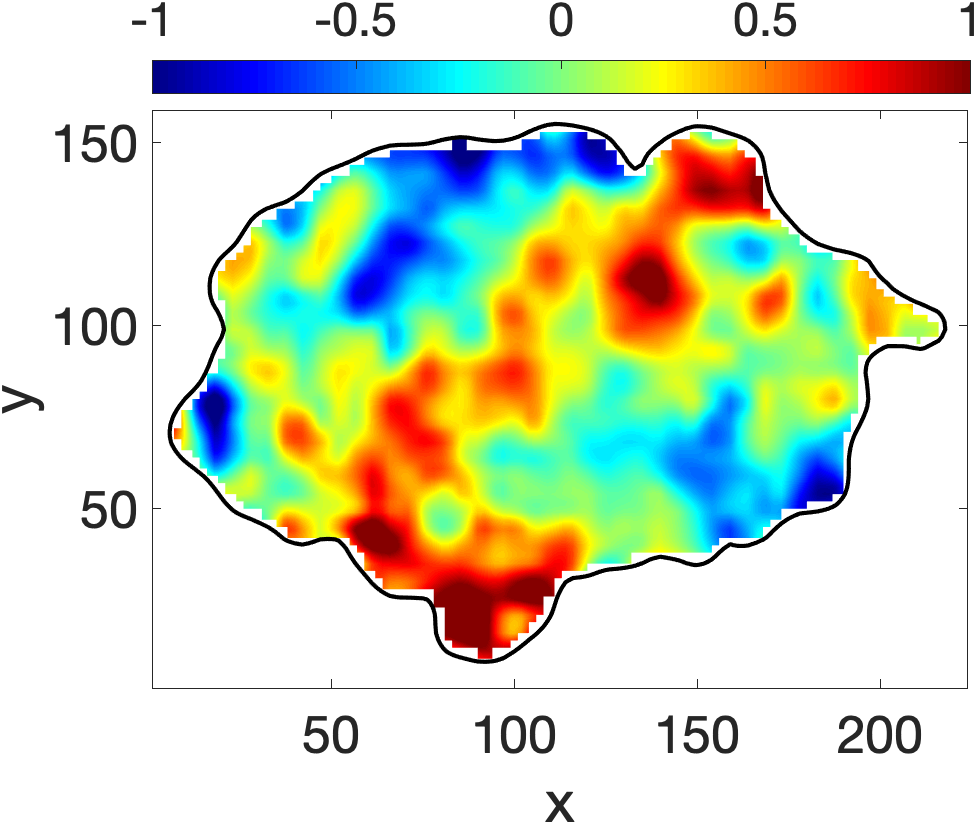}&
\hspace{-.8em}
\includegraphics[width=30mm]{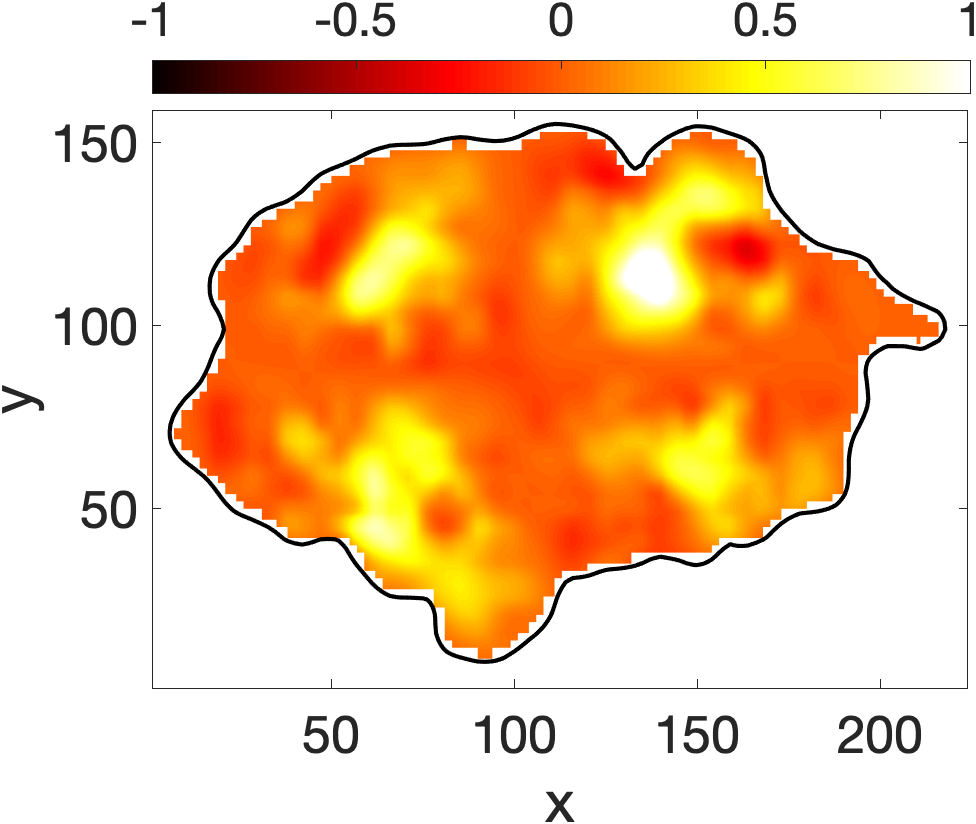}&
\hspace{-.8em}
\includegraphics[width=30mm]{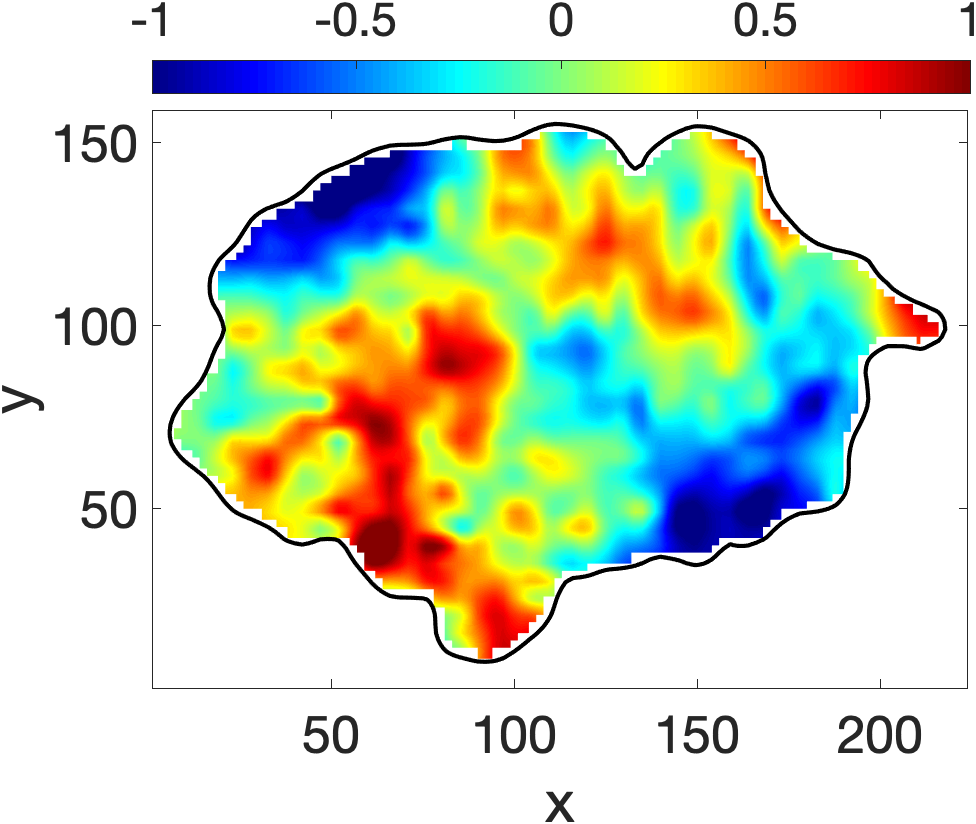}&
\hspace{-.8em}
\includegraphics[width=30mm]{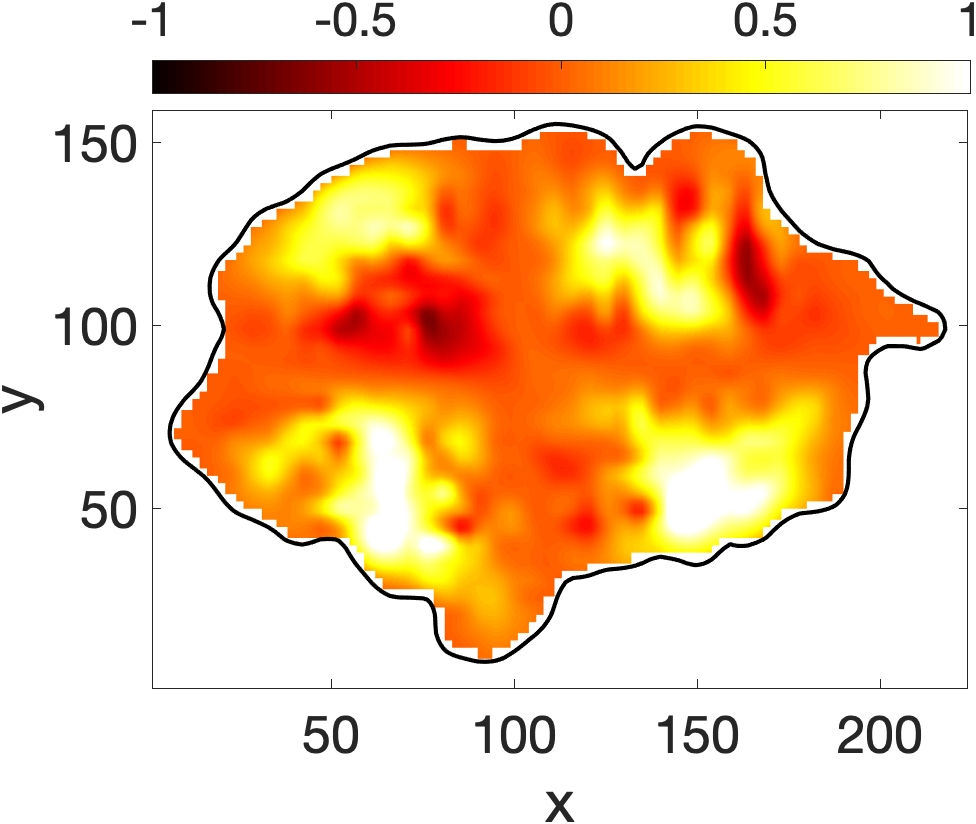}\\

\end{tabular}
  \caption{The same as Figure \ref{fig:TEE saus}, but here the first row shows the 9$^{th}$ POD and the spatial structure of DMD mode corresponding to 2.8 mHz at T$_{e1}$. The second row shows the 10$^{th}$ POD and the spatial structure of DMD mode corresponding to 4.6  mHz at T$_{e3}$. The first column shows the slow body fluting ($n=2$) mode corresponding to T$_{e1}$ (M$_4$, top panel) and T$_{e3}$ (M$_5$, bottom panel).}
  \label{fig:TEE n=2 1}
\end{figure}

For the elliptical sunspot, the first 10 POD modes and the models, at every time interval, are shown in the Appendix \ref{sec_app1}. The POD modes that show a good agreement with the theoretical predictions are presented in Table \ref{Table_E}. In Figures \ref{fig:TEE saus} to \ref{fig:TEE n=2 1}, we display examples of the observed modes obtained through the POD and DMD analysis at different time intervals, providing the cross-correlation analysis with the theoretical mode that corresponds to the realistic shape of the sunspot. The first MHD mode is the fundamental slow body sausage mode, which is observed in most time intervals, as shown in Table \ref{Table_E}. Figure \ref{fig:TEE saus} shows the fundamental slow body sausage mode at T$_{e5}$ and T$_{e10}$. The first row of Figure \ref{fig:TEE saus} shows the spatial structure of the 4$_{th}$ POD mode and the DMD mode corresponding to 3.8 mHz at T$_{e5}$, while the second row shows the spatial structure of the 5$_{th}$ POD mode and the DMD mode corresponding to 3.4 mHz at T$_{e10}$. The dominant frequency of the time coefficient of the POD modes is in the range of 3.3 to 4 mHz.

The second MHD mode that we have identified is the fundamental slow body kink mode, as shown in first row of Figure \ref{fig:TEE kink1}. The first row of Figure \ref{fig:TEE kink1} shows the 6$^{th}$ POD mode and the DMD mode that correspond to 3.3 mHz at T$_{e1}$, that has an azimuthal symmetry corresponding to the slow body kink mode. The second row displays the 6$^{th}$ POD mode and the DMD mode that corresponds to 3.5 mHz at T$_{e5}$. Since the  amplitude of the observed modes in the second row of Figure \ref{fig:TEE kink1} increases as we approach the edge of the umbra, and since the cross-correlation between the observed modes and the theoretical model gives a pattern closed to the slow body mode, the mode shown in the second row of Figure \ref{fig:TEE kink1} is identified as a fast surface kink mode. 

The last MHD mode that we have identified in the elliptical sunspot is the slow body fluting ($n=2$) mode, as shown in Figure \ref{fig:TEE n=2 1}. The first row of Figure \ref{fig:TEE n=2 1} is showing the 9$^{th}$ POD mode and the DMD mode that correspond to 2.8 mHz at T$_{e1}$, while the second row shows the 10$^{th}$ POD mode and the DMD mode that correspond to 4.6 mHz at T$_{e3}$. It is evident that the morphology of this wave is considerably changing in time as the shape of the waveguide is changing.

\begin{figure*}
    \centering
    \includegraphics[width=\textwidth]{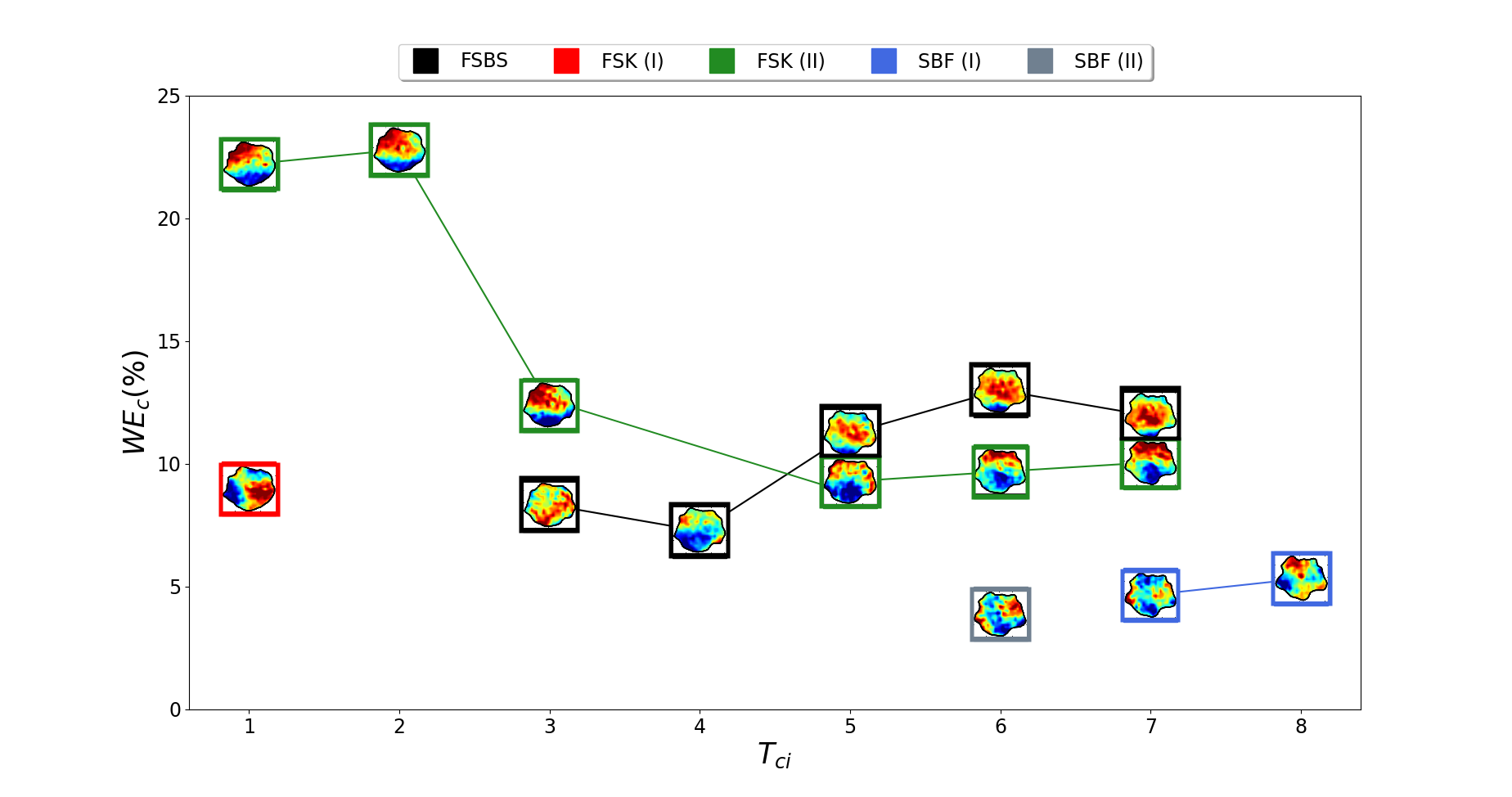}
\caption{Wave contribution from the observed MHD modes to the total variance of the signal in the case of the circular sunspot, WE$_{c}$, as function of time. The values of WE$_{c}$ correspond to the contribution of the  POD mode linked to MHD wave mode in each time interval, T$_{ci}$. Different modes are shown by different colours, as indicated by the legend in the upper part of the plot. Here, FSBS stands for the fundamental slow body sausage ($n=0$) mode, FSK for fast surface kink ($n=1$) mode and SBF for slow body fluting ($n=2$) mode, while (I) and (II) refer to two different (perpendicular) directions of the wave polarisation.} \label{fig:S_c}
\end{figure*}

\begin{figure*}
    \centering
    \includegraphics[width=\textwidth]{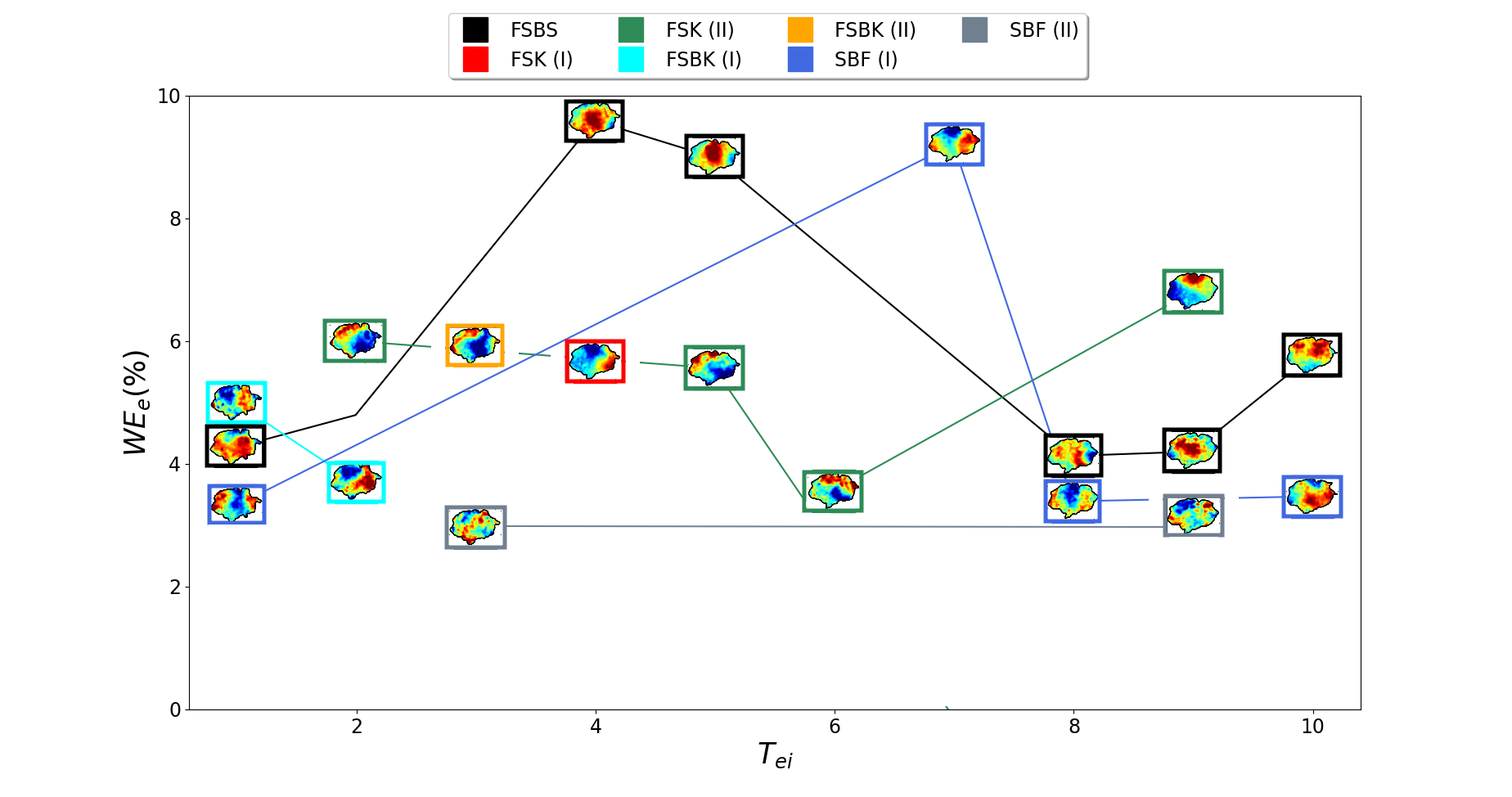}
     \caption{Wave contribution from the observed MHD wave modes to the total variance of the signal for the elliptical sunspot, WE$_{e}$, as a function of time. The values of WE$_{e}$ correspond to the contribution of the POD mode linked to MHD wave mode at each time interval, T$_{ei}$. The identified MHD modes are shown by different colours and specified by the legend in the upper part of the plot. The abbreviation of possible modes is identical with the one used in Figure \ref{fig:S_c}.} \label{fig:S_e}
\end{figure*}

The POD decomposes the modes and provides their ranking according to their contribution to the overall energy of the signal \citep[][]{albidah2020RS,albidahApJ}. The time evolution of the wave energy (WE) contribution is calculated for the case of circular sunspot (WE$_{c}$) and elliptical sunspot (WE$_{e}$). The observed WE of POD modes which were identified as MHD wave modes, i.e the modes with the highest correlation value between theoretical and observed counterparts are shown in Figures \ref{fig:S_c} and \ref{fig:S_e}. The time dependent behaviour of WE$_{c}$ and WE$_{e}$ is similar and, as expected, the main contribution to the wave energy is provided by fundamental modes.
For the case of circular sunspot (Figure \ref{fig:S_c}) the fast surface kink mode has the dominant contribution initially but then the fundamental slow body sausage mode appears with higher contribution starting from the time interval $T_{c5}$. The slow body fluting ($n=2$) mode has, approximately, twice smaller contribution. This change in behaviour could be related to the change in the wave driver from being azimuthally anti-symmetric to symmetric. The higher order wave modes in this case are less frequently excited.

In case of the elliptical sunspot illustrated in Figure \ref{fig:S_e}, the fundamental slow body sausage and fast surface kink modes have an anti-phase behaviour in time compared with the circular sunspot with regards to growth and decay in contribution.  In contrast to the previous case, the higher order modes, i.e slow body fluting ($n=2$) modes (with two different polarizations) are notably present.  
The contribution to the signal's energy attributed to these modes grows steadily in time, followed by a sudden decay. The body and surface kink ($n=1$) modes were observed with a different polarization of the wave along the major and the minor axes. Additionally, the kink mode were observed with an accidental appearance of their contribution along the time intervals as shown in Figure \ref{fig:S_e}.

\section{Summary and Conclusions}
In the present study, we have focused on the changes in the spatial structure of umbral slow body modes by taking into account the dynamical change of the shape of two sunspots' umbral boundaries in time. We employed the same theoretical model used successfully by \citet{stangalini2022large} who showed that that the perturbations in the line-of-sight velocity component decay very rapidly at the umbra/penumbra boundary. Every acoustic or MHD wave model requires a boundary condition to solve the governing equations. For $p$-mode models the Lagrangian pressure perturbation is taken to be exactly zero at the Sun’s surface. There are uncertainties in both the size of the pressure perturbation and the exact location of the Sun’s ``surface”. However, the approximation has worked well. The benefits outweigh the drawbacks. With the standard \citet{edwin1983wave} cylinder model the background quantities have a tangential discontinuity at the flux tube boundary. Again, this is an approximation of reality, but it is needed to derive dispersion relations. It was shown by \cite{aldhafeeri2022comparison} that the outer slow body mode node and boundary are very close together in space, which justifies our approximation. For our theoretical model, the employed boundary condition has the vast benefit that we can use the exact irregular observed umbral shape to model slow body mode eigenfunctions. This is the main goal of the paper, i.e. to investigate how the changing shape of sunspot umbrae affects the spatial structure of the slow body modes. The standard cylinder model is simply not applicable to study the evolution of umbral slow body modes identified by our analysis because the model prescribes that the cross-sectional shape of the waveguide is always fixed as a perfect circle.

The POD and DMD techniques used here are the same as the ones presented earlier by \citep[e.g.][]{albidah2020RS,albidahApJ}, however, in these studies the shapes of the umbra were considered to be stationary. Each analyzed Doppler velocity data set was divided equally into ten overlapping time intervals and then POD and DMD were applied separately to the data sets within each time interval. The identified modes were cross-correlated with the theoretically obtained counterparts which were determined by taking the same shape of the umbra within selected time intervals. The comparison was calculated on a pixel-by-pixel basis of the cross-correlation analysis. It was shown that even minor changes in the shape of umbrae alter the morphology and nature of wave modes, particularly the higher-order modes. 

We found that the main contribution to the overall energy of the signal is provided by fundamental modes and the energy contribution from waves may  considerably change in time.
Given the far-from-ideal spatial and temporal resolution of the HMI instrument, it is likely that some significant information on the dynamics within the sunspots cannot be acquired or resolved. As a result, there could be other higher-order modes or further overtones that cannot be identified in our analysis. Nevertheless, POD/DMD have shown their ability to identify the higher order modes and overtones for observations that have a much better spatial and temporal resolution, e.g. the H$\alpha$ time series \citep[][]{albidahApJ}. From the performed analysis it follows that the optimal time length needed to assess the impact of changes in the umbral shape on observed wave modes was 37 minutes for the circular sunspot and 60 minutes for the elliptical sunspot. 
This estimate, however, may depend on the time-spatial resolution of the sunspot data. 

Our results show that the fundamental modes are less sensitive to the change of the sunspot shape. We conclude, that wave modes variations within umbral regions may be related to changes in the nature of the driver. For higher modes, proper detection needs to consider the variation of umbral shape as their theoretical eigenfunctions present higher variability in time, affecting the identification of the POD mode and thereby their contribution to the energy of the signal. We found that the analysis of the data using optimal time intervals is essential to capture the dynamical contribution of each mode to the observed signal, allowing in future analysis to find significant statistics on the temporal evolution of wave modes. Moreover, a spectral principal component analysis based on a sequence of time intervals may allow for the correlation between the observed wave mode dynamics and the nature of the wave driver. 

\section{Acknowledgement}
V.F., G.V., and I.B. are grateful to the Royal Society, International Exchanges Schemes, collaborations with Pontificia Universidad Catolica de Chile, Chile (IES/R1/170301),  Aeronautics Institute of
Technology, Brazil, (IES/R1/191114), Monash University, Australia (IES/R3/213012) and Instituto de Astrofisica de Canarias, Spain (IES/R2/212183). V.F. and S.S.A.S. are grateful to Science and Technology Facilities Council (STFC) grant ST/V000977/1. This research has also received financial support from the ISEE, International Joint Research Program  (Nagoya University, Japan) and the European Union’s Horizon 2020 research and innovation program under grant agreement No. 824135 (SOLARNET). 
 A.A. acknowledges the Deanship of Scientific Research (DSR), King Faisal University, Al-Hassa (KSA), for financial support under grant Track (grant No.1267). 
We are also very grateful to Mr. Miguel S. A. Schiavo for his patience and encouragement.

\section{Appendix}

\subsection{POD and model time interval} \label{sec_app1}
The spatial structure of the first 10 POD modes, for every time interval (T$_{ci}$), for the circular and (T$_{ei}$) of the elliptical sunspots are shown in Figures \ref{fig:POD_C} and \ref{fig:POD_E}, respectively. The MHD modes that correspond to the exact shape of the umbra, for every time interval, are shown in Figure \ref{fig:model_circule}, for the circular sunspot,  and \ref{fig:model_elliptc}, for the elliptical sunspot. 

\subsection{Correlation} \label{sec_app2}
The cross-correlation between the observed and the theoretical modes has been calculated on a pixel-by-pixel basis for all time intervals. For every correlation matrix we have taken the summation of the pixels, and hence for every single correlation of the observed and the theoretical mode is represented as an integer. As a result, we have obtained the correlation panels shown in Figures \ref{fig:Corr_stat C} and \ref{fig:Corr_stat E} for the circular and elliptical sunspot, respectively. Every panel refers to different time interval. For example, upper left panel of Figure \ref{fig:Corr_stat C}, labelled by $T_{c1}$, refers to the correlation of the POD modes in the first row of Figure \ref{fig:POD_C} and theoretical modes in the first row of Figure \ref{fig:model_circule}. 

At the end, every column of the correlation panels are normalised by the maximum value of that column along all time intervals. This step is required to make the correlation of higher order modes visible. Without this step, the correlation of the sausage and kink mode will be the dominant and other modes will not be visible as high correlation. However, the higher correlations needs to be checked and validated in order to be considered.

\subsection{Further time intervals of POD analysis}
\begin{figure}
    \centering
    \includegraphics[width=160mm]{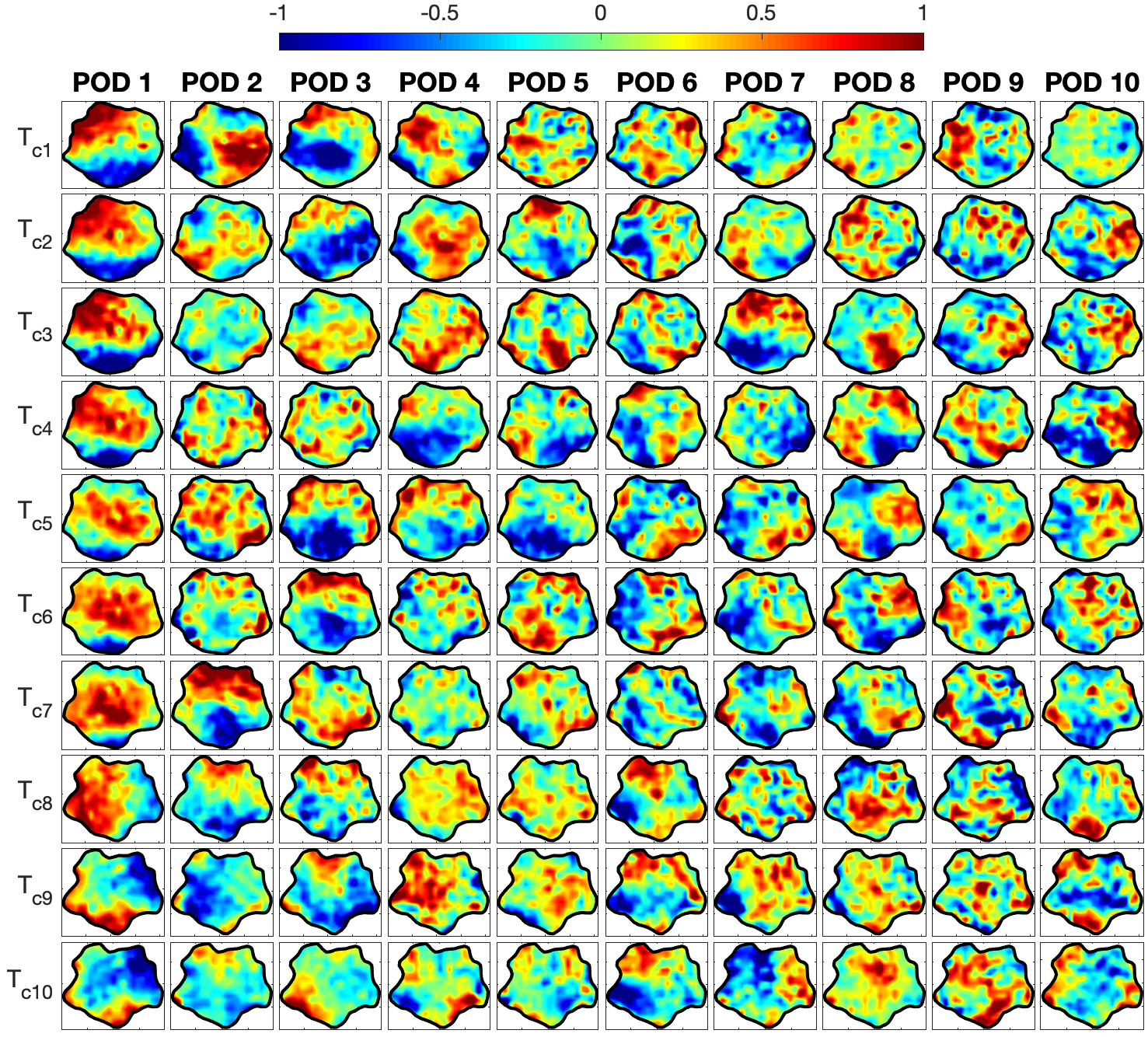}
     \caption{This figure represents the first 10 POD modes of the circular sunspot. Every column shows a POD mode and the rows shows how the modes are changing along the time intervals, T$_{ci}$, of the data time series. Every time interval contains 50 images, and has a duration of 37.5 minutes. Every time interval is shifted by 20 images, i.e. the initial time of T$_{ci+1}$ is after the initial time of T$_{ci}$ by 20 images, corresponding to 15 minutes.}
     \label{fig:POD_C}
\end{figure}

\begin{figure}
    \centering
    \includegraphics[width=160mm]{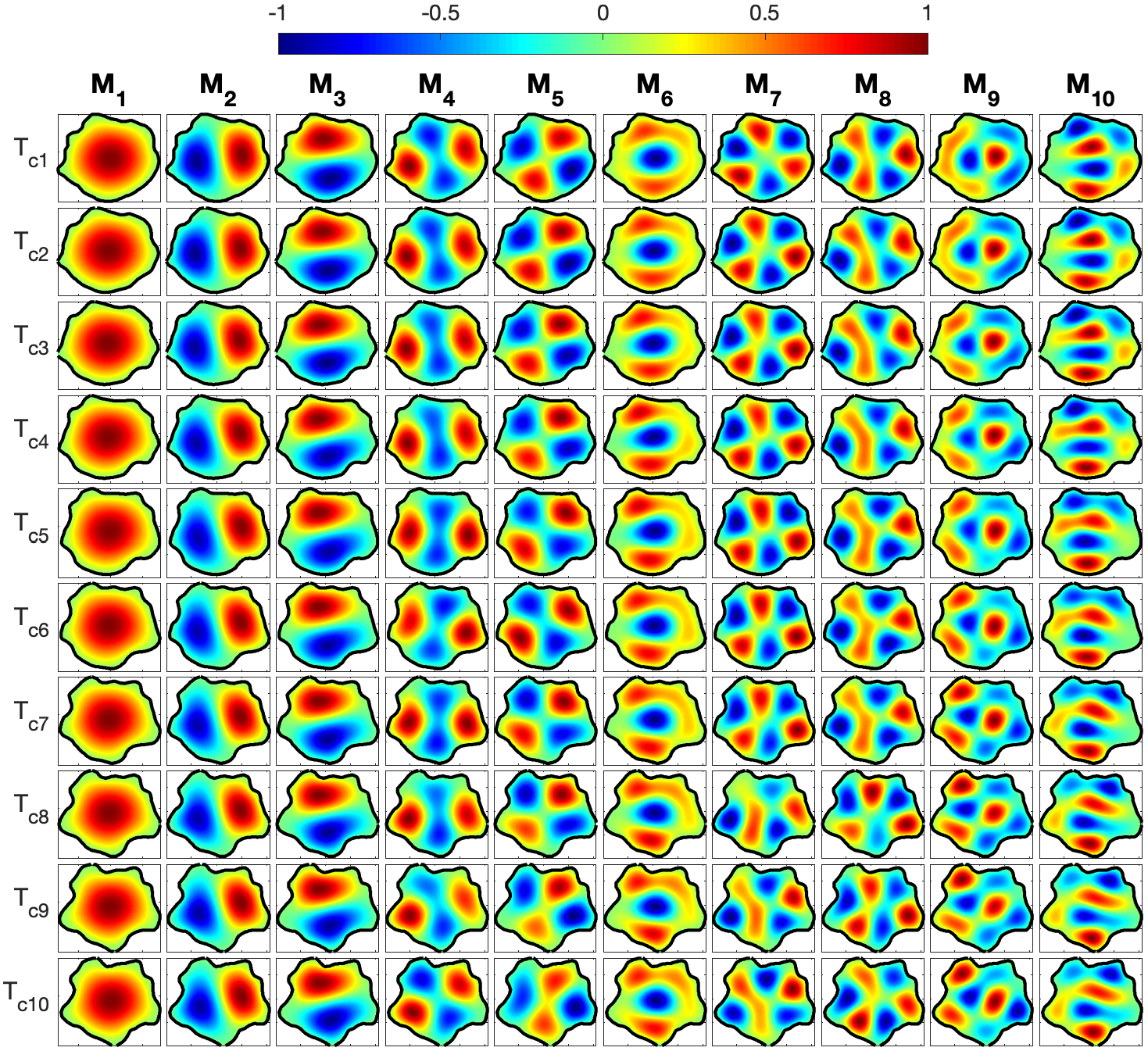}
     \caption{The theoretical eigenfunctions that correspond to the changing shapes of the observed circular sunspot (see Figure \ref{fig:POD_C}). Every row shows the spatial structure of the models at different times and changing shape. Columns represent different types of slow body modes, and they are labelled by M$_i$, where $i=1, \dots, 10$. In particular, M$_1$ stands for the fundamental sausage ($n=0$), M$_2$ and M$_3$ denote the fundamental kink ($n=1$), M$_4$ and M$_5$ are showing the fluting ($n=2$), M$_6$ is showing the sausage overtone ($n=0$), M$_7$ and M$_8$ are showing the fluting ($n=3$) and the last two columns (M$_9$ and M$_{10}$) are showing the kink overtone ($n=1$).}
     \label{fig:model_circule}
\end{figure}

\begin{figure}
    \centering
    \includegraphics[width=180mm]{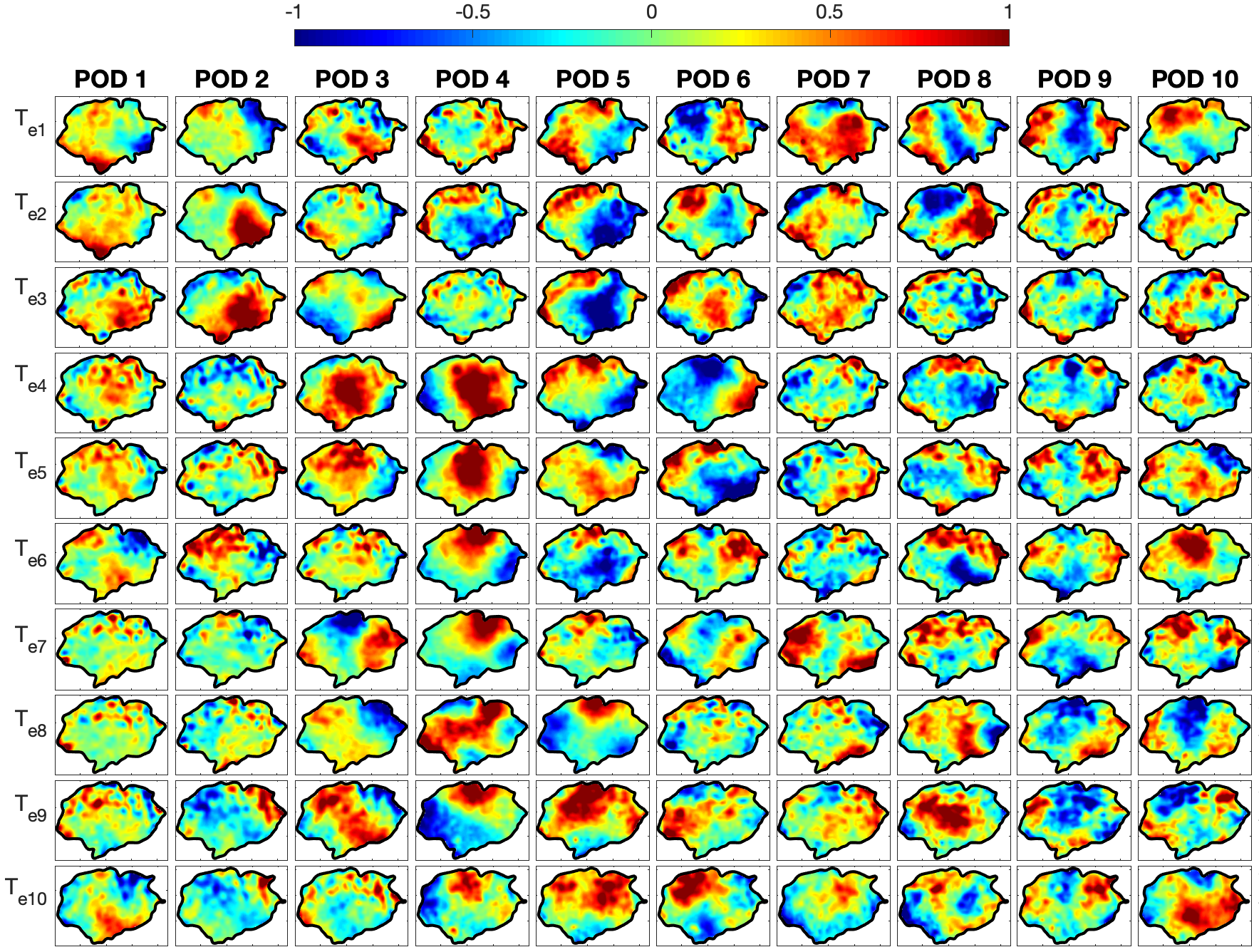}
     \caption{
    This figure represents the first 10 POD modes of the elliptical sunspot. Every column shows a POD mode and the rows shows how the modes are changing along the time intervals, T$_{ei}$, of the data time series. Every time interval contains 80 images, and has a duration of 60 minutes. Every time interval is shifted by 40 images, i.e. the initial time of T$_{ei+1}$ is after the initial time of T$_{ei}$ by 40 images, corresponding to 30 minutes.}
     \label{fig:POD_E}
\end{figure}

\begin{figure}
    \centering
    \includegraphics[width=180mm]{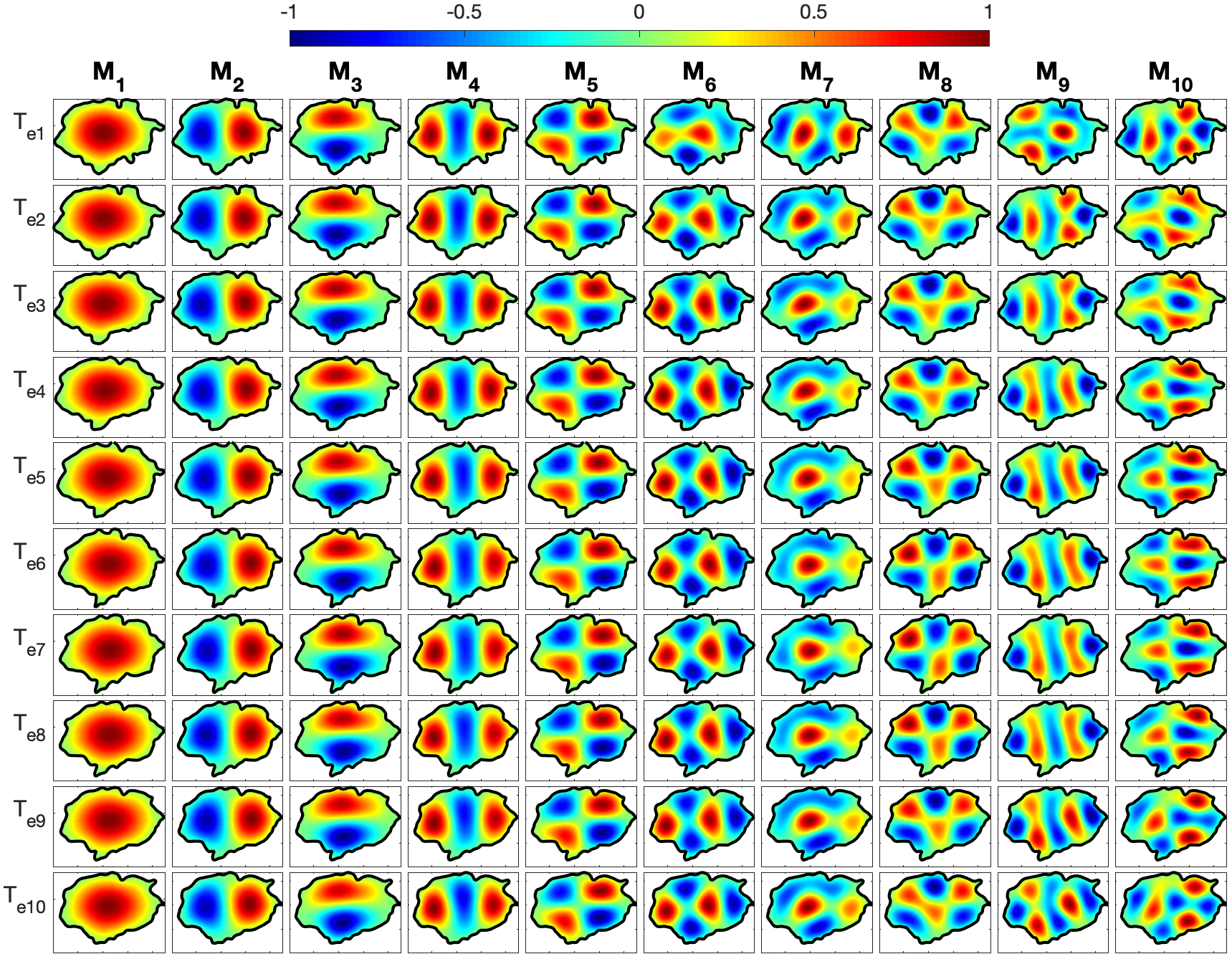}
     \caption{The theoretical eigenfunctions that correspond to the changing shapes of the observed elliptical sunspot (see Figure \ref{fig:POD_E}). Every row shows the spatial structure of the models at different times and changing shape. Columns represent different types of slow body modes, and they are labelled by M$_i$, where $i=1, \dots, 10$. In particular, M$_1$ stands for the fundamental sausage ($n=0$), M$_2$ and M$_3$ denote the fundamental kink ($n=1$), M$_4$ and M$_5$ are showing the fluting ($n=2$), M$_6$ is showing the sausage overtone ($n=0$), M$_7$ and M$_8$ are showing the fluting ($n=3$) and the last two columns (M$_9$ and M$_{10}$) are showing the kink overtone ($n=1$).}
     \label{fig:model_elliptc}
\end{figure}

\begin{figure}
\centering
\begin{tabular}{ccc}

\includegraphics[width=55mm]{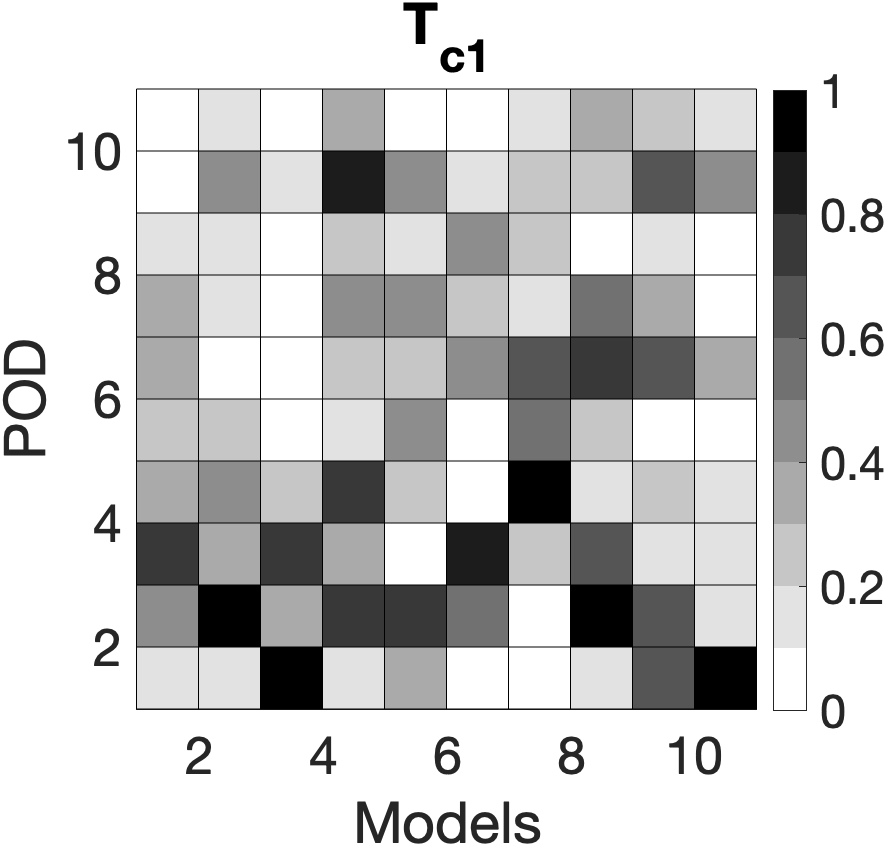}&
\hspace{1.5em}
\includegraphics[width=55mm]{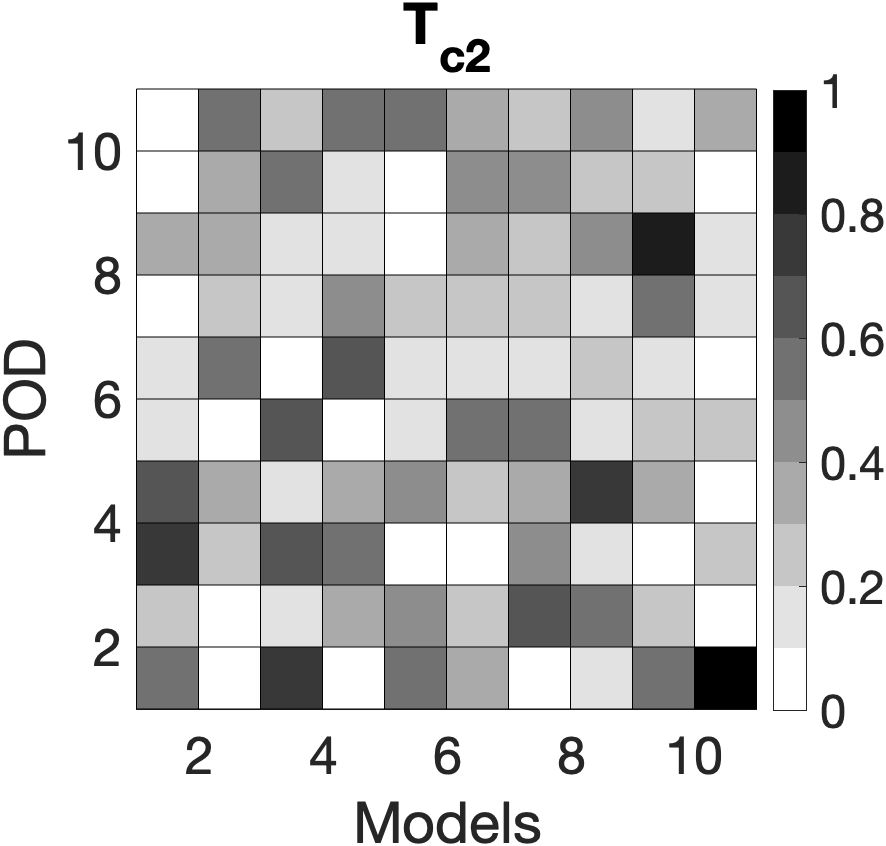}&
\hspace{1.5em}
\includegraphics[width=55mm]{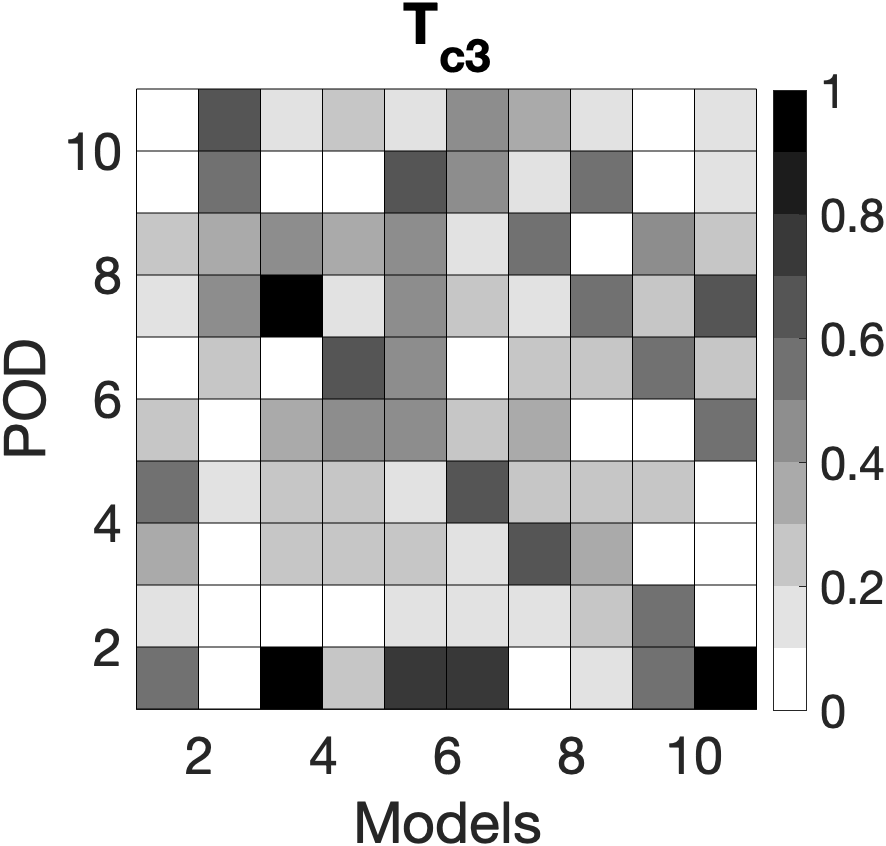}\\

\includegraphics[width=55mm]{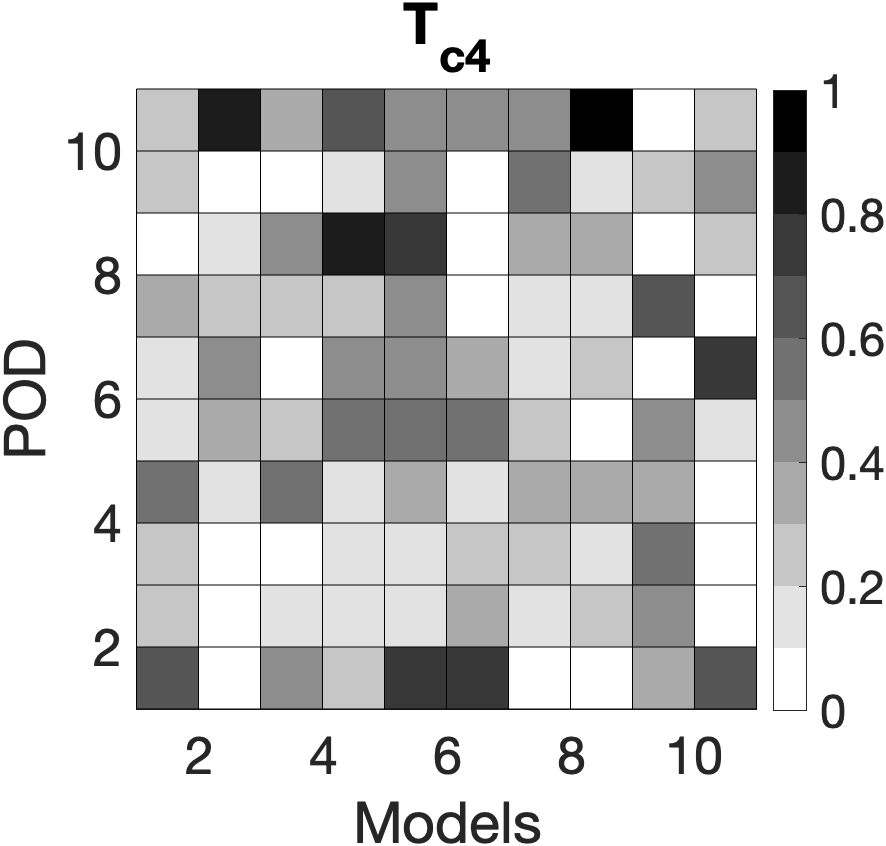}&
\hspace{1.5em}
\includegraphics[width=55mm]{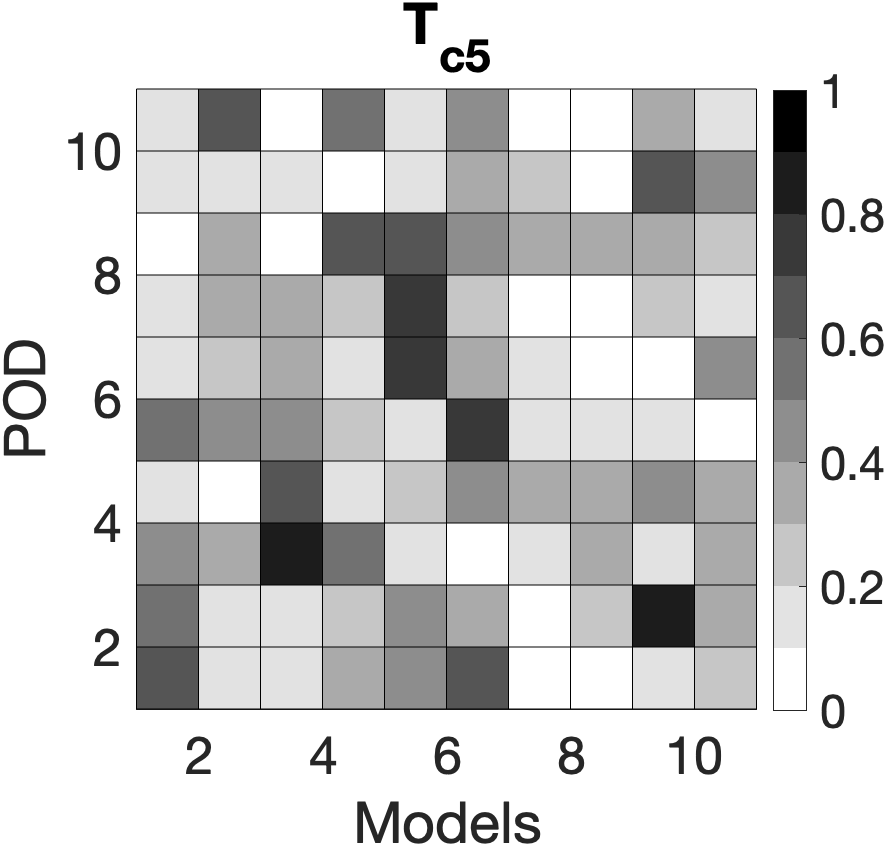}&
\hspace{1.5em}
\includegraphics[width=55mm]{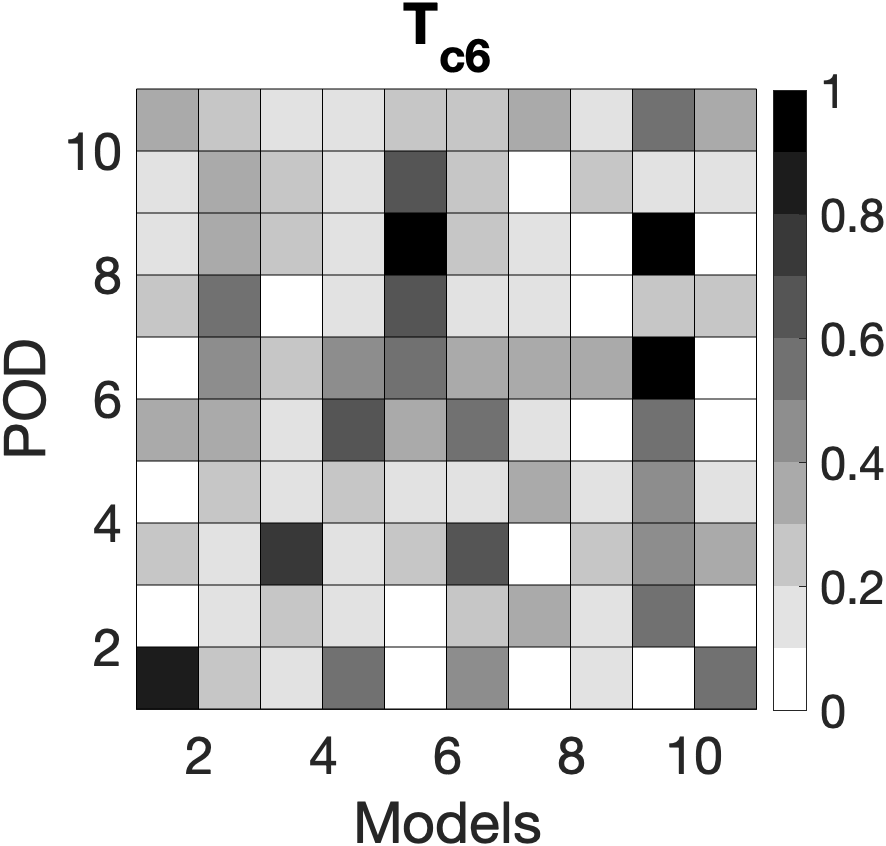}\\

\includegraphics[width=55mm]{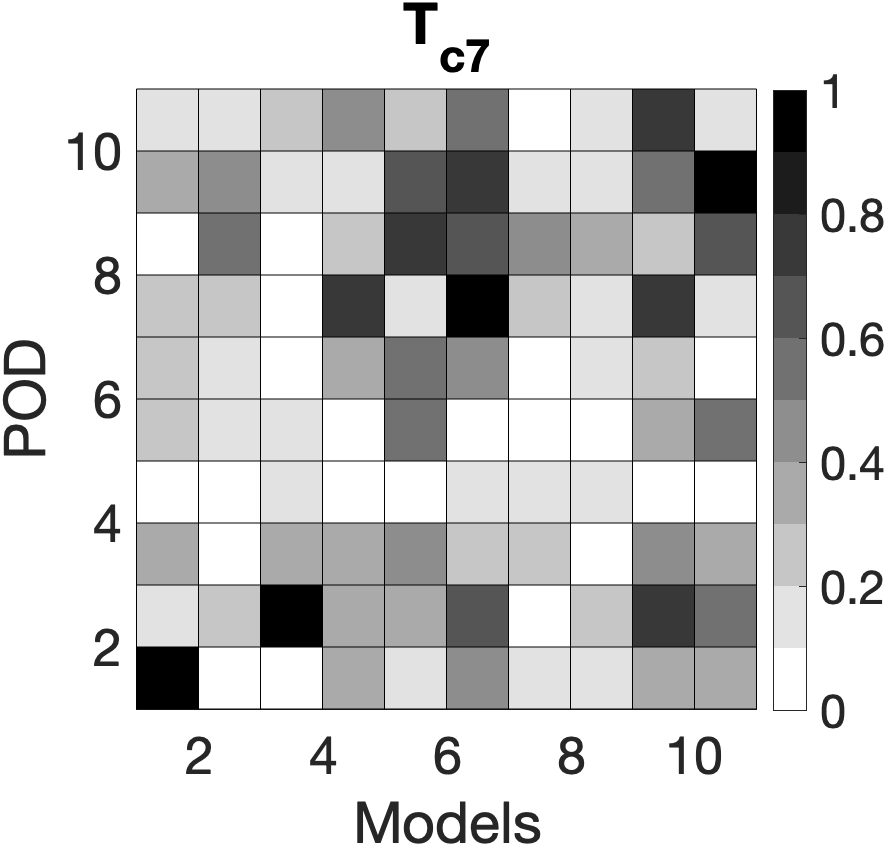}&
\hspace{1.5em}
\includegraphics[width=55mm]{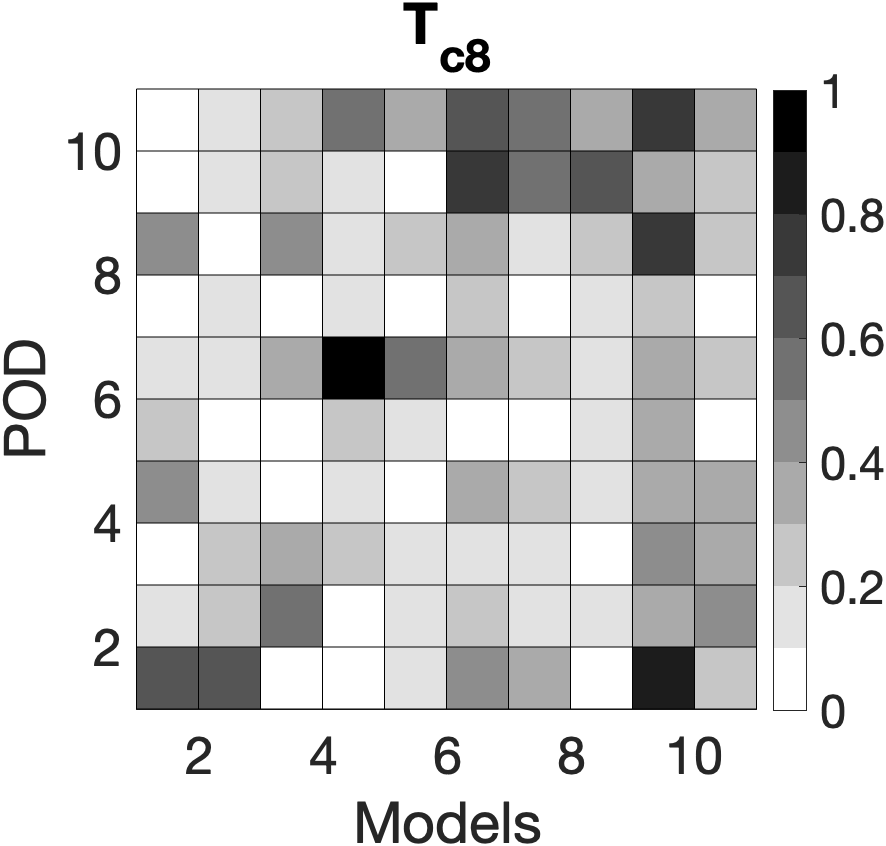}&
\hspace{1.5em}
\includegraphics[width=55mm]{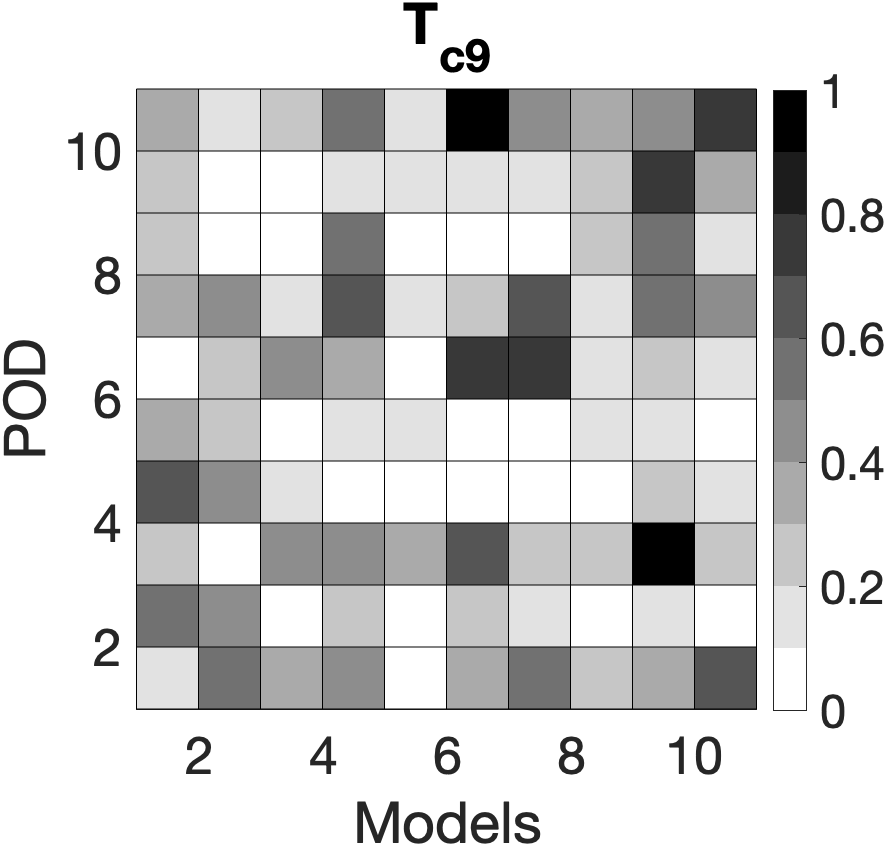}\\

\includegraphics[width=50mm]{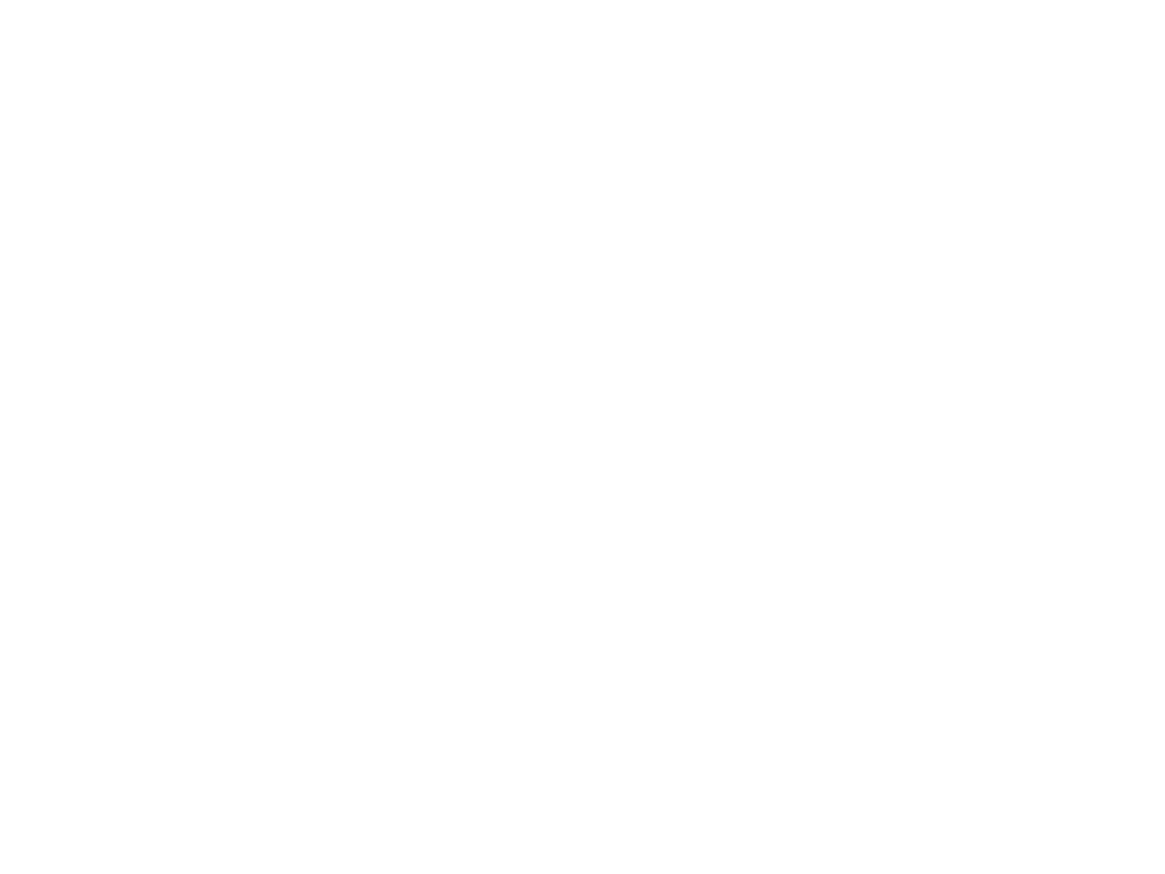}&
\hspace{1.5em}
\includegraphics[width=55mm]{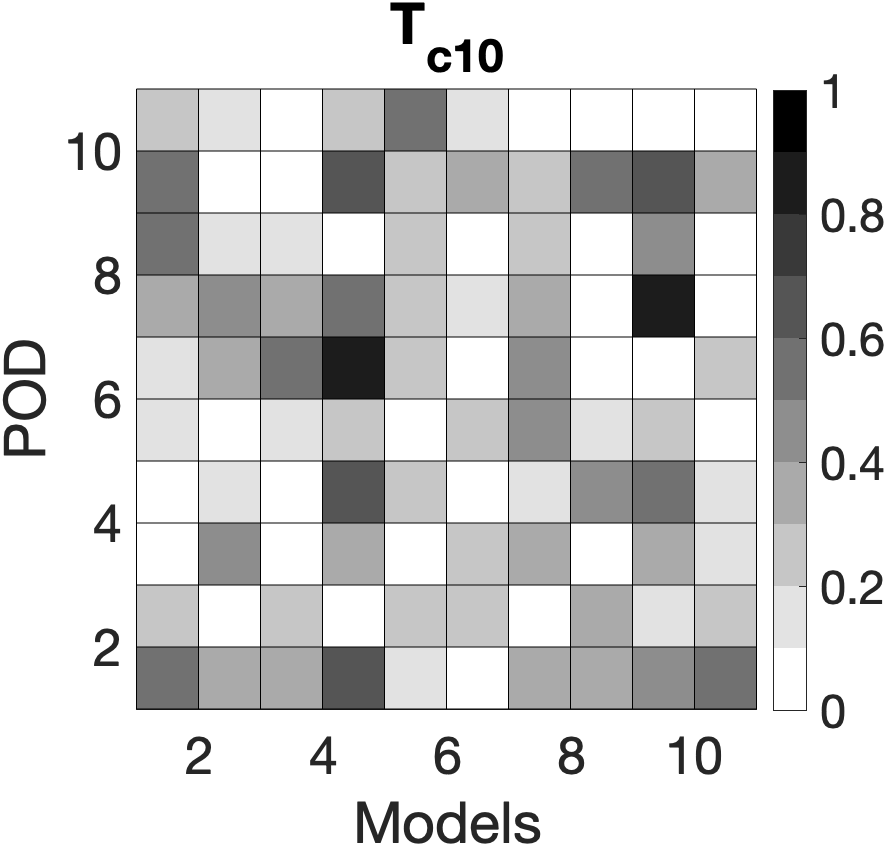}

\end{tabular}
  \caption{This figure shows the summation of pixels in the correlation matrix between POD modes in Figure \ref{fig:POD_C} and the theoretical models in Figure \ref{fig:model_circule}, for every T$_{ci}$. The colorbar shows the amplitude of the summation of the correlation matrix, while every column along the time intervals were normalised by the maximum.}
  \label{fig:Corr_stat C}

\end{figure}

\begin{figure}
\centering
\begin{tabular}{ccc}

\includegraphics[width=55mm]{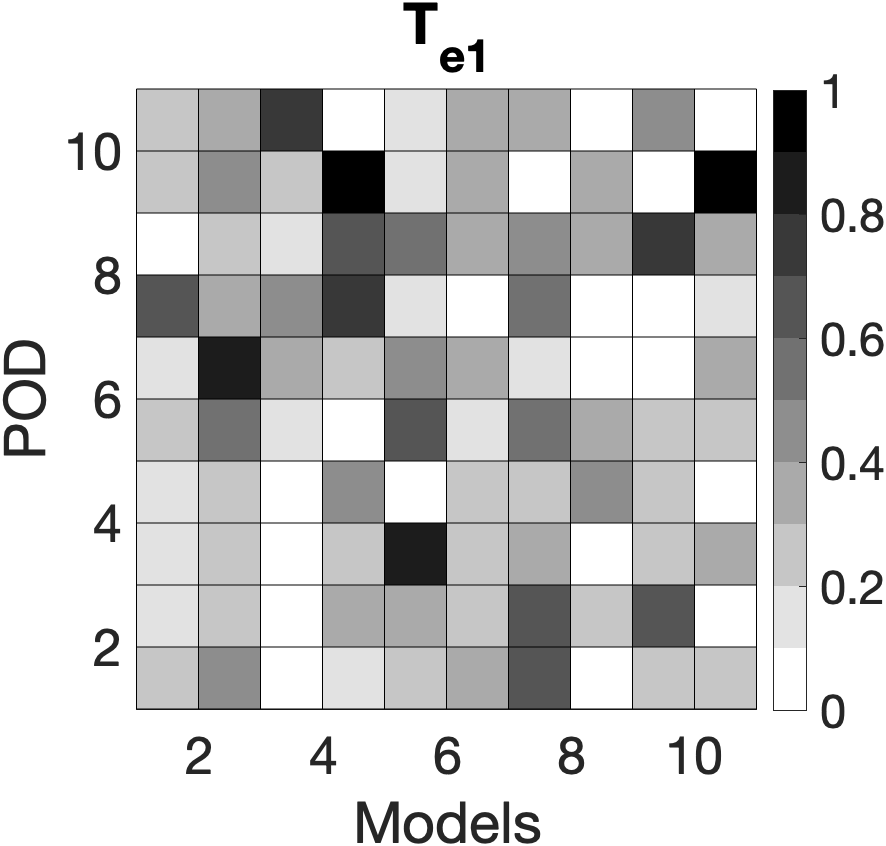}&
\hspace{1.5em}
\includegraphics[width=55mm]{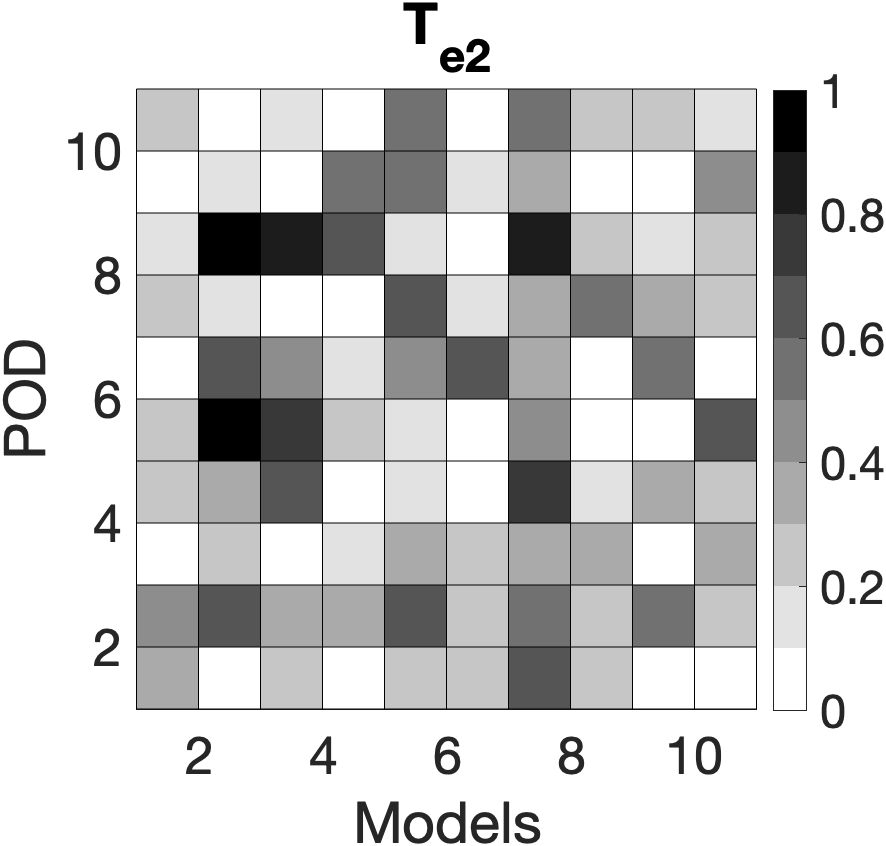}&
\hspace{1.5em}
\includegraphics[width=55mm]{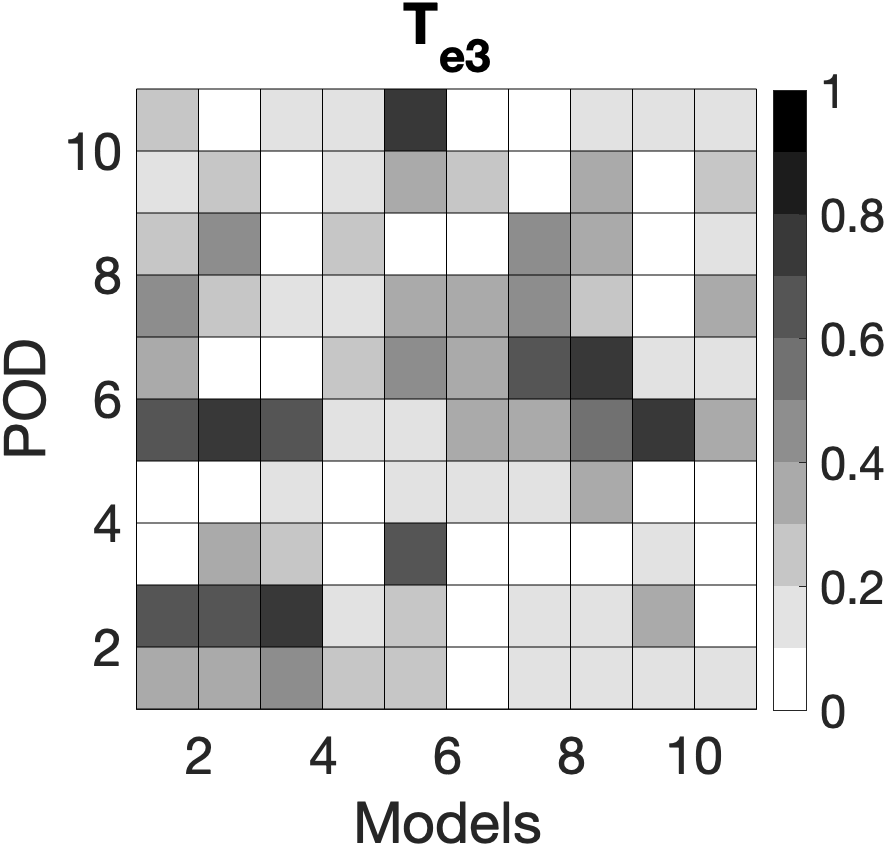}\\

\includegraphics[width=55mm]{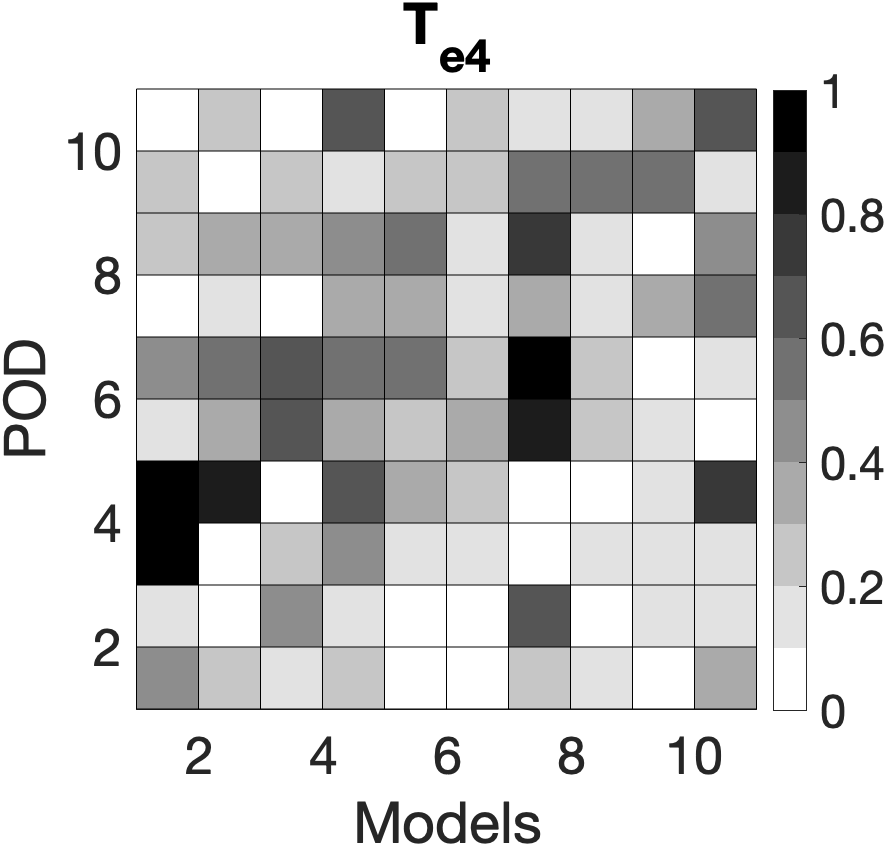}&
\hspace{1.5em}
\includegraphics[width=55mm]{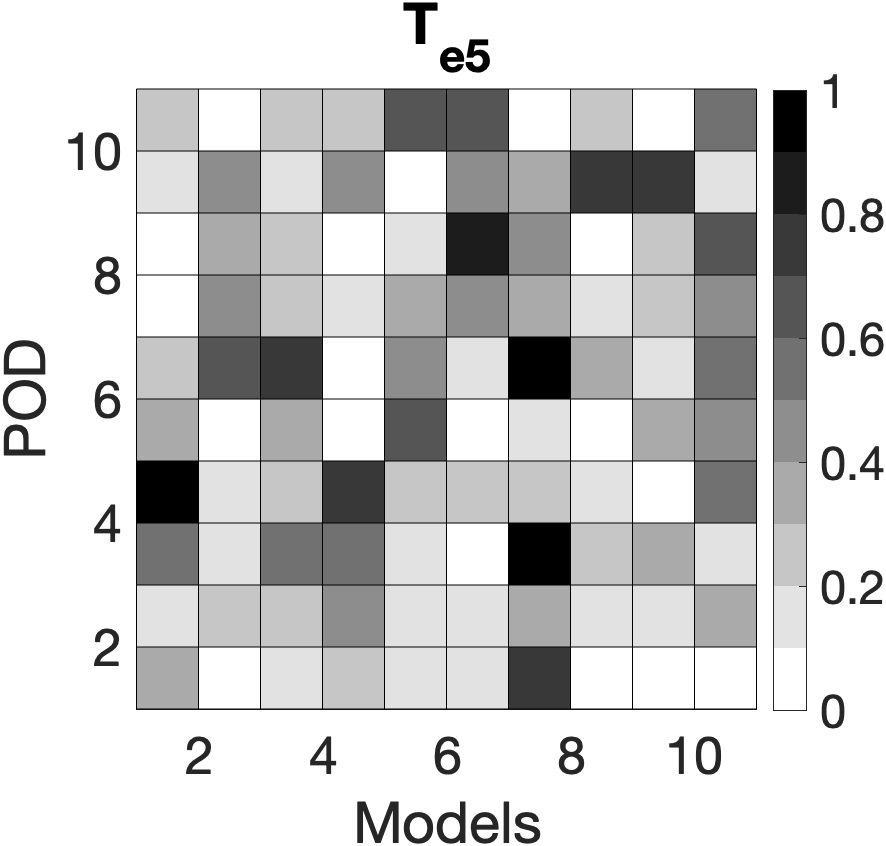}&
\hspace{1.5em}
\includegraphics[width=55mm]{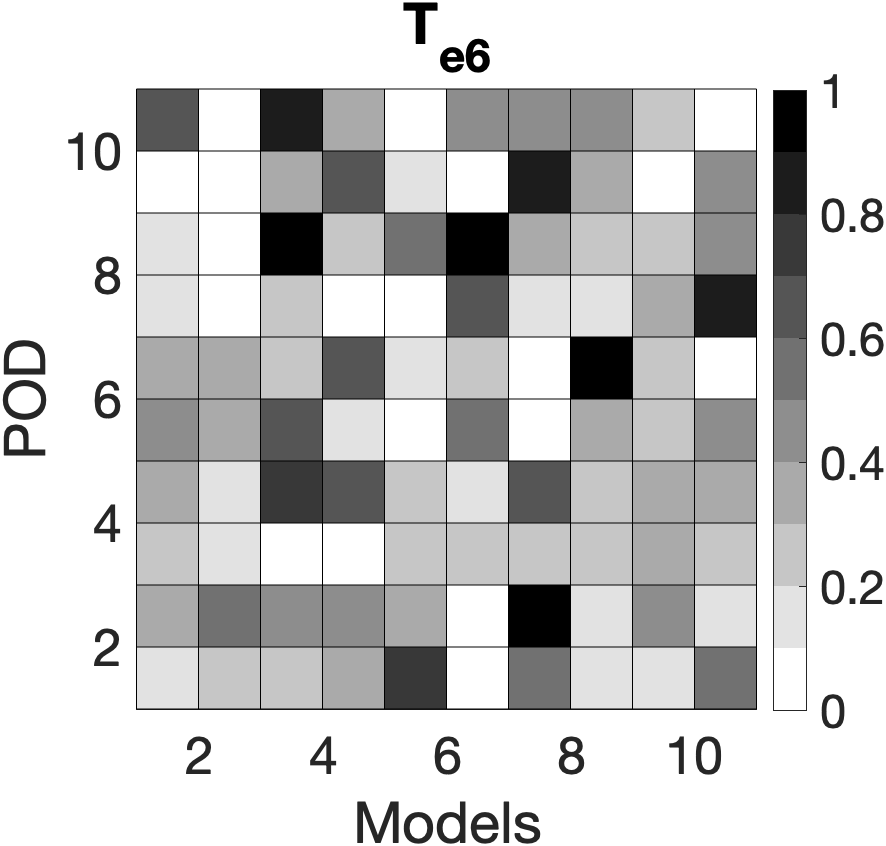}\\

\includegraphics[width=55mm]{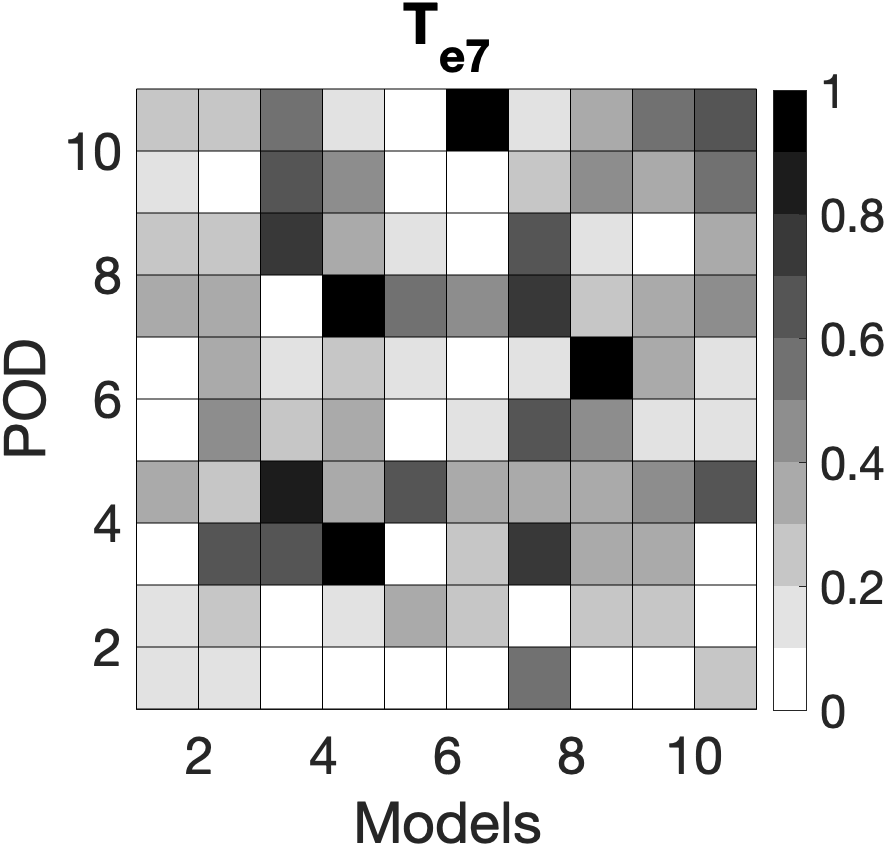}&
\hspace{1.5em}
\includegraphics[width=55mm]{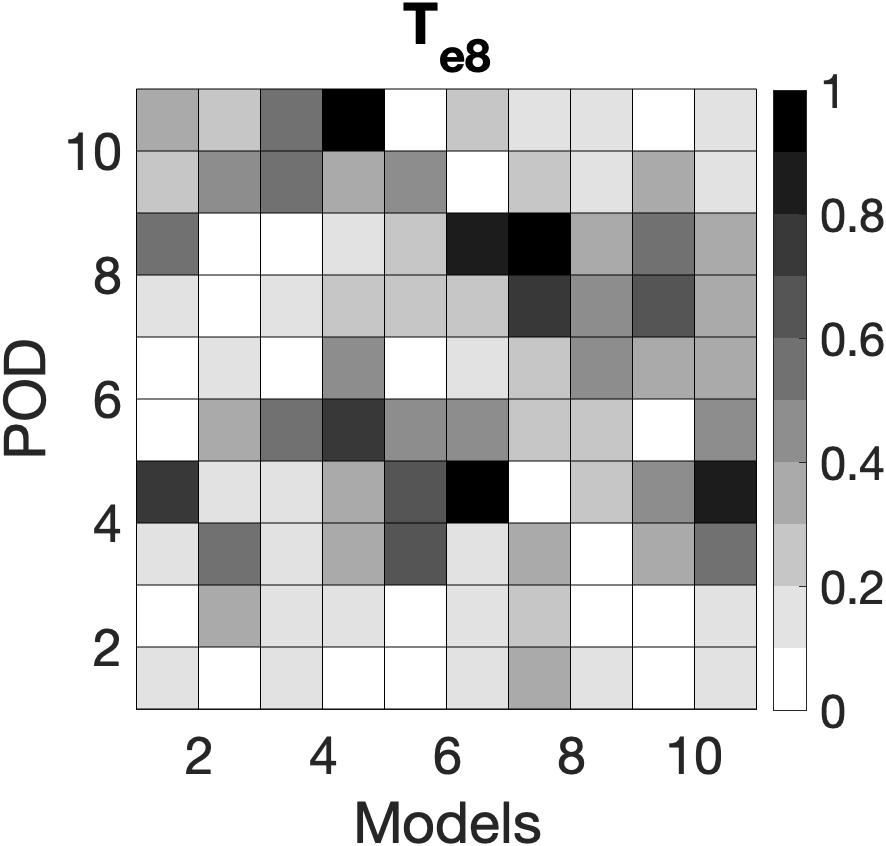}&
\hspace{1.5em}
\includegraphics[width=55mm]{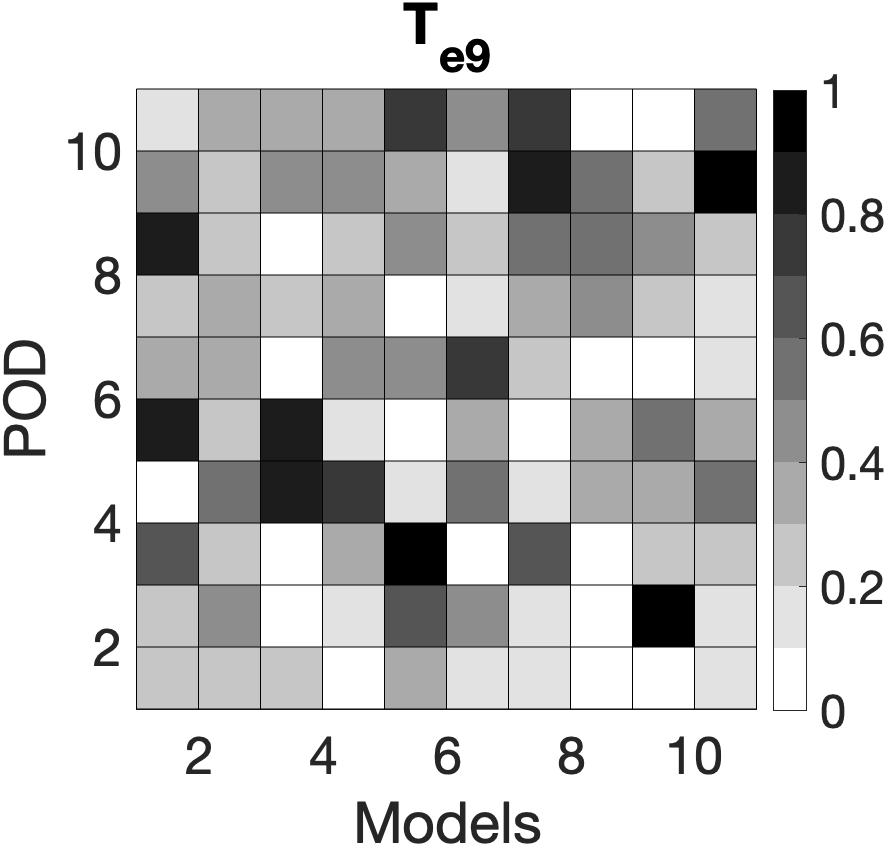}\\

\includegraphics[width=55mm]{C_blank_white.png}&
\hspace{1.5em}
\includegraphics[width=55mm]{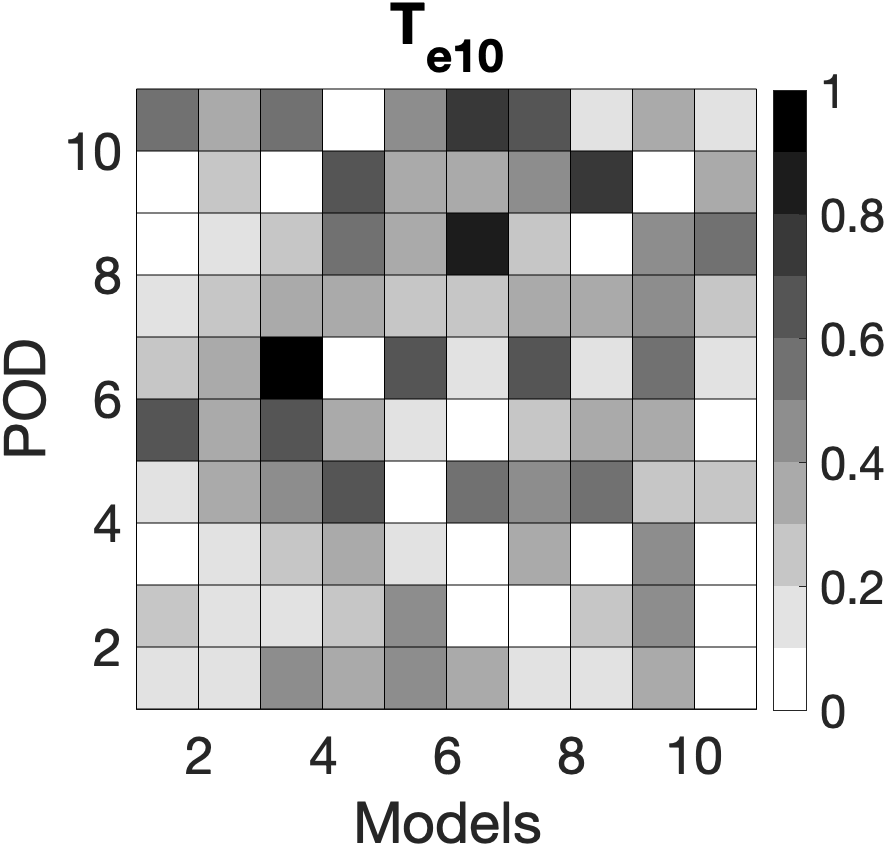}

\end{tabular}
  \caption{This figure shows the summation of pixels in the correlation matrix between POD modes in Figure \ref{fig:POD_E} and the theoretical models in Figure \ref{fig:model_elliptc}, for every T$_{ei}$. The colorbar shows the amplitude of the summation of the correlation matrix, while every column along the time intervals were normalised by the maximum.}
  \label{fig:Corr_stat E}
  
\end{figure}

Our analysis was also applied on different time intervals than presented in Figures \ref{fig:POD_C} and \ref{fig:POD_E}. This step was applied to insure that the width of the time intervals (37.5 minutes for the circular sunspot and 60 minutes for the elliptical sunspot) are working well and the results are robust. It is important to mention that, all time intervals (T$_{ci}$ and T$_{ei}$) that are mentioned in the text above, tables \ref{Table_C} - \ref{Table_E} and Figures \ref{fig: mean} - \ref{fig:TEE n=2 1} refer to that in Figures \ref{fig:POD_C} and \ref{fig:POD_E} for the circular and elliptical sunspot, respectively.

\begin{figure}
    \centering
    \includegraphics[width=16cm]{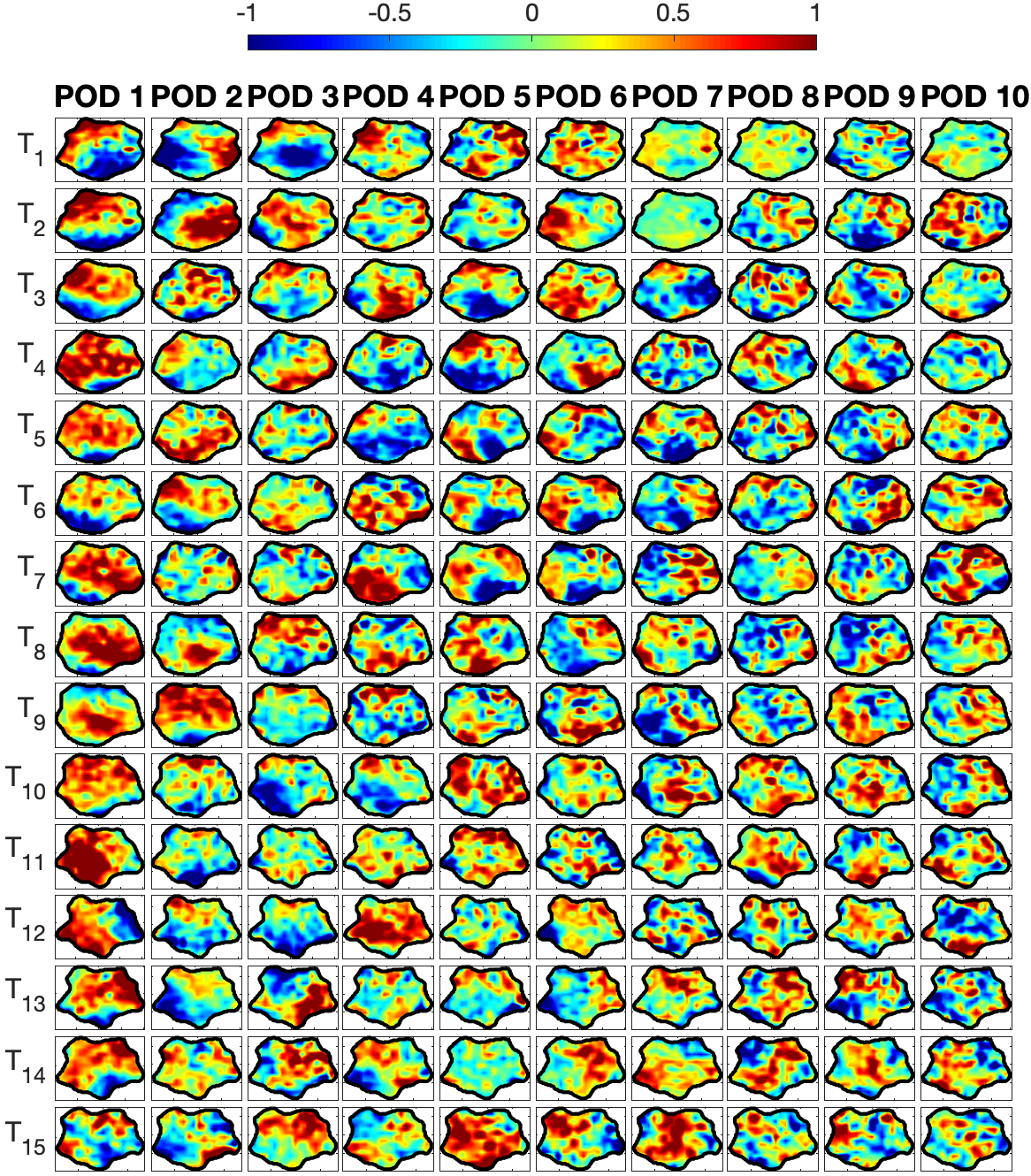}
     \caption{This figure represents the first 10 POD modes of the circular sunspot. Every column shows a POD mode and the rows shows how the modes are changing along the time intervals, T$_i$, of the data time series. Every time interval contains 30 images, and has a duration of 22.5 minutes. Every time interval is shifted by 15 images, corresponding to 11 minutes.}
     \label{fig:POD_C30}
\end{figure}

\begin{figure}
    \centering
     \includegraphics[width=160mm]{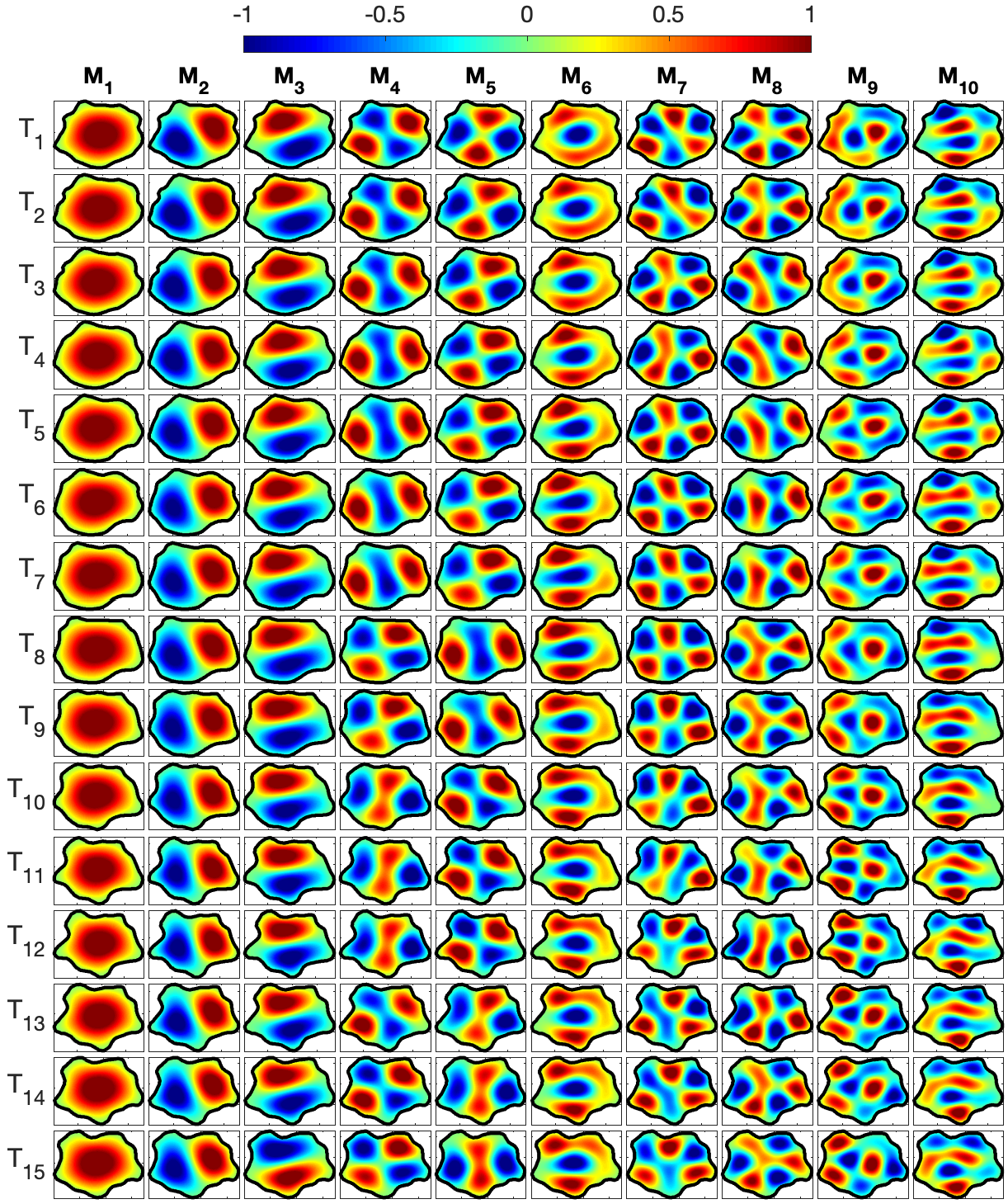}
     \caption{The theoretical eigenfunctions
     that correspond to the changing shapes of the observed circular sunspot (see Figure \ref{fig:POD_C30}). Every row shows the spatial structure of the models at different times and changing shape. Columns represent different types of slow body modes, and they are labelled by M$_i$, where $i=1, \dots, 10$. In particular, M$_1$ stands for the fundamental sausage, M$_2$ and M$_3$ denote the fundamental kink, M$_4$ and M$_5$ are showing the fluting ($n=2$), M$_6$ is showing the sausage overtone, M$_7$ and M$_8$ are showing the fluting ($n=3$) and the last two columns (M$_9$ and M$_{10}$) are showing the kink overtone.} \label{fig:Model_30C}
\end{figure}

\begin{figure}
    \centering
    \includegraphics[width=160mm]{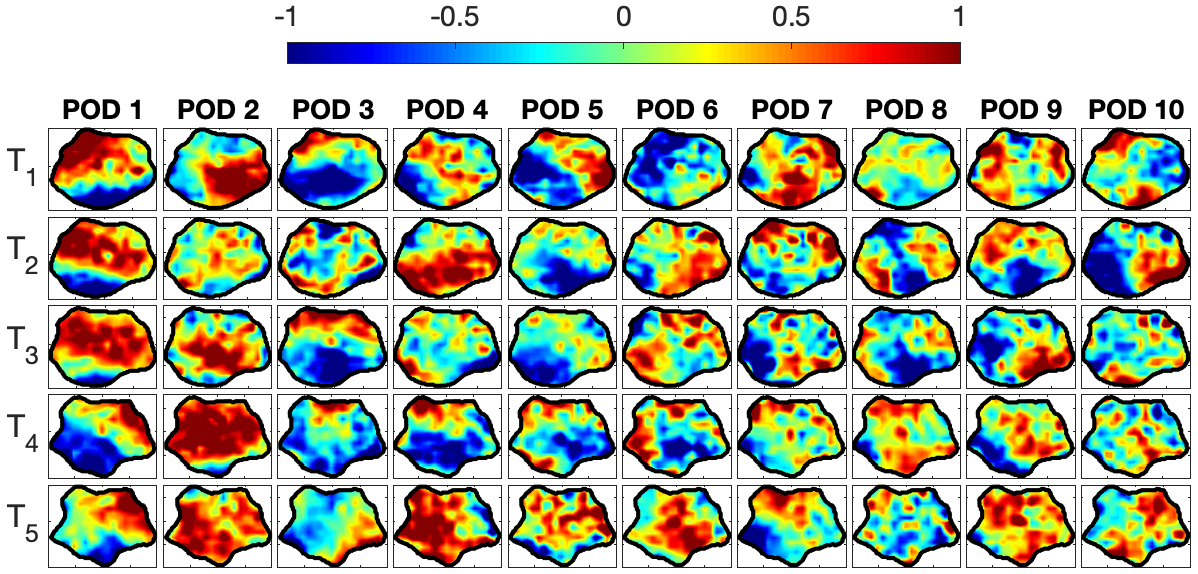}
     \caption{This figure represents the first 10 POD modes of the circular sunspot. Every column shows a POD mode and the rows shows how the modes are changing along the time intervals, T$_i$, of the data time series. Every time interval contains 80 images, and has a duration of 60 minutes. Every time interval is shifted by 40 images, corresponding to 30 minutes.}
     \label{fig:POD_C80}
\end{figure}

\begin{figure}
    \centering
    \includegraphics[width=160mm]{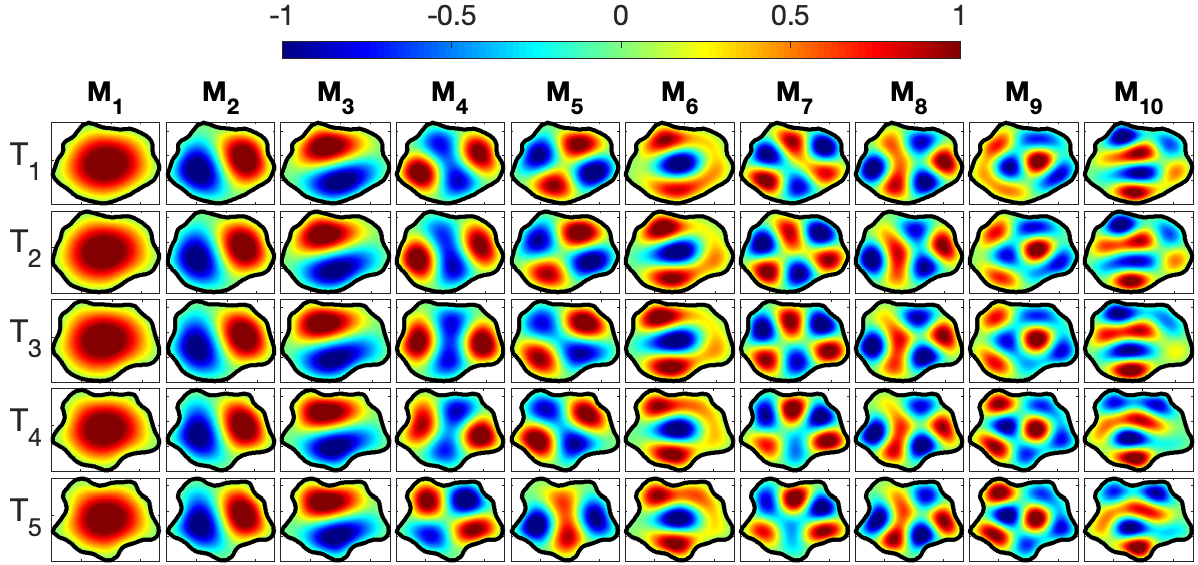}
     \caption{The theoretical eigenfunctions that correspond to the changing shapes of the observed circular sunspot (see Figure \ref{fig:POD_C80}). M$_1$ stands for the fundamental sausage, M$_2$ and M$_3$ denote the fundamental kink, M$_4$ and M$_5$ are showing the fluting ($n=2$), M$_6$ is showing the sausage overtone, M$_7$ and M$_8$ are showing the fluting ($n=3$) and the last two columns (M$_9$ and M$_{10}$) are showing the kink overtone.} \label{fig:Model_80S_C}
\end{figure}

For the circular sunspot, we have decreased the period of the time interval to be 22.5 minutes, with an overlapping time of 11 minutes and the results of the POD analysis of the time intervals is shown in Figure \ref{fig:POD_C30}, and the theoretical models that correspond to the shape are shown in Figure \ref{fig:Model_30C}. In addition, we have increased the period of the intervals to be 60 minutes with overlapping time of 30 minutes and the spatial structure of the POD analysis is shown in Figure \ref{fig:POD_C80}, and the theoretical models that correspond to the actual shape are shown in Figure \ref{fig:Model_80S_C}.

\begin{figure}
    \centering
    \includegraphics[width=18cm]{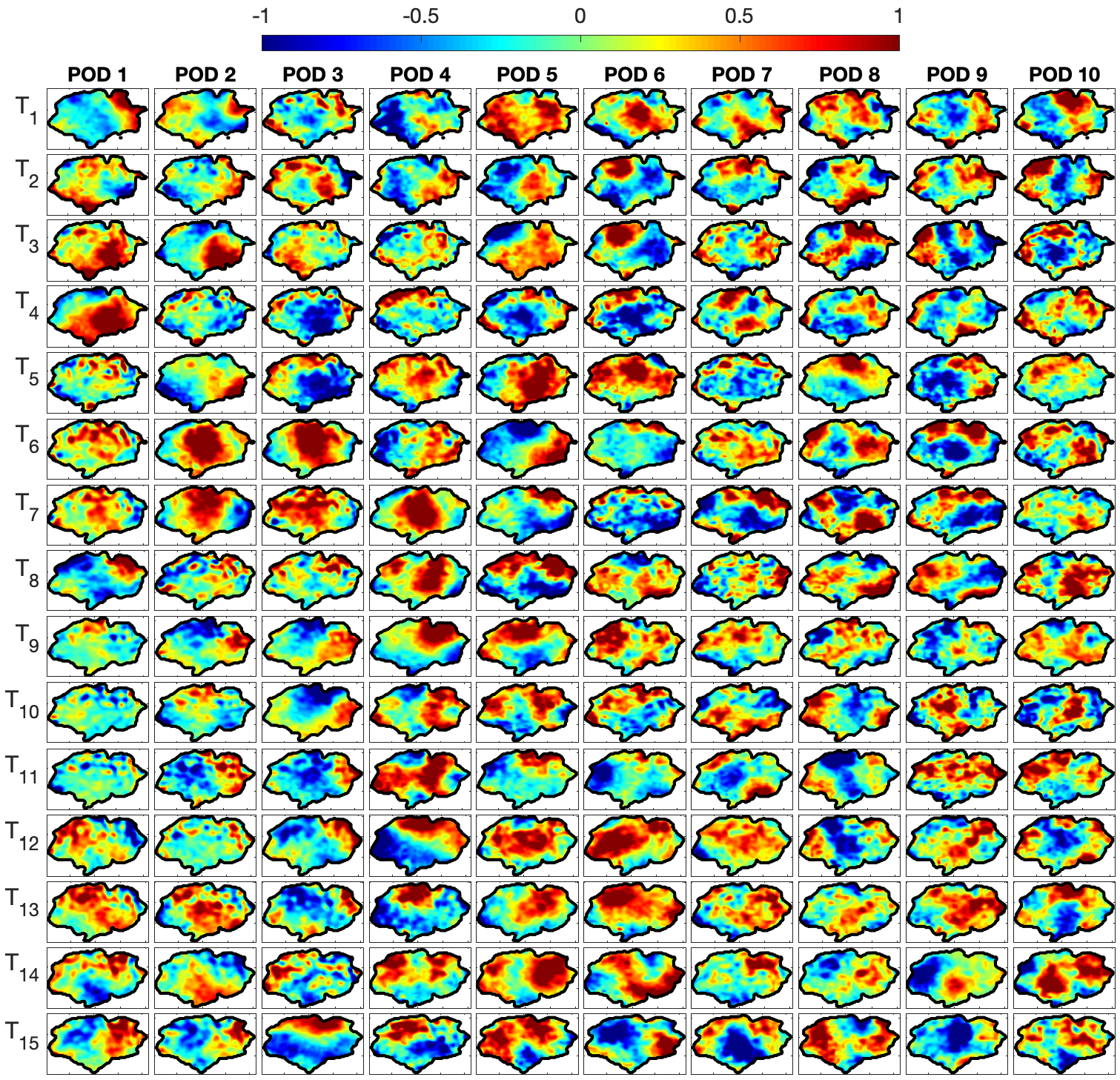}
     \caption{This figure represents the first 10 POD modes of the elliptical sunspot. Every column shows a POD mode and the rows shows how the modes are changing along the time intervals, T$_i$, of the data time series. Every time interval contains 50 images, and has a duration of 37.5 minutes. Every time interval is shifted by 30 images, corresponding to 22.5 minutes.}
     \label{fig:POD_E50}
\end{figure}

\begin{figure}
    \centering
    \includegraphics[width=18cm]{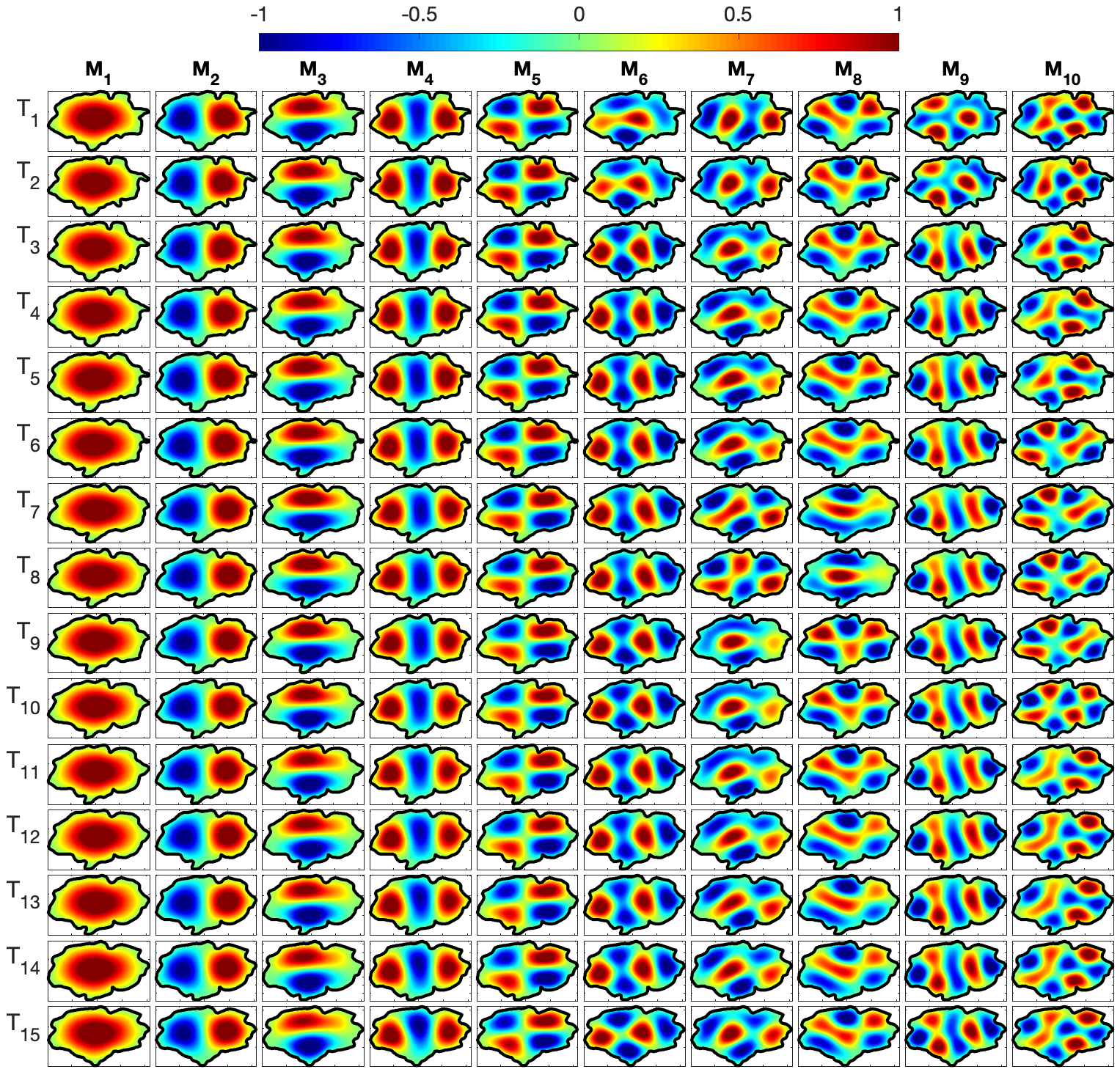}
     \caption{The theoretical eigenfunctions that correspond to the changing shapes of the observed elliptical sunspot (see Figure \ref{fig:POD_E50}). M$_1$ stands for the fundamental sausage, M$_2$ and M$_3$ denote the fundamental kink, M$_4$ and M$_5$ are showing the fluting ($n=2$), M$_6$ is showing the sausage overtone, M$_7$ and M$_8$ are showing the fluting ($n=3$) and the last two columns (M$_9$ and M$_{10}$) are showing the kink overtone.}  \label{fig:modles_E50}
\end{figure}

\begin{figure}
    \centering
    \includegraphics[width=160mm]{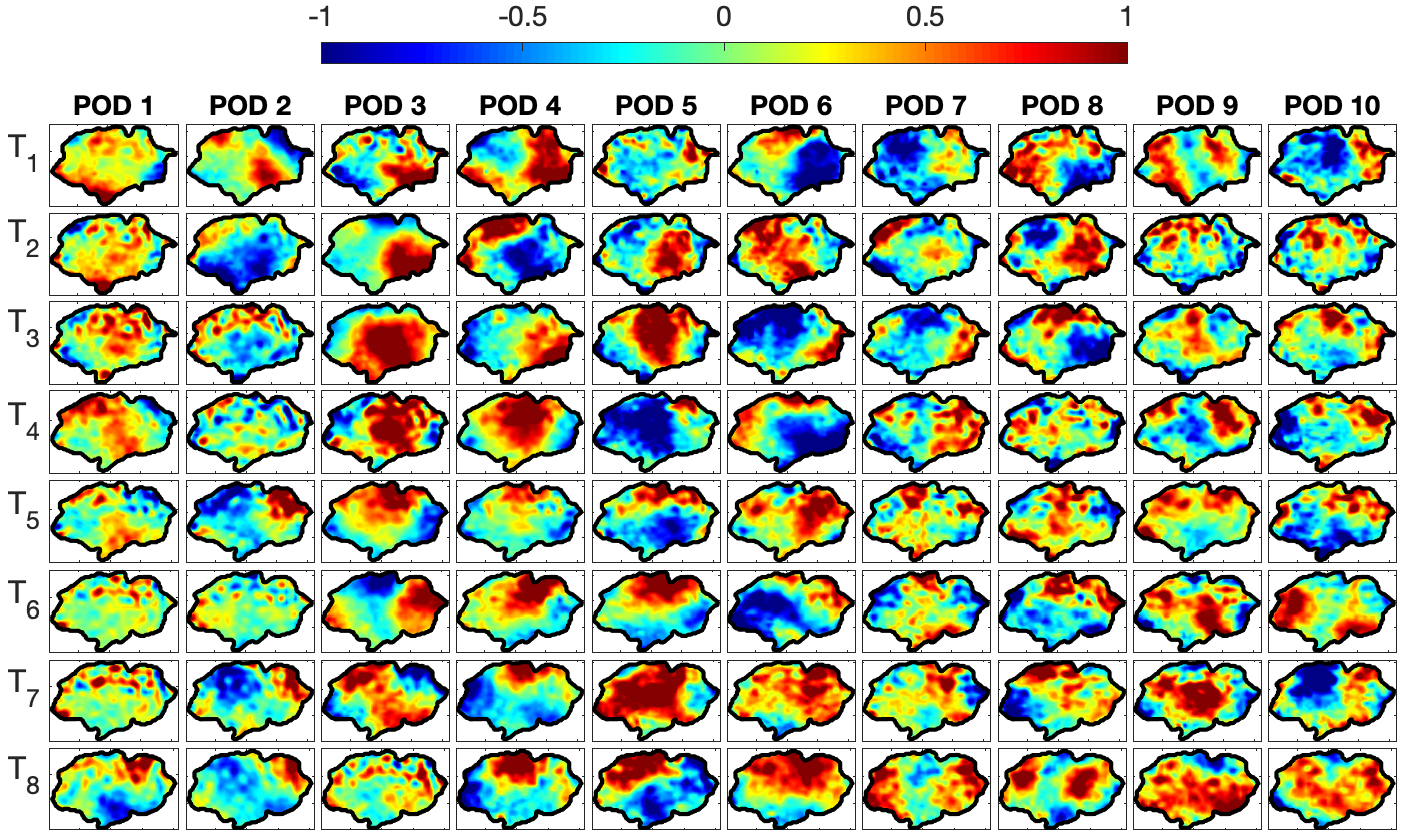}
     \caption{This figure represents the first 10 POD modes of the elliptical sunspot. Every column shows a POD mode and the rows shows how the modes are changing along the time intervals, T$_i$, of the data time series. Every time interval contains 100 images, and has a duration of 75 minutes. Every time interval is shifted by 50 images, corresponding to 37.5 minutes.}
     \label{fig:POD_E100}
\end{figure}

\begin{figure}
    \centering
    \includegraphics[width=160mm]{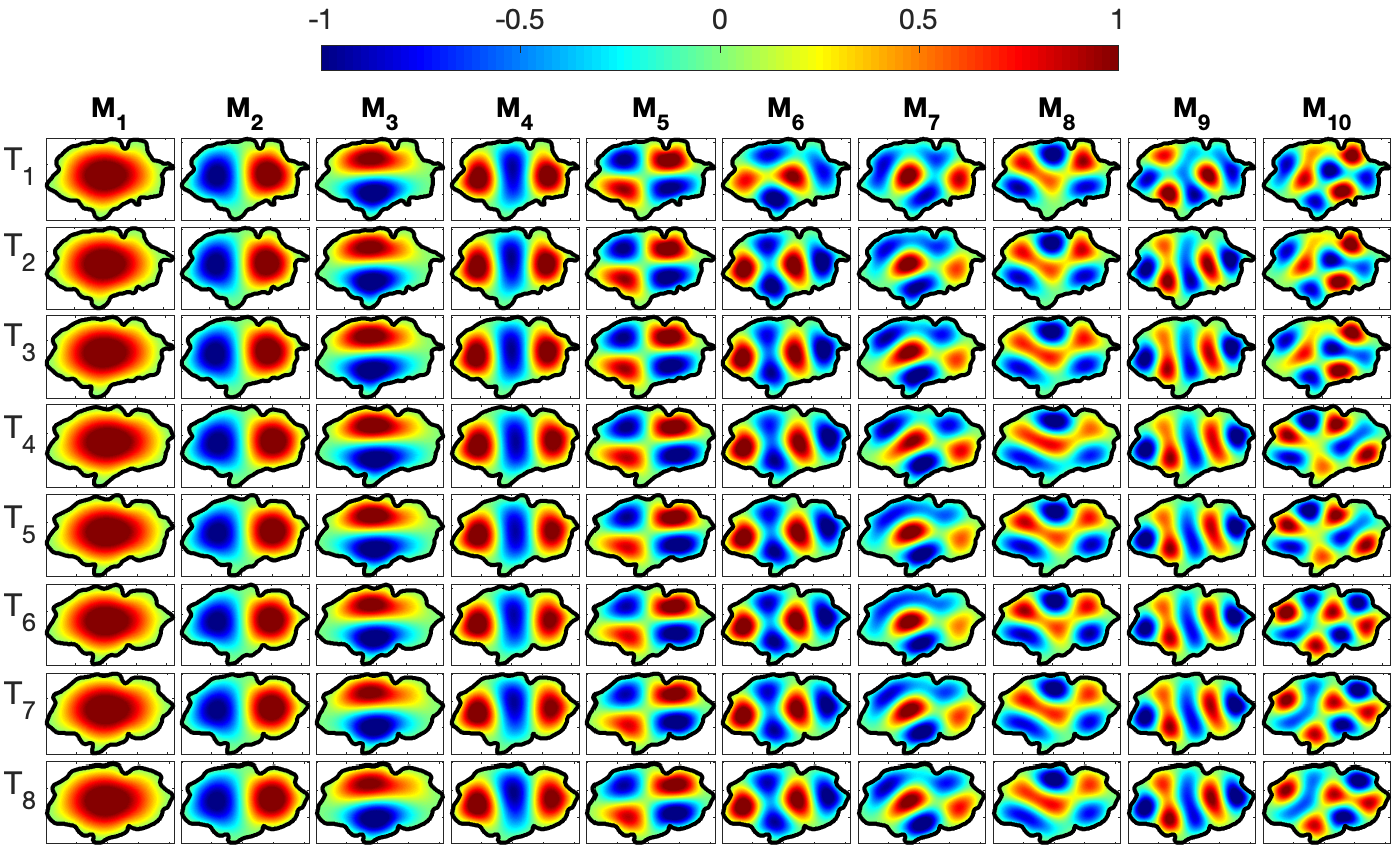}
     \caption{The theoretical eigenfunctions that correspond to the changing shapes of the observed elliptical sunspot (see Figure \ref{fig:POD_E100}). M$_1$ stands for the fundamental sausage, M$_2$ and M$_3$ denote the fundamental kink, M$_4$ and M$_5$ are showing the fluting ($n=2$), M$_6$ is showing the sausage overtone, M$_7$ and M$_8$ are showing the fluting ($n=3$) and the last two columns (M$_9$ and M$_{10}$) are showing the kink overtone.}  \label{fig:modle_E100}
\end{figure}

Similarly, for the elliptical sunspot, we have decreased the period of the time intervals to be 37.5 minutes, with an overlapping time of 22.5 minutes and the result of the POD and the theoretical models are shown in Figures \ref{fig:POD_E50} and \ref{fig:modles_E50}, respectively. We have also increased the period of the time intervals to be 75 minutes, with an overlapping time of 37.5 minutes and the result of the POD and the theoretical models are shown in Figures \ref{fig:POD_E100} and \ref{fig:modle_E100}, respectively. A different time sequencing, obviously, will modify the complexity of the recovered signal, however the patterns we obtained in the case of our investigation still persist. Shorter time sequences used for POD/DMD analysis here provide more precise information on mode changes and their individual contribution to the signal identified in the larger time sequence. For example, it is clearly visible in the first columns of Figures (\ref{fig:POD_E}) and (\ref{fig:POD_E50}) for the time intervals $T_{e1}- T_{e3}$ and $T_1 -T_6$, correspondingly.  

\begin{table}
\begin{center}
\begin{tabular}{c c c c c c}
\hline

\hline
  M$_i$   &  T$_{ci}$   &  POD & peaks of PSD [mHz]  &   DMD [mHz] & MHD mode observed \\
\noalign{\smallskip}
\hline
\hline

M$_1$ & T$_{c3}$ & 4 & 3.5 and 4.4 & 3 & FSBS ($n=0$) \\
M$_1$ & T$_{c4}$ & 4 & 3.5 and 4.4 & 3.4 & FSBS ($n=0$) \\
M$_1$ & T$_{c5}$ & 1 & 3.5 and 4.8 & 3.6 & FSBS ($n=0$)\\
M$_1$ & T$_{c6}$ & 1 & 3.1 and 4.8 & 3.04 & FSBS ($n=0$)\\
M$_1$ & T$_{c7}$ & 1 & 3.5 and 4.4 & 4.3 & FSBS ($n=0$)\\
\hline
M$_2$ & T$_{c1}$ & 2 & 4 & 3.1  & FSK ($n=1$) (I)\\
M$_3$ & T$_{c1}$ & 1 & 3.5 & 3.4  & FSK ($n=1$) (II)\\
M$_3$ & T$_{c2}$ & 1 & 3.5 & 3.2  & FSK ($n=1$) (II)\\
M$_3$ & T$_{c3}$ & 1 & 2.6 and 4 & 4.22  & FSK ($n=1$) (II)\\
M$_3$ & T$_{c5}$ & 3 & 3.1 and 4.4 & 4.23  & FSK ($n=1$) (II)\\
M$_3$ & T$_{c6}$ & 3 & 3.1 and 4.4 & 3.44  & FSK ($n=1$) (II)\\
M$_3$ & T$_{c7}$ & 2 & 3.1 and 4.4 & 3.9  & FSK ($n=1$) (II)\\
\hline
M$_4$ & T$_{c7}$ & 7 & 3.5 and 4.8 & 3 & SBF ($n=2$) (I)\\
M$_4$ & T$_{c8}$ & 6 & 3.5 and 4.4 & 3.17 & SBF ($n=2$) (I)\\
M$_5$ & T$_{c6}$ & 8 & 3.5  & 4 & SBF ($n=2$) (II)\\

\hline

\hline
\end{tabular}\\
 \caption{A summary of all possible MHD modes that observed in the circular sunspot in the selected time intervals. The first column represents the theoretical mode and they are labelled according to the  Figure \ref{fig:model_circule}. The second column shows the time interval of the sub-data in which the mode was observed. The POD mode numbers are presented are displayed in the third column, similar to Figure \ref{fig:POD_C}. The fourth column contains the  frequencies (in mHz) corresponding the peaks in the power spectrum density (PSD) of the time coefficient of the POD mode. The fifth column shows the frequency (in mHz) corresponding to the DMD mode. Finally, the last column display the MHD wave mode that the POD mode and DMD mode have a good agreement with, while (FSBS) is a notation for fundamental slow body sausage ($n=0$) mode, (FSK) fast surface kink ($n=1$) mode and (SBF) slow body fluting ($n=2$) mode. Here, (I) and (II) refer to two different (perpendicular) directions of the wave polarisation, and $n$ refer to the number of nodes along the azimuthal direction.}
 \label{Table_C}
\end{center}
\end{table}


\begin{table}
\begin{center}
\begin{tabular}{c c c c c c}
\hline

\hline
 M$_i$   &  T$_{ei}$   &  POD & peaks of PSD [mHz]  &   DMD [mHz] & MHD mode observed \\
\noalign{\smallskip}
\hline
\hline

M$_1$    & T$_{e1}$    & 7     & 3.6       & 4     & FSBS ($n=0$)\\
M$_1$    & T$_{e4}$    & 3     & 3.6       & 3.4    & FSBS ($n=0$)\\
M$_1$    & T$_{e5}$    & 4     & 3.3       & 3.8     & FSBS ($n=0$)\\
M$_1$    & T$_{e8}$    & 8     & 3.3       & 3     & FSBS ($n=0$)\\
M$_1$    & T$_{e9}$    & 8     & 3.6 and 4.1       & 3.3      & FSBS ($n=0$)\\
M$_1$    & T$_{e10}$    & 5     & 3.3       & 3.4      & FSBS ($n=0$)\\

\hline

M$_2$    & T$_{e1}$    & 6     & 3.6 and 4.1      & 3.3 & FSBK ($n=1$) (I)\\
M$_2$    & T$_{e2}$    & 8     & 3.6             & 3.3 & FSBK ($n=1$) (I)\\
M$_2$    & T$_{e4}$    & 6     & 3 and 3.8             & 3.4 & FSK ($n=1$) (I)\\

\hline
M$_3$    & T$_{e2}$    & 5     & 3.6             & 3.1 & FSK ($n=1$) (II)\\
M$_3$    & T$_{e3}$    & 5     & 3.6             & 3.3 & FSBK ($n=1$) (II)\\
M$_3$    & T$_{e5}$    & 6     & 3.3 and 3.8     & 3.5 & FSK ($n=1$) (II)\\
M$_3$    & T$_{e6}$    & 8     & 2.8 and 3.8     & 2.8 & FSK ($n=1$) (II)\\
M$_3$    & T$_{e9}$    & 4     & 3.3             & 3.6 & FSK ($n=1$) (II)\\
\hline
M$_4$    & T$_{e1}$     & 9     & 3 and 3.6   & 2.8 & SBF ($n=2$) (I)\\
M$_4$    & T$_{e7}$     & 3     & 3 and 3.6   & 2.86 & SBF ($n=2$) (I)\\
M$_4$    & T$_{e8}$     & 10    & 3 and 3.6   & 2.59 & SBF ($n=2$) (I)\\
M$_4$    & T$_{e10}$  & 9     & 3.6         & 2.8 & SBF ($n=2$) (I)\\
\hline
M$_5$    & T$_{e3}$  & 10     & 4.4        & 4.6 & SBF ($n=2$) (II)\\
M$_5$    & T$_{e9}$  & 10     & 4.1        & 3.8 & SBF ($n=2$) (II)\\

\hline

\hline
\end{tabular}
 \caption{This table displays all possible MHD wave modes that observed in the elliptical sunspot along the time intervals. The first column represents the theoretical mode and they are labelled as they are in Figure \ref{fig:model_elliptc}. The second column shows the time interval of the sub-data that the mode was observed. In the third column, the POD mode numbers are presented and displayed as they are in Figure \ref{fig:POD_E}. While, the fourth column gives information about the corresponding frequencies (in mHz) to the pecks in the power spectrum density (PSD) of the time coefficient of the POD mode. The fifth column shows the frequency (in mHz) that corresponding to the DMD mode. The last column display the MHD wave mode that the POD mode and DMD mode have a good agreement with, while (FSBS) is a notation for fundamental slow body sausage ($n=0$) mode, (FSBK)fundamental slow body kink ($n=1$) mode, (FSK) fast surface kink mode and (SBF) slow body fluting ($n=2$) mode. Here, (I) and (II) refer to two different (perpendicular) directions of the wave polarisation, and $n$ refer to the number of nodes along the azimuthal direction.}
 \label{Table_E}
\end{center}
\end{table}

\bibliography{ApJ}{}
\bibliographystyle{aasjournal}

\end{document}